\documentclass[aps, pra, nofootinbib, onecolumn, 11pt, tightenlines, notitlepage, superscriptaddress, longbibliography]{revtex4-1}
\usepackage[letterpaper, total={6in, 9in}]{geometry}
\usepackage{amssymb,mathtools}
\usepackage{amsmath,dsfont,physics}
\usepackage{mathrsfs}
\usepackage{graphicx}
\usepackage{dcolumn}
\usepackage[colorlinks=true,allcolors=blue]{hyperref}
\usepackage{siunitx}
\usepackage{verbatim}
\usepackage{natbib}
\usepackage{cleveref}
\usepackage{float}
\usepackage{color,xcolor}
\usepackage[utf8]{inputenc}
\usepackage[english]{babel}
\usepackage{amsthm}
\usepackage{appendix}
\usepackage{braket}

\begin{document}

\title{Optimizing lateral quantum dot geometries for reduced exchange noise}

\author{Brandon Buonacorsi}
\affiliation{Institute for Quantum Computing, University of Waterloo, Waterloo, Ontario N2L 3G1, Canada}
\affiliation{Waterloo Institute for Nanotechnology, University of Waterloo, Waterloo, Ontario N2L 3G1, Canada}
\affiliation{Department of Physics and Astronomy, University of Waterloo, Waterloo, Ontario N2L 3G1, Canada}

\author{Marek Korkusinski}
\affiliation{Quantum Theory Group, Security and Disruptive Technologies, National Research Council, Ottawa, Ontario K1A 0R6, Canada}

\author{Bohdan Khromets}
\affiliation{Institute for Quantum Computing, University of Waterloo, Waterloo, Ontario N2L 3G1, Canada}
\affiliation{Waterloo Institute for Nanotechnology, University of Waterloo, Waterloo, Ontario N2L 3G1, Canada}
\affiliation{Department of Physics and Astronomy, University of Waterloo, Waterloo, Ontario N2L 3G1, Canada}

\author{Jonathan Baugh}
\affiliation{Institute for Quantum Computing, University of Waterloo, Waterloo, Ontario N2L 3G1, Canada}
\affiliation{Waterloo Institute for Nanotechnology, University of Waterloo, Waterloo, Ontario N2L 3G1, Canada}
\affiliation{Department of Chemistry, University of Waterloo, Waterloo, Ontario N2L 3G1, Canada}

\begin{abstract}
For electron spin qubits in quantum dots, reducing charge noise sensitivity is a critical step in achieving fault tolerant two-qubit gates mediated by the exchange interaction. This work explores how the physical device geometry affects the sensitivity of exchange to fluctuations in applied gate voltage and interdot bias due to charge noise. We present a modified linear combination of harmonic orbitals configuration interaction (LCHO-CI) method for calculating exchange energies that is applicable to general quantum dot networks. In the modified LCHO-CI approach, an orthogonal set of harmonic orbitals formed at the center of the dot network is used to approximate the many-electron states. This choice of basis significantly reduces the computation time of the full CI calculation by enabling a pre-calculated library of matrix elements to be used in evaluating the Coulomb integrals. The resultant many-electron spectra are mapped onto a Heisenberg Hamiltonian to determine the individual pairwise electronic exchange interaction strengths, $J_{ij}$. The accuracy of the modified LCHO-CI method is further improved by optimizing the choice of harmonic orbitals without significantly lengthening the calculation time. The modified LCHO-CI method is used to calculate $J$ for a silicon MOSFET double quantum dot occupied by two electrons. Two-dimensional potential landscapes are calculated from a 3D device structure, including both the Si/SiO$_2$ heterostructure and metal gate electrodes. The computational efficiency of the modified LCHO-CI method enables systematic tuning of the device parameters to determine their impact on the sensitivity of $J$ to charge noise, including plunger gate size, tunnel gate width, SiO$_2$ thickness and dot eccentricity. Generally, we find that geometries with larger dot charging energies, smaller plunger gate lever arms, and symmetric dots are less sensitive to noise. 
\end{abstract}

\maketitle

\section{Introduction}
\label{sec:intro}

Electron spin qubits in lateral, gate-defined quantum dots are a promising candidate for scalable quantum computing architectures \cite{veldhorst2014addressable,zajac2018resonantly,watson2018programmable,yoneda2018quantum,xue2019benchmarking,huang2019fidelity,sigillito2019coherent,nichol2017high,cerfontaine2019closed}.
%In these architectures, single spin rotations are driven by an on resonance microwave magnetic field \cite{} while two-spin rotations are realized through the spin exchange interaction \cite{}.
In arrays of singly-occupied quantum dots, the interdot Coulomb interaction together with fermionic statistics leads to the effective spin exchange interaction. The dynamics of exchange enable SWAP and $\sqrt{\rm SWAP}$ quantum logic gates, key ingredients for universal quantum control of spin qubits \cite{loss1998quantum, divincenzo2000universal}. In weak spin-orbit materials like silicon, exchange between spins labeled $i$ and $j$ can be described by an effective Heisenberg Hamiltonian \cite{divincenzo2000universal} with strength $J_{ij}$. The orbital wavefunctions, especially the overlap between orbitals in adjacent dots, can be manipulated via the gate electrodes that define the dots, so that $J_{ij}$ is determined by the set of applied gate voltages $\vec{V}$ (we use vector notation to indicate the set of voltages applied on $n$ gates, $\{V_1, V_2,...,V_n\}$). Electrostatic control of the exchange interaction in multi-dot systems is routine in lateral GaAs quantum dots \cite{petta2005coherent,nowack2011single,nichol2017high,kandel2019coherent}, and is becoming routine in silicon dots \cite{veldhorst2015two,watson2018programmable,zajac2018resonantly,huang2019fidelity,xue2019benchmarking} in both MOSFET and Si/SiGe devices. However, it remains a challenge to realize two-spin quantum gates with the high fidelities required for fault-tolerant quantum computing, mainly due to the sensitivity of exchange to charge noise \cite{dial2013charge, paladino20141}. The relationship between $J_{ij}$ and $\vec{V}$ for a dot network depends on $\vec{V}$ and the physical device geometry in ways that can be challenging to predict. Accurate numerical calculation of $J_{ij}(\vec{V})$ for arbitrary device layouts and network topologies is critical to realistic modeling of spin qubit processors. Such realistic modeling is necessary for optimizing device geometries and voltage control sequences that will mitigate the impacts of charge noise on the fidelity of quantum logic gates. 

In order to accurately model $J_{ij}(\vec{V})$, techniques borrowed from quantum chemistry must be used. Generally, methods for determining $J_{ij}$ offer a trade-off between computational complexity and quantitative accuracy when evaluating the energy spectra of many-electron systems. Approximate computational methods including Hubbard \cite{korkusinski2007topological,deng2018negative}, Heitler-London \cite{burkard1999coupled,calderon2006exchange}, and Hund-M\"{u}lliken \cite{hu2000hilbert,van2006semiconductor,hatano2008manipulation} techniques use only the localized, lowest energy $s$-~orbitals to construct the many-electron state. However, these approaches are only accurate over a limited range of device parameters \cite{pedersen2007failure} and fail more easily in Si compared to GaAs, due to the larger effective mass in Si \cite{li2010exchange}. Exact diagonalization of the many-electron Hamiltonian using a full configuration interaction (CI) formalism \cite{delgado2007theory, shim2008gate, hsieh2010quantum, nielsen2010implications, barnes2011screening, deng2020interplay} produces more accurate modeling. In a full CI calculation, the many-electron basis set is constructed by including all configurations of the $s$-, $p$-, $d$-, $f$-, etc., orbital states for the Hamiltonian diagonalization. A convergent spectrum requires a sufficient number of these excited orbital states. The full-CI method is applicable, in principle, to any quantum dot network; however, these calculations are computationally intensive due to the need to evaluate Coulomb matrix elements for all configurations. Calculating the dependence of exchange strength $J_{ij}$ on varying device parameters, such as gate voltages and device layout, tends to be impractical for large parameter spaces. 

Methods for determining $J_{ij}$ that are both computationally efficient and numerically accurate are key to designing devices with improved robustness to charge noise. Such optimization would complement standard charge noise reduction techniques such as dynamical decoupling \cite{buterakos2018crosstalk}, composite pulses \cite{setiawan2014robust,wang2014robust,zhang2017randomized}, and symmetric point operation \cite{reed2016reduced,martins2016noise,yang2017suppression}.
Furthermore, quantum optimal control techniques such as GRAPE \cite{khaneja2005optimal,yang2019silicon} and effective Hamiltonian engineering \cite{haas2019engineering}, which require many repeated estimations of $\nabla J_{ij}(\vec{V})$, could be applied to exchange operations in quantum dots if efficient CI computations were available. Finally, the dynamics of coupled spins could be simulated directly in terms of the applied gate voltages using such tools. 

Section~\ref{sec:mathMethods} of this paper presents a modification of the linear combination of harmonic orbitals and configuration interaction (LCHO-CI) approach introduced by Gimenez \textit{et al.}~\cite{gimenez2007linear}.
The modified LCHO-CI provides calculations of $J_{ij}$ with significantly improved efficiency, while retaining quantitative accuracy. In Section~\ref{sec:SE_states}, we show how to construct single electron states in a quantum dot network using a large orthogonal basis of harmonic orbitals. Section~\ref{sec:manyelectronHam} describes the full CI calculation that accounts for all electron-electron correlations in the system. By using an orthogonal basis of harmonic orbitals, evaluation of the Coulomb matrix elements (the most computationally intensive part of the calculation) is reduced to a scalar multiplication and subsequent basis transformation of a pre-calculated library of Coulomb matrix elements. This strategy significantly reduces the resources needed to evaluate the Coulomb interactions, reducing the total computation time of the LCHO-CI calculation. The resulting many-electron spectra found with the LCHO-CI method are then mapped to the effective Heisenberg Hamiltonian \cite{gimenez2007linear} to obtain $J_{ij}$ for the quantum dot network. Section~\ref{sec:optimalOmega} describes how the harmonic orbital basis can be optimized to improve the accuracy of the LCHO-CI calculations without increasing the computation time.

Section~\ref{sec:device_sims} uses the modified LCHO-CI method to study the sensitivity of $J$ with respect to charge noise in a realistic double quantum dot geometry. In Section~\ref{sec:device_model}, we introduce the full 3D device structure and model it with a self-consistent Poisson solver. This allows us to map out how the 2D electronic potential landscape varies with the geometric parameters of the physical gate layout, as well as the applied gate voltages.
In Section~\ref{sec:device_chargeNoise}, these 2D potentials are used to study how the sensitivity of $J$ to charge noise is influenced by the physical device parameters including dot size, tunnel gate width, gate oxide thickness and dot eccentricity. Sensitivity to charge noise is determined by calculating $\frac{\partial J}{\partial V_{bias}}$ as a function of the bias voltage $V_{\rm bias}$ applied across the plunger gates of the double dot. We also convert this bias to an effective inter-dot detuning $\epsilon$ to obtain $\frac{\partial J}{\partial \epsilon}$. Overall, the results confirm that dots with larger charging energies and smaller plunger gate lever arms show less sensitivity to charge noise. The tools developed here are applicable to to optimizing the design of quantum dot networks for robustness to charge noise. 

\section{Mathematical methods}
\label{sec:mathMethods}

In this section we outline a variation of the LCHO-CI approach \cite{gimenez2007linear} for determining many-electron states and energies in a quantum dot network. We present the method using double quantum dot networks, but the approach generalizes to more quantum dots. Each quantum dot (QD) is assumed to be formed electrostatically by surface gate electrodes. In order to directly compare our method with the Heitler-London and Hund-M{\"u}lliken approaches~\cite{burkard1999coupled, li2010exchange}, the confining 2D electric potential is approximated by a quartic model:
\begin{align}
\label{eq:quartic_pot}
    V(x,y) = \frac{m^*\omega_0^2}{2}\left[\frac{1}{4d^2}(x^2 - d^2)^2 + y^2\right]
\end{align}
where $m^*$ is the effective mass ($0.067m_0$ for GaAs and $0.191m_0$ for Si where $m_0$ is the free electron mass), $2d$ is the separation between the QDs, and $\omega_0$ is the harmonic frequency of both wells.
The characteristic width of each potential minimum is given by $l_0 = \sqrt{\hbar/m^*\omega_0}$.

Later, in Section~\ref{sec:device_sims}, we provide a more realistic model of the potential landscape by simulating a Si MOSFET double QD device structure using self-consistent 3D Poisson calculations.
Throughout this work we assume that there is no magnetic field; however, this can be included by adding a vector potential term to the Hamiltonian in Eq.~\ref{eq:basisHam}.
We note that adding a magnetic field will impact the convergence of the single-electron calculations discussed below in Section~\ref{sec:SE_states}. The magnetic field introduces a varying phase component in the orbital wave functions which may require more harmonic orbital states in order to accurately approximate the single-electron orbitals.

%%%%%%%%%%%%%%%%%%%%%%%%%%%%%%%%%%%%%%%
\subsection{Constructing single-electron states with harmonic orbitals}
\label{sec:SE_states}

We begin the LCHO-CI calculation by evaluating the single-electron states for the Hamiltonian
\begin{equation}
    \label{eq:basisHam}
    H = -\frac{\hbar^2}{2m^*}\left[\frac{\partial^2}{\partial x^2} + \frac{\partial^2}{\partial y^2}\right] + V(x,y),
\end{equation}
where $\hbar$ is the reduced Planck's constant.
The single-electron states $\ket{\xi_j}$ are eigenfunctions of $H$ with corresponding eigenenergies $\epsilon_j$ that satisfy $H\ket{\xi_j} = \epsilon_j\ket{\xi_j}$.
In order to simplify the LCHO-CI calculation later, it is useful to approximate the single-electron states~$\ket{\xi_j}$ using a basis of radially symmetric 2D harmonic orbitals (HOs) centered at the origin of the quantum dot network. The explicit form of the 2D HO states is $\phi_{nm}(x,y)$ $= \phi_n(x)\phi_m(y)$, where $\phi_q(s)$ $= \frac{1}{\sqrt{2^q q!}}\left(\frac{m^*\omega}{\pi\hbar}\right)^{1/4}\exp(-\frac{m^*\omega s^2}{2\hbar}) H_q\left(\sqrt{\frac{m^*\omega}{\hbar}}s\right)$, $H_q$ are the Hermite polynomials and $\omega$ is the harmonic frequency. The full 2-dimensional HO basis $\{\phi_{nm}(x,y)\}$ is found by taking the Cartesian product of two 1-dimensional HO bases $\{\phi_n(x)\}$ and $\{\phi_m(y)\}$, where each 1D basis is composed of the lowest $M_x$ and $M_y$ energy states, respectively.
The total number of 2D HO states $\{\phi_{i}(x,y)\}$ is $M = M_xM_y$ ($M_x=M_y$ throughout this work). 

\begin{figure}
    \centering
    \includegraphics[width = 0.9\textwidth,trim = {0 14cm 0 15cm},clip]{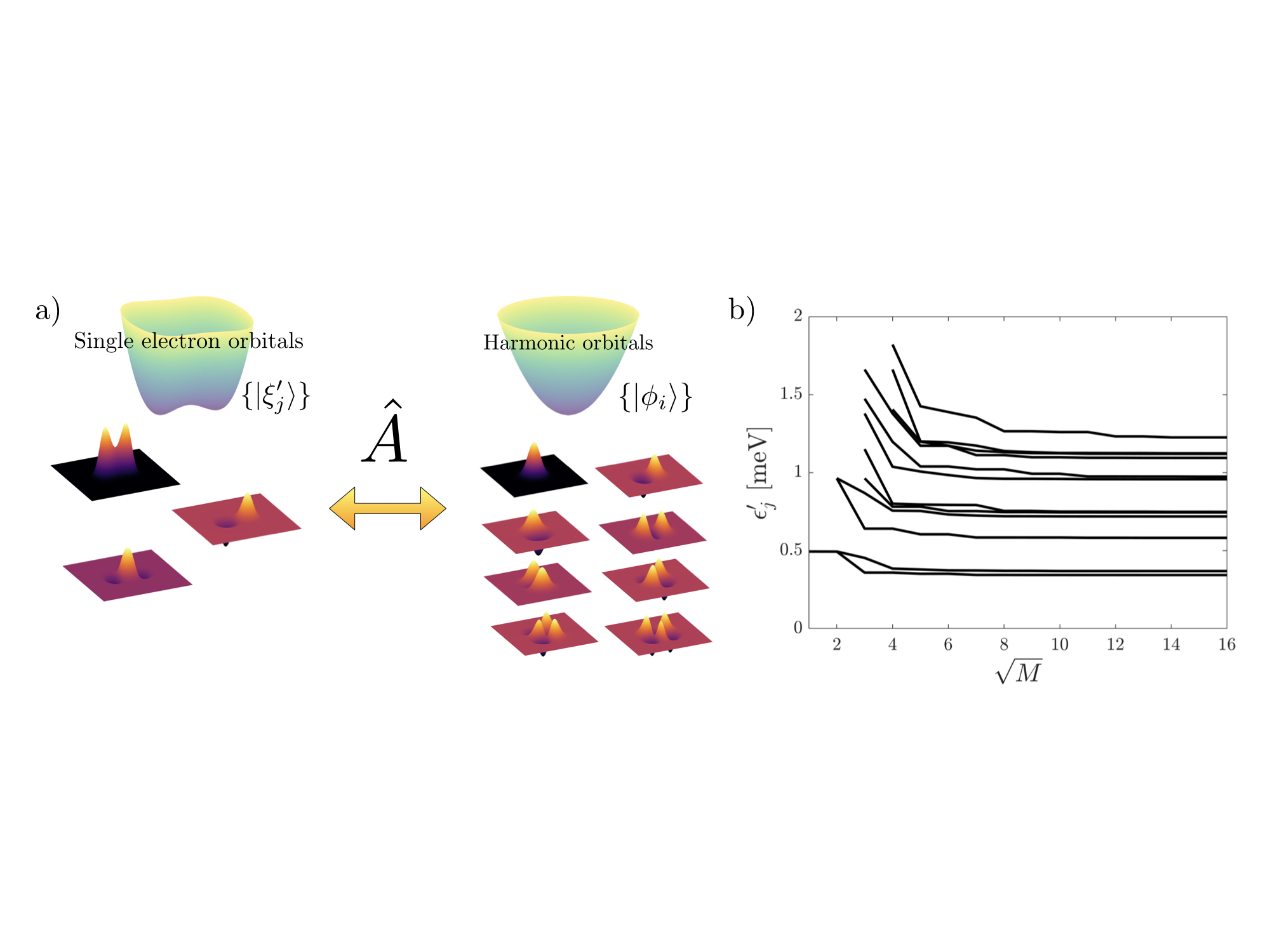}
    \caption{Using harmonic orbitals (HO) to approximate single-electron states. a) The operator $\hat{A}$ transforms between the HO basis $\{\ket{\phi_i}\}$ and the approximate single-electron orbitals $\ket{\xi_j'}$. The first three single-electron orbitals for a quartic potential are shown at  left, and the eight lowest energy HO states are shown at right. b) Convergence of the first twelve approximate single-electron energies $\epsilon_j'$ versus the number of HOs, $M$, used to compose the basis $\{\ket{\phi_i^\omega}\}$. Here, $M = M_xM_y$ and $M_x=M_y$ are the numbers of 1-dimensional HOs taken along the $x$ and $y$ axes, respectively, to construct the 2-dimensional HOs.}
    \label{fig:SEconv}
\end{figure}

Next, we find approximations $\ket{\xi'_j}$ to the first $N$ single-electron states $\ket{\xi_j}$ using a linear combination of harmonic orbitals (LCHO)
\begin{equation}
    \ket{\xi'_j} = \sum_{i=1}^M A_{ij} \ket{\phi_i},
\end{equation}
where $i$ is a composite index describing the $n,m$ indices of the HO state and $A_{ij}$ are expansion coefficients.
$H$ is rewritten in the 2D HO basis $H^\phi$, and we obtain the generalized eigenvalue problem
\begin{equation}
    H^\phi \hat{A} = \epsilon' \hat{A},
\end{equation}
where $H^\phi$ has matrix elements $H^\phi_{ij} = \bra{\phi_i}H\ket{\phi_j}$, $\hat{A} = (\vec{A}_1,\vec{A}_2,\dots,\vec{A}_M)$ describes the unitary transformation between $\{\ket{\phi_i}\}$ and $\{\ket{\xi'_i}\}$, and $\epsilon'$ are approximations to the single-electron state energies $\epsilon$.
The basis $\{\ket{\xi'_j}\}$ converges to $\{\ket{\xi_j}\}$ as $M$ increases and more HO basis states are included in the set.
A schematic of the transformation between $\{\ket{\phi_i}\}$ and $\{\ket{\xi'_j}\}$ via $\hat{A}$ is shown in Figure~\ref{fig:SEconv}a.
The three lowest energy single-electron orbitals are shown on the left using the quartic potential given in Eq.~\ref{eq:quartic_pot}, where $m^* = 0.191m_0$ (Si), $\hbar\omega_0=$ 0.375~meV ($l_0 = 32.6$ nm), and $d = 50$ nm.
Several of the lowest energy HO states used in approximating $\{\ket{\xi'_j}\}$ are shown on the right for a harmonic frequency of $\hbar\omega_0=$ 0.188~meV ($l_0 = 46.1$ nm).
Figure~\ref{fig:SEconv}b shows convergence of the 12 lowest energies $\epsilon_j'$ as a function of $M$ for the quartic and harmonic parameters used in Figure~\ref{fig:SEconv}a. The lowest two levels converge rapidly compared to the higher levels, which do not fully converge until $M > 12^2$. The three lowest energies are $\epsilon'_0=0.3436$ meV, $\epsilon'_1=0.3692$ meV, and $\epsilon'_2=0.5822$ meV.

Typically, $M \gg N$ is required for $\{\ket{\xi'_j}\}$ to accurately approximate $\{\ket{\xi}\}$.
After $\hat{A}$ is calculated, it is trimmed to have dimension $M \times N$, making $\hat{A}$ semi-unitary ($\hat{A}^\dagger\hat{A} = I$ but $\hat{A}\hat{A}^\dagger \neq I$).
This reduces the computational complexity when transforming the Coulomb matrix element basis in Section~\ref{sec:manyelectronHam}. 
So far, the choice of $\omega$ used to construct $\{\ket{\phi_i}\}$ is arbitrary; we will show in Section~\ref{sec:optimalOmega} how to optimize the choice of $\omega$ to best approximate $\{\ket{\xi_j}\}$.

%%%%%%%%%%%%%%%%%%%%%%%%%%%%%%%%%%%%%%
\subsection{Many-electron Hamiltonian}
\label{sec:manyelectronHam}

Here we focus on the construction of the general many-body Hamiltonian for the quantum dot network. Written in second quantization form, the Hamiltonian is
\begin{equation}
\label{eq:2ndQHam}
    H = \sum_i \epsilon_i c_i^\dagger c_i + \frac{1}{2}\sum_{ijkl}\bra{ij}v\ket{kl} c_i^\dagger c_j^\dagger c_k c_l,
\end{equation}
where $c^\dagger_i$ and $c_i$ are the fermionic creation and annihilation operators acting on an electron in the $i^{\rm th}$ spin-orbital state $\ket{i} = \ket{\chi_{m_s}}\ket{\xi_j}$. Here,
$\ket{\chi_{m_s}}$ is the spin component of the spin-orbital state, taking one of two values $m_s = \pm 1/2$.
The $\epsilon_j$ are single-electron energies as described in Section~\ref{sec:SE_states}.
The labels $i$, $j$, $k$, and $l$ are composite indices for the spin and orbital components of the corresponding spin-orbital state.
Lastly, $v = \frac{e^2}{4\pi\epsilon_0\epsilon_r}\frac{1}{|\vec{r}_2 - \vec{r}_1|}$ is the standard Coulomb potential, where $\epsilon_0$ is the vacuum permittivity and $\epsilon_r$ is the relative permittivity of the material (12.4 for GaAs and 7.8 for Si/SiO$_2$ where $\epsilon_{\rm Si/SiO_2} = $ $[\epsilon_{\rm Si} + \epsilon_{\rm SiO_2}]/2$).

Evaluating the Coulomb Matrix Elements (CMEs) $\bra{ij}v\ket{kl}$ in the single-electron basis is numerically challenging due to the divergent $\frac{1}{|\vec{r}_2 - \vec{r}_1|}$ potential.
However, the CMEs can be calculated by decomposing them into the HO basis using the transformation $\hat{A}$: 
\begin{equation}
    \bra{ij}v\ket{kl} = \braket{\chi_i|\chi_l}\braket{\chi_j|\chi_k} \sum_{\alpha=1}^M\sum_{\beta=1}^M\sum_{\gamma=1}^M\sum_{\delta=1}^M A_{i\alpha}^* A_{j\beta}^* A_{k\gamma} A_{l\delta} \braket{\alpha\beta|v| \gamma\delta}, 
\end{equation}
where latin indices correspond to single-electron states $\ket{\xi}$, greek indices correspond to 2D HO states $\ket{\phi}$, and $^*$ denotes the complex conjugate.
In the HO basis, the CMEs have a fully analytic solution (see Appendix~\ref{app:CMEs_symHOs} for a derivation): 
\begin{align}
\label{eq:CMEEquation}
\bra{\alpha\beta}v\ket{\gamma\delta} &= \bra{n_\alpha m_\alpha n_\beta m_\beta}v\ket{n_\gamma m_\gamma n_\delta m_\delta} \nonumber \\
&= \frac{e^2}{4\pi\epsilon_0\epsilon_r} \int d\vec{r}_1\int d\vec{r}_2 \,\,\phi^*_\alpha(\vec{r}_1) \phi^*_\beta(\vec{r}_2) \frac{1}{|\vec{r}_2 - \vec{r}_1|} \phi_\gamma(\vec{r}_2) \phi_\delta(\vec{r}_1)\nonumber \\
&= \sqrt{\omega}\frac{e^2}{4\pi\epsilon_0\epsilon_r}\frac{\sqrt{\pi}(-1)^{n_\beta + m_\beta + n_\gamma + m_\gamma}}{\sqrt{n_\alpha! m_\alpha! n_\delta! m_\delta! n_\beta! m_\beta! n_\gamma! m_\gamma!}} \sum_{p_1 = 0}^{\min(n_\alpha,n_\delta)} p_1! \binom{n_\alpha}{p_1}\binom{n_\delta}{p_1} \nonumber \\
&\quad \times \sum_{p_2 = 0}^{\min(m_\alpha,m_\delta)} p_2! \binom{m_\alpha}{p_2}\binom{m_\delta}{p_2} \sum_{p_3 = 0}^{\min(n_\beta,n_\gamma)} p_3! \binom{n_\beta}{p_3}\binom{n_\gamma}{p_3} \\
&\quad\times \sum_{p_4 = 0}^{\min(m_\beta,m_\gamma)} p_4! \binom{m_\beta}{p_4}\binom{m_\gamma}{p_4} (-1)^p \, \frac{(2p-1)!! (2p-a-1)!! (a-1)!!}{2^{2p} \, p!} \nonumber
\end{align}
Here, $\omega$ is the harmonic frequency for the HO basis, $n$ and $m$ label the $x$- and $y$- axis modes of the 2D HO state $\phi_{nm}(x,y) = \phi_n(x)\phi_m(y)$, $a = n_\alpha + n_\delta + n_\beta + n_\gamma - 2p_1 - 2p_3$, $2p = a + m_\alpha  + m_\delta + m_\beta + m_\gamma - 2p_2 - 2p_4$, and
$k!! = k(k-2)\cdots3\cdot1$ is the double factorial for odd $k$.
The expression above only holds when both $a$ and $2p$ are even, and $\bra{\alpha\beta}v\ket{\gamma\delta} = 0$ otherwise.
The analytical solution is possible due to the use of an orthogonal HO basis.
This is a key distinction from reference \cite{gimenez2007linear}, which used a non-orthogonal set of localized HOs taken from each QD.
Eq.~\ref{eq:CMEEquation} is similar to the equations presented in references \cite{hawrylak1993far,kyriakidis2002voltage} that use a Fock-Darwin basis set.
We denote the full $M^2 \times M^2$-dimensional matrix of CMEs when written in the HO basis with harmonic frequency $\omega$ as $C_{\rm HO,\omega}$.
The matrix of CMEs written in the single-electron basis is denoted as $C_{\rm SE}$ and has dimension $N^2 \times N^2$.
$C_{\rm SE}$ can be obtained via the basis transformation $C_{\rm SE} = (A^\dagger \otimes A^\dagger) C_{\rm HO,\omega} (A \otimes A)$.

After finding $C_{\rm SE}$, we use a full configuration-interaction (CI) approach to calculate the many-electron energy spectra and eigenstates.
In this approach, a basis of $K$-electron states is constructed out of all possible $K$-electron configurations of single-electron spin-orbital states.
After this configuration basis is constructed, the Hamiltonian from Eq.~$\ref{eq:2ndQHam}$ is rewritten in the configuration basis and subsequently diagonalized to find the corresponding eigenenergies and eigenstates.
The total number of configurations $n_\textsc{c}$ of $K$-electron states grows according to $n_\textsc{c} = \binom{2N}{K} = \frac{(2N)!}{K!(2N-K)!}$, where $2N$ is the total number of single-electron spin-orbital states.
Because $H$ conserves the total spin $S_z$, rather than diagonalizing the Hamiltonian using all $K$-electron spin-orbit configurations, the configuration basis can be restricted to subspaces of fixed $|S_z|$ value to ease computational requirements.

The lowest energy states of the many-electron energy spectra can be mapped to the effective Heisenberg Hamiltonian $H = \sum_{ij}J_{ij}\vec{\sigma}_i\cdot\vec{\sigma}_j$, where indices label QD sites and $\vec{\sigma}$ are vectors of the 2-level Pauli spin operators.
For a 2-electron system, $J$ is the energy difference between the singlet and triplet $\ket{T_0}$ eigenstate.
Restricted to the $|S_z|=0$ spin subspace and with no external magnetic field applied, the ground and first excited eigenstates are the singlet and $\ket{T_0}$ states, respectively, giving $J>0$ according to the Lieb-Mattis theorem \cite{lieb1962ordering}.
An example of parameterizing the Heisenberg Hamiltonian for a 3-electron system is given in reference \cite{gimenez2007linear}.

%%%%%%%%%%%%%%%%%%%%%%%%%%%%%%%%%%%%%%%
\subsection{Choosing an optimal harmonic orbital basis}
\label{sec:optimalOmega}

The accuracy of the $K$-electron energy spectra relies on two parameters. The first parameter is $N$, the number of approximate single-electron states $|\xi'_j\rangle$ used in the construction of the Hamiltonian in Eq.~\ref{eq:2ndQHam}.
As $N$ is increased, higher energy single-electron states can add important corrections to the $K$-electron energy spectra until $N$ is sufficiently large for the energies to converge.
The second parameter is how close the approximate single-electron states $\{|\xi'_j\rangle\}$ are to $\{\ket{\xi_j}\}$.
If a sufficiently large basis set of HOs is used, quantified by $M$, then $\{|\xi'_j\rangle\}$ will converge to $\{\ket{\xi_j}\}$.
However, it is not computationally practical to use an arbitrarily large $M$ in order to accurately approximate $\{\ket{\xi_j}\}$, as the total size of $C_{\rm HO,\omega}$ scales as $M^4$.
For a fixed $M$, an improved approximation of $\{\ket{\xi_j}\}$ can be achieved by using an optimal value of $\omega$ when constructing the HO basis states.
So far, there has been no discussion on the choice of $\omega$ used when building $\{\ket{\phi^\omega_i}\}$ (here we adopt a new notation for the 2D HOs that specifies the choice of $\omega$ used to construct the basis).
In the LCHO-CI method laid out in reference \cite{gimenez2007linear}, $\{\ket{\phi_i^\omega}\}$ is constructed by taking localized HOs centered at each respective QD in the network.
For the localized HOs, $\omega$ is chosen by fitting the minima of the QD potentials to a radially symmetric harmonic potential well.
In our modified LCHO-CI approach, a single collection of HOs centered at the origin constitutes the full HO basis, and there is not a direct analogue for choosing $\omega$.
Naively, $\omega$ could be chosen by fitting the potential minimum of each QD in the network to a harmonic well and using the average $\omega$ determined from each fit; however, there is nothing to suggest that this choice $\{\ket{\phi^\omega_i}\}$ will best approximate $\{\ket{\xi_j}\}$.

The idea of optimizing basis orbitals used to approximate $\{\ket{\xi_j}\}$ was also used in reference \cite{nielsen2010implications}, which optimized the relative spacing and width of Gaussian orbitals to improve the accuracy of the full CI calculation. The optimal choice of $\omega$ should maximize the overlap between bases $\{\ket{\xi'_j}\}$ and $\{\ket{\xi_j}\}$, i.e. $F = \sum_{j=1}^N |\braket{\xi_j|\xi'_j}|^2$.
If $\{\ket{\xi'_j}\}$ perfectly describes $\{\ket{\xi_j}\}$, then $F=N$.
We can optimize $\omega$ by recalculating $\{\ket{\xi_j'}\}$ for a given choice of $\omega$ and subsequently minimizing $1-\frac{1}{N}F$.
However, evaluating $\{\ket{\xi'_j}\}$ during each optimization step means $H_{\phi}$ must be constructed and subsequently diagonalized as described in Section~\ref{sec:SE_states}.
The construction of $H_\phi$ alone requires the evaluation of $M(M+1)/2$ inner products $H^\phi_{ij} = \bra{\phi_i}H\ket{\phi_j}$.
Optimizing $\omega$ this way can be very slow due to the large value of $M$ typically required for convergent LCHO-CI calculations.
We note that the exact length of time it takes to calculate $H_\phi$ strongly depends on the number of grid points used in constructing the 2D potentials. More grid points increases the computation cost of each individual inner product, a numerical integration over the 2D grid.

To reduce the computational complexity of this optimization, we take a different approach. If the single-electron basis states $\{\ket{\xi_j}\}$ can be accurately decomposed into the HO basis $\{\ket{\phi^\omega}\}$, then for each state $\ket{\xi_j}$, we have $\sum_{i=1}^M|\braket{\xi_j|\phi_i^\omega}|^2 \approx 1$.
If instead $\{\ket{\phi^\omega}\}$ poorly describes $\ket{\xi_j}$, then $\sum_{i=1}^M|\braket{\xi_j|\phi_i^\omega}|^2 < 1$.
Therefore for a choice of $N'$ single-electron states $\ket{\xi_j}$, an optimal $\omega$ can be found via the following minimization problem
\begin{equation}
    \min_\omega \,\, f_{\min}(\omega) = \min_\omega \,\, 1-\frac{1}{N'}\sum_i^{N'}\sum^M_j|\langle\xi_i|\phi_j^\omega\rangle|^2, 
\end{equation}
%%%%%%%%%%%%%%%%
where we refer to the optimization function as $f_{\min}(\omega)$.
Note that we have specified using a smaller subset $N'$ of the single-electron orbitals compared to the full number of $N$ orbitals used in the main LCHO-CI calculation.
Using $N'<N$ does not significantly alter the final $\omega$ value and allows for a faster minimization.
The $N'$ single-electron states $\ket{\xi_j}$ only need to be evaluated once at the beginning of the minimization, and only $N' \times M$ inner products are calculated during each minimization step.
Typically $N'\ll M$, so this is much faster than directly calculating $H^\phi$ and does not bottleneck the full LCHO-CI calculation.
Non-optimal $\omega$ values can yield orders of magnitude worse values of $f_{\min}(\omega)$ compared to the optimal $\omega$.
Minimizations in this paper were done using a BFGS quasi-Newton search with a first-order optimality tolerance of $1\times10^{-6}$ \cite{fletcher2013practical}.
%%%%%%%%%%%%%%%%%

\begin{figure}
    \centering
    \includegraphics[width = 0.6\textwidth]{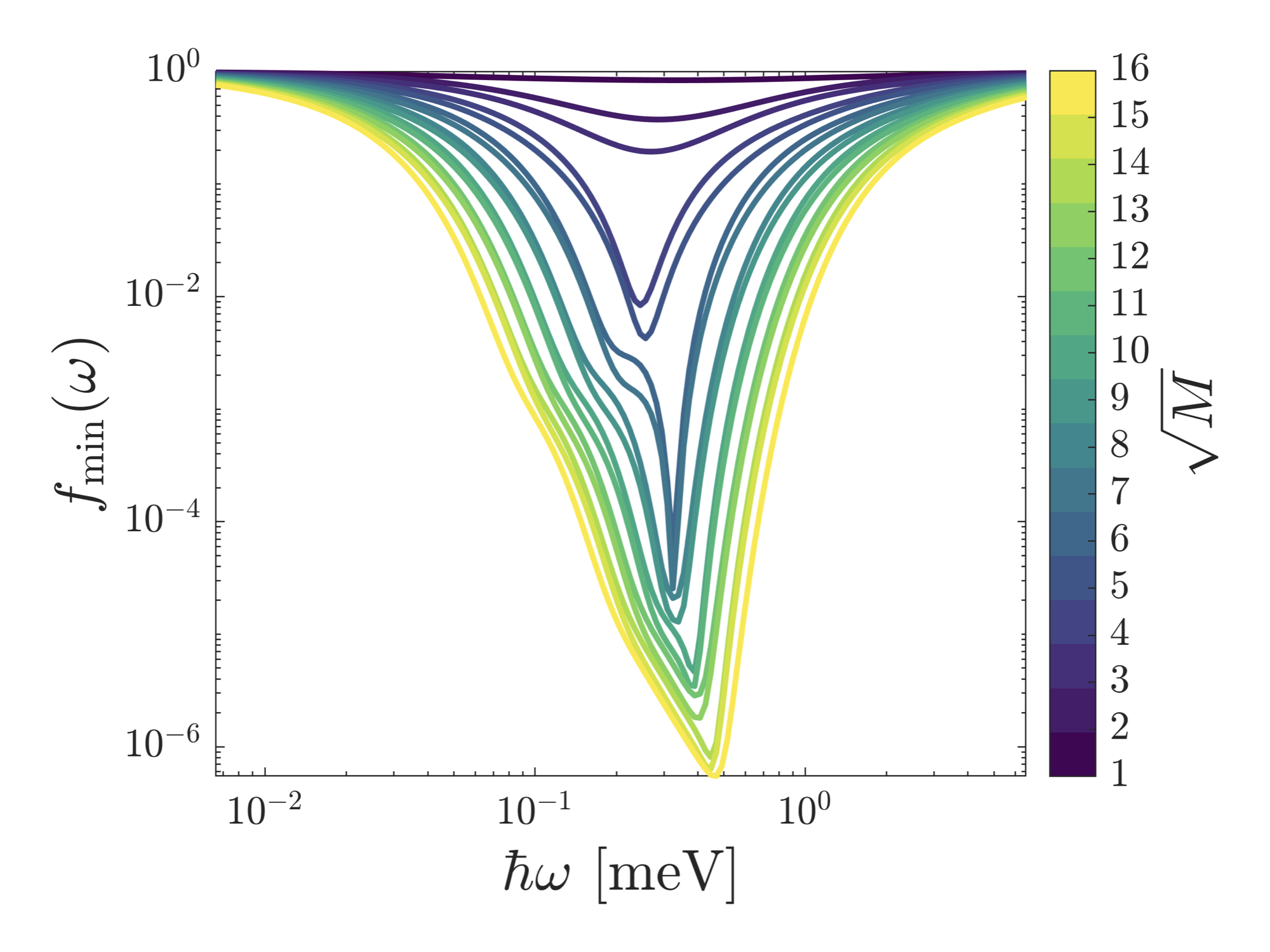}
    \caption{Dependence of the optimization function $f_{\min}(\omega)$ on $\omega$ and the size of the harmonic orbital basis ($M_xM_y=M$). $N'=6$ for a quartic potential with parameters $m^* = 0.191m_0$, $\hbar\omega_0=$ 0.375 meV, and $d = 50$ nm. Color indicates a different number of HO basis states ranging from $M_x=M_y=1$ (purple) to $M_x=M_y$=16 (yellow).}
    \label{fig:opt_omega}
\end{figure}

Figure~\ref{fig:opt_omega} shows how the optimization function $f_{\min}(\omega)$ depends on $\omega$ and the size $M$ of the HO basis $\{\ket{\phi_i^\omega}\}$.
The single-electron orbitals $\ket{\xi_j}$ are found using a quartic potential where $m^* = 0.191m_0$ (Si/SiO$_2$ system), $\hbar\omega_0=$ 0.375 meV ($l_0 = 32.6$ nm), and $d = 50$ nm.
The first $N'=6$ single-electron states are used in the minimization, and $M$ is stepped from $1$ to $16^2$.
Recall that $M=M_xM_y$ and $M_x=M_y$.
For a fixed $M$, there is a single optimal value of $\omega$ that minimizes $f_{\min}$.
Non-optimal $\omega$ values can yield orders of magnitude worse values of $f_{\min}(\omega)$ compared to the optimal $\omega$.
Thus, optimizing $\omega$ is a useful way to improve the accuracy of the approximated orbitals $\{\ket{\xi'_j}\}$ in the LCHO-CI calculation.
At a fixed $\omega$, as $M$ is increased, $f_{\min}(\omega)$ always decreases; this is in accordance with the fact that as the size of $\{\ket{\phi_i^\omega}\}$ increases, better approximations of $\{\ket{\xi_j}\}$ can be obtained irrespective of the choice of~$\omega$.
As $M$ increases, the optimal $\omega$ appears to converge towards a fixed value.

After the optimal $\omega$ is found, the CMEs ($C_{\rm HO}$) must be evaluated, typically the most computationally intensive part of a CI calculation. However, we make use of the fact that in Eq.~\ref{eq:CMEEquation} the only dependence on $\omega$ is a scalar $\sqrt{\omega}$ term that can be factored out.
A large matrix of CMEs can be pre-calculated for a unit choice of $\omega = 1$ ($C_{\rm HO,1}$) and then scaled by $\sqrt{\omega}$ to obtain the matrix of CMEs for the optimal HO basis: 
\begin{equation}
    C_{\rm HO,\omega} = \sqrt{\omega}C_{\rm HO,1}.
\end{equation}
Using the pre-calculated library $C_{\rm HO,1}$ allows for numerically fast, quantitatively accurate full LCHO-CI calculations across a set of quantum dot network potentials.
This represents an important computational speed-up in this modified LCHO-CI approach.
The longest part of the calculation (evaluating $C_{\rm SE}$) becomes a scalar multiplication of $C_{\rm HO,1}$ and subsequent rotation of $C_{\rm HO,\omega}$ into the single-electron basis.
The idea of using a pre-calculated library of CMEs to speed up CI calculations was used previously in reference \cite{pedersen2007failure} for a Gaussian orbital basis; however, the orbital basis was not optimized to improve the accuracy of the results.

In Section~\ref{sec:device_sims}, we perform modified LCHO-CI calculations of a double quantum dot where $M=16^2$, $N=18$, and $N'=6$.
The grid spacing of the 2D potential along $\hat{x}$ and $\hat{y}$ axes is 0.5~nm.
The choice of $M$ gives a total of $16^8$ CMEs to calculate using Eq.~\ref{eq:CMEEquation}.
In this paper, the CME calculations were done with MATLAB using a computer with an Intel Xeon E5-2650 processor and parallelized across 24 cores.
It took $\sim$10 hours to evaluate all the CMEs (only half of the CMEs must be explicitly calculated, as $C_{\rm HO,1}$ is Hermitian, and 3/4 of the matrix elements are zero).
The scalar multiplication that converts $C_{\rm HO,1}$ to $C_{\rm HO,\omega}$ is relatively fast, taking only a few seconds.
The basis rotation of $C_{\rm HO,\omega}$ into $C_{\rm SE}$ takes tens of seconds and depends on the size of both $N$ and $M$.
For the $N$ and $M$ values used, after the initial long calculation in evaluating $C_{\rm HO,1}$, the total evaluation of $C_{\rm SE}$ took $\sim$25 seconds.
The other LCHO-CI steps include optimization of $\omega$, evaluation of $A$, and construction of the second quantization Hamiltonian, which took approximately 20, 90, and 30 seconds respectively.
This gives a total calculation time of $~\sim$3 minutes for modified LCHO-CI on a double quantum dot problem with approximately $300 \times 100$ grid points in the 2D potential. 
Larger grids will increase the time required to optimize $\omega$ and evaluate $A$.
The other steps in the modified LCHO-CI calculation are not affected by the grid size.

For some QD networks, such as a linear chain of QDs, it may be desirable to use an elliptical set of 2D HOs, where $\omega_x \neq \omega_y$, in order to better approximate the single-electron states $\ket{\xi_j}$.
If $\omega_x \neq \omega_y$, then the HO CMEs still have a closed analytical form; however, $\omega_x$ and $\omega_y$ cannot be factored out of the CME expression (this is shown in Appendix~\ref{app:CMEs_elipHOs}). 
Thus, an elliptical HO basis does not provide the same computational speedup as the radially symmetric case, and this is why we have chosen to use HOs with $\omega_x = \omega_y$.
We note that it is possible to pre-calculate a discrete set of elliptical CME matrices $C_{\rm HO,1,\kappa}$, where $\omega_x=1$ and $\kappa=\omega_x/\omega_y$ is the HO eccentricity.
One can then optimize over a continuous choice of $\omega_x$ as well as a discrete set of $\kappa$ values and subsequently find $C_{{\rm HO},\omega_x,\omega_y} = \sqrt{\omega_x}C_{\rm HO,1,\kappa}$ (see Appendix~\ref{app:CMEs_elipHOs}).
This comes at the cost of storing a pre-calculated $C_{\rm HO,1,\kappa}$ matrix for each choice of $\kappa$ which, due to the large choice of $M$ typically required for these calculations, may render this approach impractical.
An alternative idea is to use an asymmetric choice of $M_x \neq M_y$ when building a 2D HO basis to approximate $\{\ket{\xi_j'}\}$. These ideas are not explored in this work, as we focus on a small (double dot) system where symmetrical HOs and $M_x=M_y$ are satisfactory. 
%%%%%%
\subsection{Comparison to Heitler-London and Hund-M\"{u}lliken methods}
\label{sec:compare_HL_HM}

We conclude the discussion of the modified LCHO-CI approach by comparing it with the Heitler-London (HL) \cite{burkard1999coupled,calderon2006exchange} and Hund-M\"{u}lliken (HM) \cite{hu2000hilbert,van2006semiconductor,hatano2008manipulation} methods for evaluating the exchange strength~$J$.
We consider a double quantum dot system with a quartic potential given by Eq.~\ref{eq:quartic_pot}, occupied by 2 electrons.
In both the HL and HM approaches, the 2-electron singlet and triplet states are constructed using localized $s$-orbitals taken from each dot. $(n,m)$ denotes the electron occupancy in each quantum dot. The HL approach includes only the singly occupied $S(1,1)$ and $T(1,1)$ states, while the HM method extends the basis set to include the doubly-occupied singlet $S(0,2)$ and $S(2,0)$ states. The localized $s$-orbitals for the HL and HM methods are found by approximating both potential wells has having harmonic confinement $\omega_0$, and minima located at $\pm d$.
Approximating the orbital states this way gives rise to analytical expressions for $J$ when the quartic potential is used, for both the HL and HM methods \cite{burkard1999coupled}.
The analytical expressions make these methods useful for exploring qualitative behavior under varying parameters, including magnetic field $B$ and interdot detuning.
However, both methods are known to break down at small interdot separations $2d$ and give a nonphysical result with $J<0$ at zero magnetic field.
Quantitatively, this breakdown occurs when the ratio of the Coulomb and confinement energies $c=\sqrt{\pi/2}(e^2/4\pi\epsilon_0\epsilon_r l_0)/\hbar\omega_0 > 2.8$.
$c$ is inversely proportional to $l_0 = \sqrt{\hbar/m^*\omega_0}$ and therefore is directly proportional to $m^*$.
This means that the HL and HM methods break down more easily in Si/SiO$_2$, which has a larger $m^*$ compared to GaAs, and for small dots with small interdot separations. 
The interdot separation at which breakdown occurs increases with dot radii \cite{calderon2006exchange}.
This limits the parameter space over which the HL and HM methods can be used to predict $J$, especially for electrons in silicon.

\begin{figure}[ht!]
    \centering
    \includegraphics[width = 0.85\textwidth]{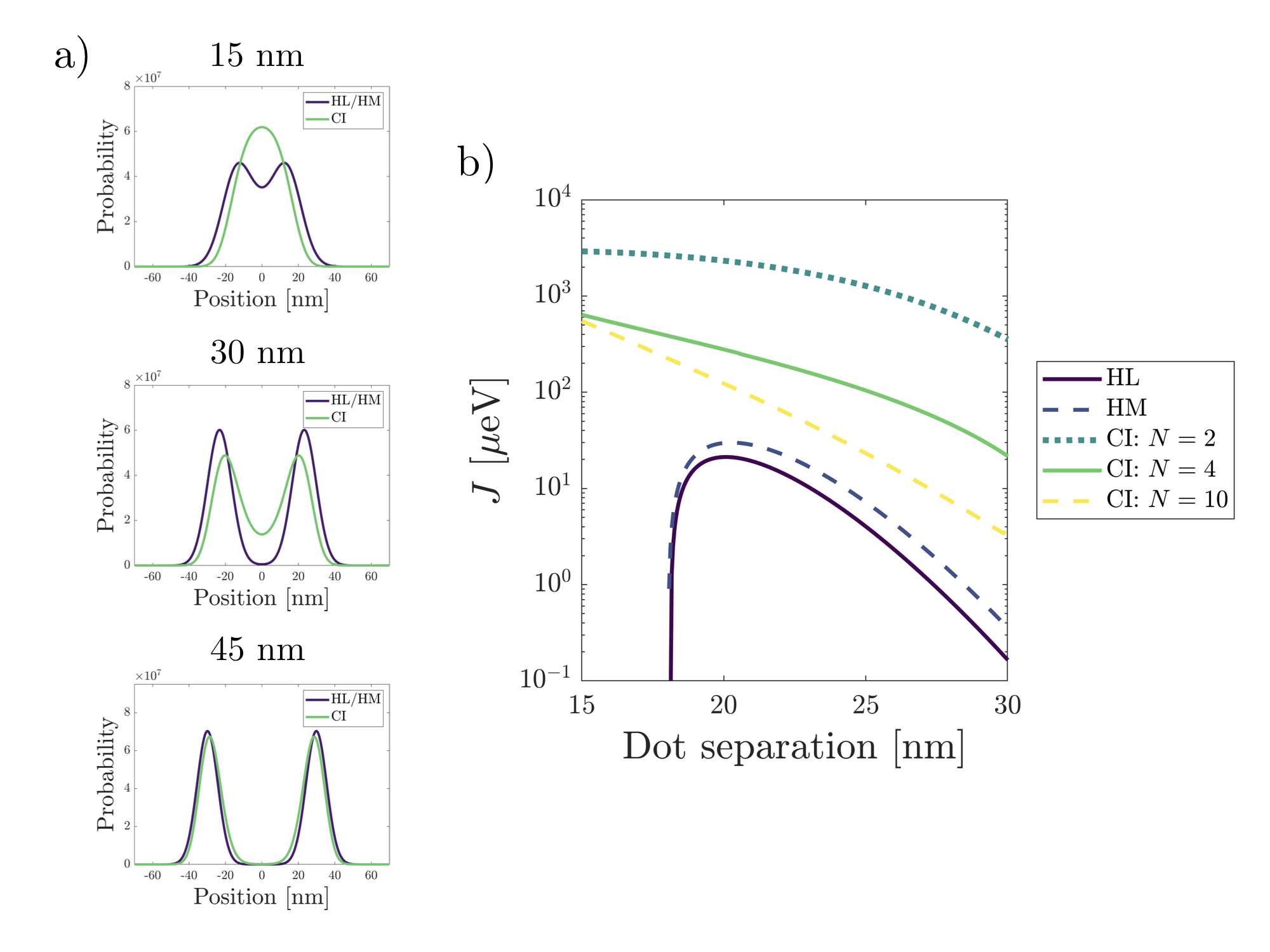}
    \caption{Comparison of the HL (purple) and HM (dark blue) methods to the modified LCHO-CI method for evaluating $J$. The system is a double QD in Si/SiO$_2$ described by a quartic potential, where $l_0=$ 6 nm. a) Probability densities of the ground single-electron orbital states for the HL/HM and LCHO-CI calculations at different dot separations. At small dot separations, the HL/HM states overestimate the separation of the two localized wave functions compared to the numerically calculated orbitals. As dot separation increases, the approximate HL/HM states converge to the numerical results. b) $J$ versus the interdot separation $2d$ is plotted. Three different LCHO-CI calculations are done for $N=2$ (light blue), $N=4$ (green), and $N=10$ (yellow). For all LCHO-CI calculations, $M=15^2$.}
    \label{fig:HL_HM_comp}
\end{figure}

The failure of the HL/HM at small dot separations occurs due to poor approximations of the localized $s$-orbitals.
Each localized $s$-orbital is assumed to be taken from a harmonic well (i.e. Gaussian orbitals) and separated by $2d$.
However, at small interdot separations, the tunneling barrier in the quartic potential is lowered, and the electrons delocalize towards the center of the double well potential.
This effect is shown in Figure~\ref{fig:HL_HM_comp}a which compares the ground single-electron orbital state for the HL/HM and LCHO-CI methods at different dot separations $2d$.
We consider a Si/SiO$_2$ material system with $l_0=6$~nm for the quartic potential. 
The ground eigenstate for the HL/HM methods is $\ket{\psi_0}=(\ket{R}+\ket{L})/\sqrt{2}$ where $\ket{R}$ and $\ket{L}$ are the localized $s$-orbitals taken at $\pm d$ respectively.
The ground eigenstate for the LCHO-CI method is obtained numerically using the methods outlined in Section~\ref{sec:SE_states}.
At small dot separations $2d$, the HL/HM methods overestimate the true localization of the electron orbitals.
We expect that when $d/l_0$ is large, the numerical and HL/HM approaches should produce similar results for $J$.
When $d/l_0>10$, we find that the overlap between the approximate HL/HM orbital and the numerical LCHO-CI ground state orbital is $\sim$0.99, for our chosen parameters. However, a convergent LCHO-CI calculation at those dot separations is computationally intensive as it requires a larger set of orbitals, so we did not compare $J$ values in that regime.  
%%%%%%%

Figure~\ref{fig:HL_HM_comp}b shows the calculated $J$ versus the dot separation $2d$.
In addition to the HL (purple) and HM (dark blue) calculations, we performed three modified LCHO-CI calculations where $N=2$ (blue), 4 (green), and 10 (yellow) single-electron orbitals are used to construct the two-electron configuration basis.
For all LCHO-CI calculations, $M=15^2$.
$J$ increases (roughly exponentially) as the dots are brought closer together, as expected.
The most striking feature is the breakdown of the HL and HM methods around a dot separation of 18 nm for the chosen $l_0$, whereas in contrast, none of the LCHO-CI calculations show a similar breakdown. This is not surprising, as the LCHO-CI performs an exact diagonalization of the many-electron Hamiltonian. $J$ decreases with the number of orbitals $N$ for the LCHO-CI calculations until it converges around $N=10$. For larger $N$ values, $J$ changes by $<3\%$.

For the quartic potential, the HL and HM methods underestimate the exchange energies found using the LCHO-CI approach.
The LCHO-CI calculation with $N=2$ uses a 2-electron configuration basis similar to the HL and HM methods, with the exception that the doubly occupied triplet $T(0,2)$ and $T(2,0)$ states are added to the basis set.
Even though the $N=2$ and HL/HM basis functions are similar, there are large discrepancies in the calculated $J$ values. This is attributed to the approximate nature of the HL/HM orbitals and that they overestimate the actual dot separation, as discussed above. In this section, we have demonstrated the necessity of using numerical approaches, such as the modified LCHO-CI method, to calculate $J$ for general QD networks. Approximate methods like the HL and HM approaches are computationally simpler, but suffer from breakdown in some configurations and tend to underestimate the exchange strength $J$.

\section{Charge noise sensitivity of a double quantum dot}
\label{sec:device_sims}

In this section, the modified LCHO-CI method is used to investigate the charge noise sensitivity of the exchange interaction in a 2-electron double quantum dot. Charge noise originates from two sources: fluctuations in the applied gate electrode voltages and the fluctuations of background charge traps. Both sources perturb the potential minima and tunnel barrier in the double dot. The potential fluctuations in turn perturb the electron orbitals, modulating the exchange strength $J$. We apply the tools described previously investigate how device geometry affects the susceptibility of $J$ to charge noise of this type. As charge noise is a primary decoherence mechanism for spin qubits, optimizing device geometries for robustness to charge noise is critical to achieving high fidelity multi-spin logic gates.
The calculations are done for a MOSFET (Si/SiO$_2$) dot system, which is of particular relevance as SiO$_2$ is known to host charge traps. The silicon valley states are assumed to have a large splitting (100s of meV) that is uniform between the two QDs.
If the valley splitting $\Delta$ is greater than the thermal broadening energy $k_B T$, electrons loaded into the double QD will populate only the lower valley eigenstate.
Since Coulomb interactions between opposite valley eigenstates are weak \cite{culcer2010quantum}, the Hamiltonian in Eq.~\ref{eq:2ndQHam} does not couple the electrons to excited valley states, and a single-valley system can be assumed for our purposes. Furthermore, we neglect the small but nonzero spin-orbit interaction in silicon, noting that it can be tuned to zero with an appropriately oriented external magnetic field \cite{tanttu2019controlling}.

\subsection{Device model}
\label{sec:device_model}

Rather than use an analytical form to describe the electrostatic potential, we simulate a 3D device structure using a self-consistent Poisson solver, including the Si/SiO$_2$ heterostructure and the metal gate electrodes used to define the quantum dots (software package nextnano++ \cite{birner2007nextnano}). One challenge with Poisson or Schr\"{o}dinger-Poisson calculations is that the electron density changes continuously as gate voltages are varied. This is in contrast to the behaviour in QDs, where a fixed electron number is maintained due to a finite charging energy, and charge can also be fixed if the QDs are tunnel-decoupled from leads. We approximate the two-electron regime by tuning the gate voltages just below the threshold of charge accumulation, i.e. the zero-electron regime. Discussion on the impact of this approximation can be found in reference~\cite{buonacorsi2020simulated}. There it was shown with a Schr\"{o}dinger-Poisson calculation that for a double QD, the presence of one electron in a double well potential reduces the tunnel barrier and increases the orbital spacing.
Because the exchange interaction is very sensitive to variations in the electrostatic potential, this approximation is important to consider when discussing our results. However, we expect that a proper accounting for this effect would only shift the exchange energies in a systematic way, and not change any of the qualitative results.

\begin{figure}[ht]
    \centering
    \includegraphics[width = 0.9\textwidth,trim = {0 5cm 0 2cm},clip]{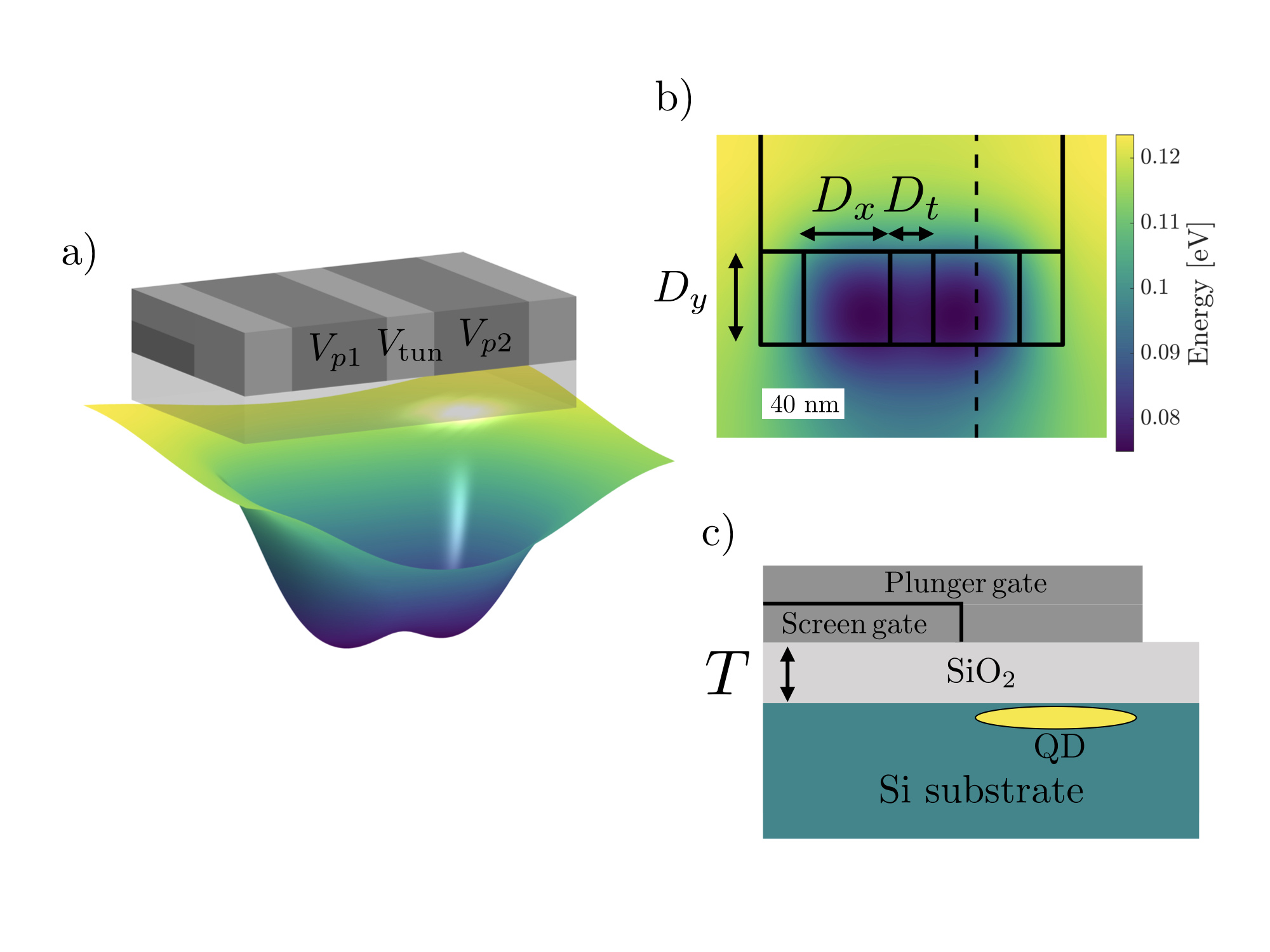}
    \caption{Schematic of a double QD device formed by plunger, tunnel, and screening gates. The geometric parameters are $D_x=$ $D_y=$ 40 nm, $D_t=$ 20 nm, and $T=$ 15 nm. Applied voltages in (a) and (b) are $V_{p1}=$ $V_{p2}= $ 0.150 V, and $V_{\rm tun}=$ 0.09455 V. a) 3D device model, where the plunger, tunnel and screening gates are shaded differently for contrast, along with a semi-transparent SiO$_2$ layer. A 2D electrostatic potential obtained by the self-consistent 3D Poisson calculation is plotted beneath. b) The 2D potential overlaid with an outline of the gate structure, showing the heads of the plunger and tunnel gates as well as the screening gate. The gate lengths (extending beyond the screening gate) are denoted $D_y$, and the widths of the plunger and tunnel gates are $D_x$ and $D_t$, respectively. The 2D potential is taken 1 nm below the Si/SiO$_2$ interface. c) Side profile of the gate structure taken along the dashed line in~b). The SiO$_2$ layer thickness is labelled $T$. The QD (yellow ellipse) is formed underneath the head of the plunger gate.}
    \label{fig:device_geom}
\end{figure}

Figure~\ref{fig:device_geom}a shows a 3D render of the double QD device structure with an example 2D electrostatic potential taken 1 nm below the Si/SiO$_2$ interface. Two plunger gates with corresponding voltages $V_{p1}$ and $V_{p2}$ form a double well potential, along with a tunnel gate with voltage $V_t$ that is used to tune the tunnel barrier. The gate voltages for this particular 2D potential are $V_{p1} = V_{p2} = 0.15$ V and $V_{\rm tun}=$ 0.09455 V, which given an exchange interaction strength $J\approx$ 1 $\mu$eV. Two outer barrier gates are included in the device structure and are kept grounded for all simulations ($V=$ 0 V).
The outer gates are included to better model a realistic device, which is typically surrounded by other metallic gates.
A grounded screening gate ($V=$ 0 V) partially underneath the plunger and barrier gates restricts the formation of the potential wells to below the unscreened portions of the plunger gates. Figure~\ref{fig:device_geom}b shows the 2D potential overlaid with an outline of the plunger, tunnel and screening gates. $D_x$ and $D_t$ indicate the plunger and tunnel gate widths, respectively, and $D_y$ indicates the gate lengths (the unscreened portions). When $D_x$, $D_y$ or $D_t$ is varied in a simulation, it is changed for all the corresponding gates together so that the gate layout is always symmetric about the central $y$-axis. The potential wells form slightly off-center along the $y$ direction due to the asymmetry of having no screening gate on the bottom half of the device. Figure~\ref{fig:device_geom}c shows a side view along the dashed black line in Figure~\ref{fig:device_geom}b, indicating the vertical structure of the device, including the SiO$_2$ layer with thickness $T$ and the location of the QD. The geometric parameters used in Figure~\ref{fig:device_geom} are $D_x=$ $D_y=$ 40 nm, $D_t=$ 20 nm, and $T=$ 15 nm.

\begin{figure}
    \centering
    \includegraphics[width = 0.6\textwidth]{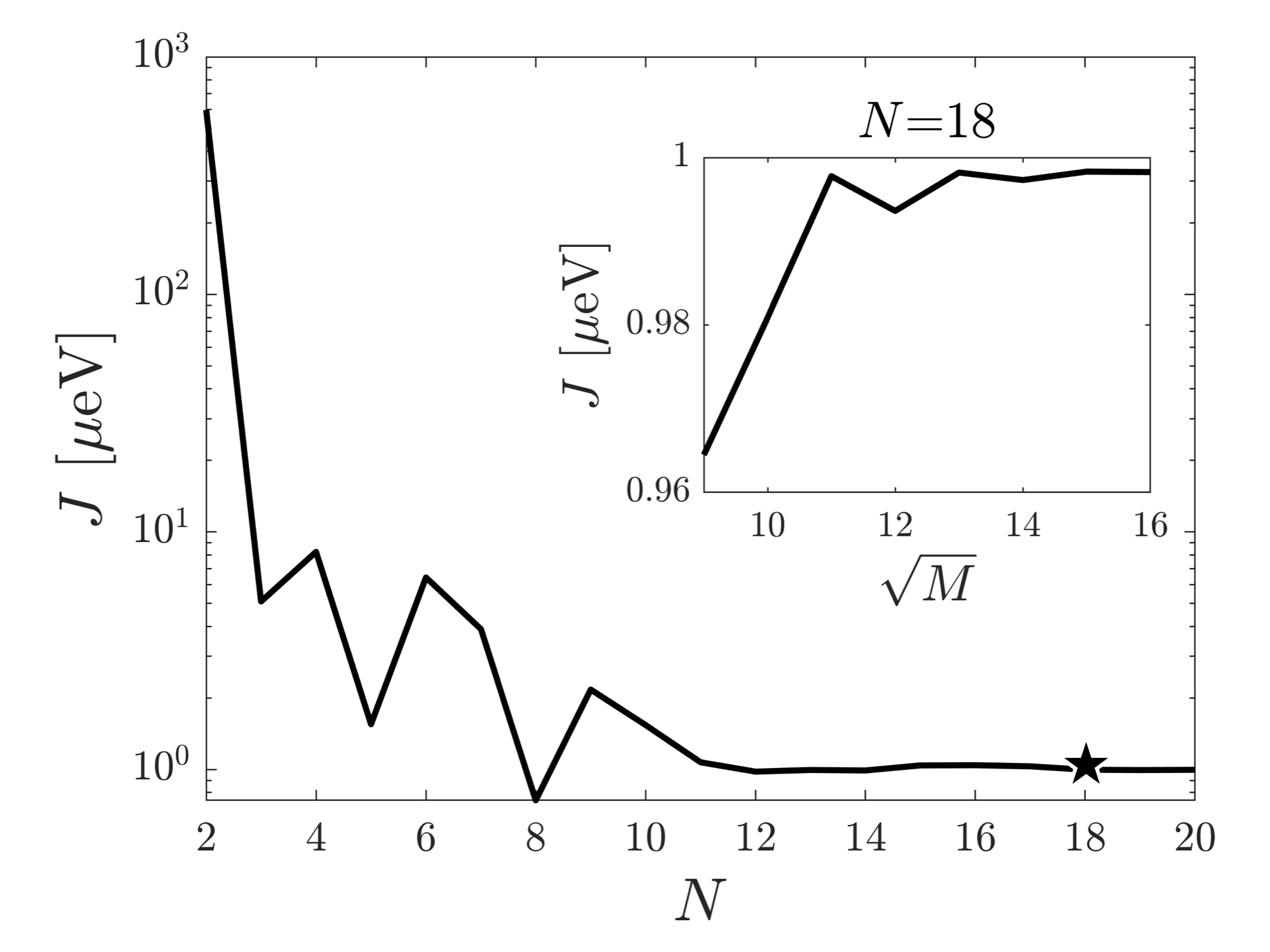}
    \caption{Convergence of $J$ with respect to the number of single-electron orbitals $N$ when $M=16^2$. The device parameters are $V_{p1}=$ $V_{p2}=$ 0.15 V, $V_t=$ 0.09455 V, $D_x=$ $D_y=$ 40 nm, $D_t=$ 20 nm, and $T=$ 15 nm. \textit{Inset:} The convergence of $J$ with respect to $M$ when $N=18$ (star in the main figure).}
    \label{fig:J_convergance}
\end{figure}

The 2D potential shown in Figure~\ref{fig:device_geom} is used to demonstrate the typical convergence behavior of $J$ the LCHO-CI method. Figure~\ref{fig:J_convergance} shows the calculated exchange strength $J$ versus the number of single-electron orbitals $N$ (main figure) and the number of harmonic orbitals $M$ (inset panel). The harmonic frequency $\omega$ is separately optimized at each data point, and in the main figure, $M=16^2$. As more single-electron orbitals are included, the higher order electron-electron correlations reduce the exchange splitting. Around $N=12$, $J$ begins to stabilize but does not stop varying by $<1\%$ until $N=18$, where $J\approx 1$~$\mu$eV. The inset shows how $J$ converges at $N=18$ with respect to the number of harmonic orbitals $M$. Good convergence ($<1\%$ variation) is achieved when $M>15^2$. Figure~\ref{fig:J_convergance} is an accurate representation of the qualitative convergence behaviour throughout the following work, where the device parameters are varied. Thus, all further exchange calculations use $N=18$ and $M=16^2$. These values give $n_C=$ 630 two-electron spin-orbital configuration states, $C_{\rm HO,\omega}$ of size $65536 \times 65536$, and $C_{SE}$ of size $324 \times 324$. 

\subsection{Exchange calculations with different gate geometries}
\label{sec:device_chargeNoise}

The 3D device structure discussed in Section~\ref{sec:device_model} allows us to study how the sensitivity of the exchange interaction to charge noise depends on the physical gate layout. This dependence is quantified by $\partial J/\partial V_{\rm bias}$, the derivative of the exchange strength $J$ with respect to a bias voltage $V_{\rm bias}$ applied between the two plunger gates that form the double QD. The bias can also be converted into an effective inter-dot detuning, $\epsilon$.
Here, $\epsilon = \epsilon_2 - \epsilon_1$, where $\epsilon_j$ is the ground state energy in the $j^{\rm th}$ QD. Fluctuations in $\epsilon$ can arise both from gate voltage noise and from charge traps in the surrounding material, and so $\partial J/\partial \epsilon$ is a general measure of the sensitivity of exchange to electrostatic noise.  

%The quantity $\partial J/\partial \epsilon$ is a measure of the sensitivity of $J$ to both fluctuations in the applied gate voltages, as well as electrostatic fluctuations due to charge traps in the surrounding material. 

We now outline the general procedure used to compute exchange energies throughout this section. First, for a series of varying device geometries, the tunneling gate voltage $V_{\rm tun}$ is varied while keeping the plunger gate voltage $V_{p} = V_{p1} = V_{p2}$ fixed. For each set of geometrical parameters, $J(V_{\rm tun})$ is linearly interpolated to find the value $V_{\rm tun}$ that gives $J = 1$ $\mu$eV (corresponding to a $\sim$2 ns SWAP quantum operation). At each interpolated $V_{\rm tun}$ value, a symmetric bias voltage $V_{\rm bias}$ is applied to the plunger gates such that $V_{p1} \rightarrow V_{p}-V_{\rm bias}/2$ and $V_{p2} \rightarrow V_{p}+V_{\rm bias}/2$. From the biased potentials, the plunger gate lever arm $\alpha$ that connects $V_{\rm bias}$ and $\epsilon$ as $\epsilon = \alpha V_{\rm bias}$ is calculated (refer to Appendix~\ref{app:parameter_extraction} for details). This gives two exchange derivatives for each device geometry: $\partial J/\partial V_{\rm bias}$ and $\partial J/\partial \epsilon$. Additionally, for each device geometry, we calculate the charging energy $U$ of each QD using the 2D potentials that correspond to $J=$ 1 $\mu$eV at zero bias. The exchange derivatives represent a non-trivial relationship between the bias and tunnel gate voltages and the device geometry.
This work focuses on exploring the impact that the device geometry has on the sensitivity of $J$ to charge noise. In order to make a systematic comparison and remove the effect of $V_{\rm tun}$ on $J$, we tune all geometries to the same reference $J$ value at zero bias. As a reminder, we use $N=18$ and $M=16^2$ for all modified LCHO-CI calculations. In this 2-electron system with no magnetic field, $J$ is the energy difference between the two lowest eigenstates and is strictly non-negative.

\subsubsection{Plunger gate size}

\begin{figure}
    \centering
    \includegraphics[width = 0.9\textwidth]{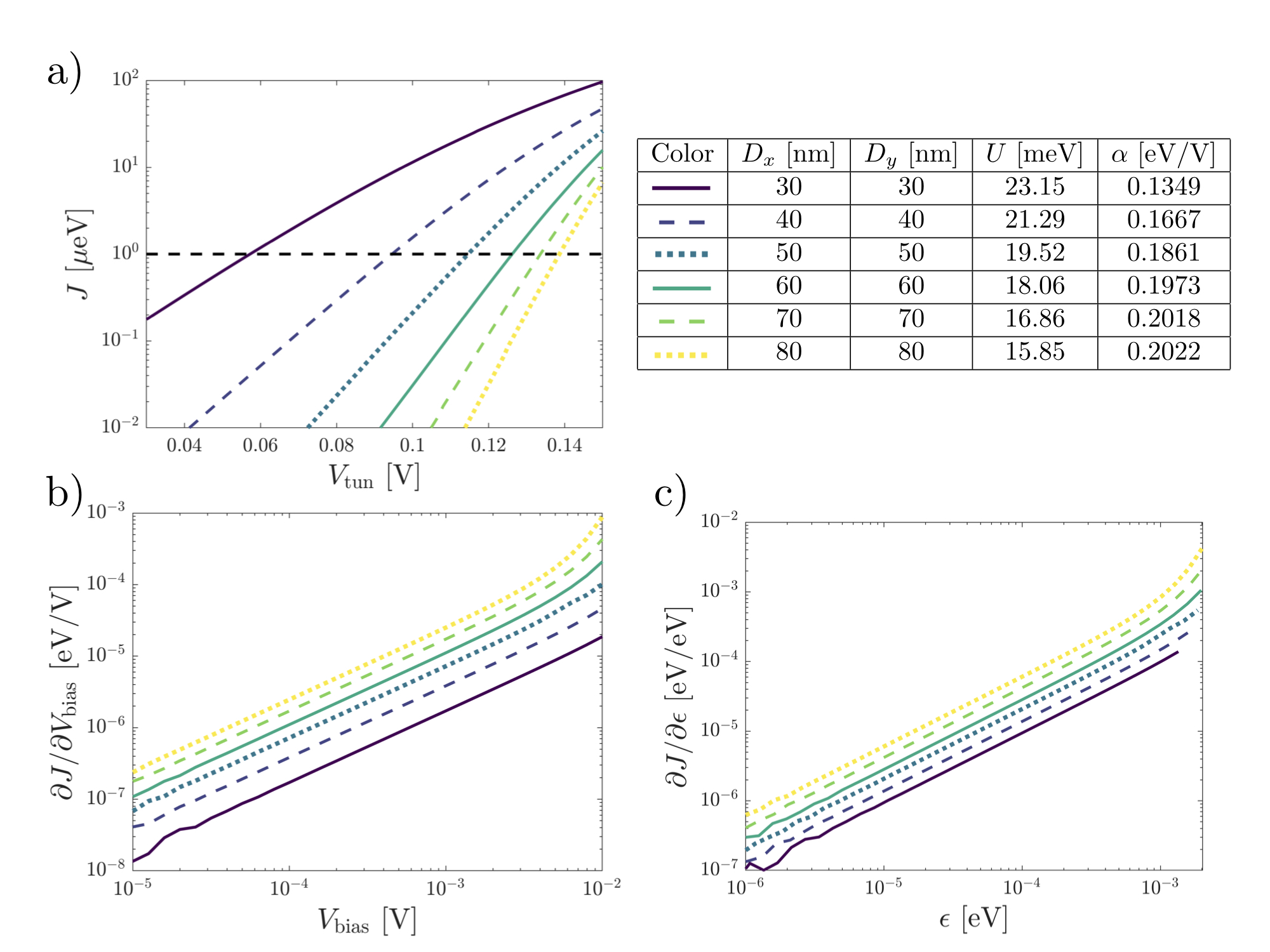}
    \caption{Dependence of $J$ on tunnel gate voltage as the dot size $D=D_x=D_y$ is varied. a) $J$ versus $V_{\rm tun}$, where the latter is varied from 0.03-0.150 V. The fixed device parameters are $V_{p}=0.150$ V, $V_{\rm bias}=0$ V, $D_t=$ 20 nm, and $T=$ 15 nm. The dashed black line indicates $J=$ 1 $\mu$eV. b) Derivative of $J$ with respect to $V_{\rm bias}$. For each device geometry, $V_{\rm tun}$ is tuned so that $J=$ 1 $\mu$eV at $V_{\rm bias}=0$. c) Derivative of $J$ with respect to $\epsilon$, where $\epsilon = \alpha V_{\rm bias}$ and $\alpha$ is the lever arm between the applied bias voltage and the effective inter-dot detuning. The upper-right table provides a legend and includes the calculated charging energy $U$ and lever arm for each geometry.}
    \label{fig:dot_size}
\end{figure}

We first study how varying the plunger gate size $D=D_x=D_y$ affects sensitivity to charge noise. The results are summarized in Figure~\ref{fig:dot_size}. $D_x$ and $D_y$ are varied together from 30~nm (purple) to 80 nm (yellow) in steps of 10 nm so that the plunger gate head remains square at each step. The tunnel gate width is $D_t =$ 20 nm, and the oxide thickness is $T =$ 15~nm. Figure~\ref{fig:dot_size}a shows how $J$ depends on $V_{\rm tun}$ when $V_{p} =$ 0.150 V. As $V_{\rm tun}$ increases, the tunnel barrier height decreases, and the localized electron orbitals have a larger overlap in the tunnel barrier region. This enhances the Coulomb interaction and increases $J$. For a fixed $V_{\rm tun}$, $J$ decreases with dot size. As the area of the plunger gate head increases, additional electric field contributions from the edge of the plunger gate push the double well minima lower, effectively raising the tunnel barrier. In other words, the capacitive coupling between each plunger gate and its QD increases with gate area. This has the effect of increasing the separation between electron orbitals, decreasing $J$. The $V_{\rm tun}$ values where $J=$ 1 $\mu$eV for each device geometry (indicated by the dashed black line) are given in Appendix~\ref{app:zero_bias_pots}.

Figures~\ref{fig:dot_size}b-c show the derivatives of $J$ with respect to $V_{\rm bias}$ and $\epsilon$, when $J=$ 1 $\mu$eV at zero bias. In both figures, the derivative increases with bias. 
This is due to the accumulated dipole characteristic of the singlet-like ground state with respect to the triplet-state \cite{li2010exchange,ramon2010decoherence}. At $V_{\rm bias}=$ $\epsilon=0$, both the low energy singlet- and triplet-like states are in the $(1,1)$ charge configuration. As a bias is applied, the singlet $S(1,1)$ superposes with the $S(0,2)$ charge configuration and acquires a dipole moment. However, the $T(1,1)$ state remains unaffected by the bias because the $T(0,2)$ state is energetically unavailable, so the triplet state acquires no dipole moment. The dipole moment in the singlet state makes the ground state more sensitive to electrostatic fluctuations as the bias increases. In Figures~\ref{fig:dot_size}b-c where $D_x=D_y\geq 60$ nm, the curve has an upturn near $V_{\rm bias} =$ $10^{-2}$~V; this is due to the singlet state being biased near the $S(1,1)-S(0,2)$ anticrossing.

Figures~\ref{fig:dot_size}b-c show that sensitivity with respect to fluctuations in $V_{\rm bias}$ and $\epsilon$ increases with dot size $D=D_x=D_y$. The physical reason can be understood by considering a Hubbard model for a double QD:
\begin{equation}
  H_{\rm Hub} = \sum_{j=1,2; \chi} \epsilon_{j} n_{j,\chi} + Un_{j,\chi}(n_{j,\chi} - 1) + \frac{t_c}{\sqrt{2}}(c_{1,\chi}^\dagger c_{2,\chi} + c_{2,\chi}^\dagger c_{1,\chi}),   
\end{equation}
where $c_{j,\chi}^\dagger$ creates an electron in the $j^{\rm th}$ QD with spin state $\chi$ and $n = c^\dagger c$.
$\epsilon_j$ is the QD ground state energy, $U$ is the QD charging energy (dots are assumed identical, so $U_1 = U_2$), and $t_c$ is the inter-dot tunnel coupling. $J$ is the difference between the ground and first excited states of $H_{\rm Hub}$ in the 2-electron basis. Typically, $t_c \ll U$, and in the small detuning limit where $\epsilon = \epsilon_2 - \epsilon_1 \ll U$, the ground state energy difference is given by \cite{reed2016reduced}
\begin{equation}
    J = \frac{2t_c^2\,U}{U^2 - \epsilon^2}.
    \label{eq:hub_J}
\end{equation}
This yields the derivatives
\begin{equation}
    \frac{\partial J}{\partial \epsilon} = \frac{4t_c^2\,U\,\epsilon}{(U^2  - \epsilon^2)^2} \quad,\quad \frac{\partial J}{\partial V_{\rm bias}} = \frac{4t_c^2\,U\,\alpha^2\,V_{\rm bias}}{(U^2  - \alpha^2V_{\rm bias}^2)^2}
    \label{eq:hub_dJ}
\end{equation}
where in the second expression $\epsilon$ is converted into the applied bias voltage via $\epsilon = \alpha V_{\rm bias}$. From Eq.~\ref{eq:hub_dJ}, we can see that both derivatives vanish at zero bias, $V_{\rm bias} = \epsilon = 0$.
The derivatives increase with bias until either $\epsilon \approx U$ or $\alpha V_{\rm bias} \approx U$. The width of this [0,$U$] interval depends on the magnitude of $U$ in the case of $\epsilon$, and on both $U$ and $\alpha$ in the case of $V_{\rm bias}$. If $U$ is large, then $\partial J/\partial \epsilon$ grows more slowly with bias, giving reduced sensitivity to charge noise compared to smaller $U$. For small $U$, the singlet state can more easily tunnel into the $S(0,2)$ charge configuration, increasing the sensitivity of $J$. For robustness to fluctuations in $V_{\rm bias}$, it is helpful for $\alpha$ to be small in addition to $U$ being large.

The table in Figure~\ref{fig:dot_size} shows that $U$ decreases and $\alpha$ increases with dot size. Both effects are unsurprising, as
the charging energy is inversely proportional to the QD radius \cite{kouwenhoven1997electron}, and the capacitive coupling between the plunger gate and QD increases with the gate area.
Even when the capacitive coupling is accounted for by converting $V_{\rm bias}$ into $\epsilon$ using $\alpha$, we see that $\partial J/\partial \epsilon$ still increases with dot size, which we ascribe to decreasing charging energy. In summary, the results of Figures~\ref{fig:dot_size}b-c show that smaller QDs are less susceptible to charge noise from both $V_{\rm bias}$ and $\epsilon$.

\begin{figure}[t]
    \centering
    \includegraphics[width = 0.9\textwidth]{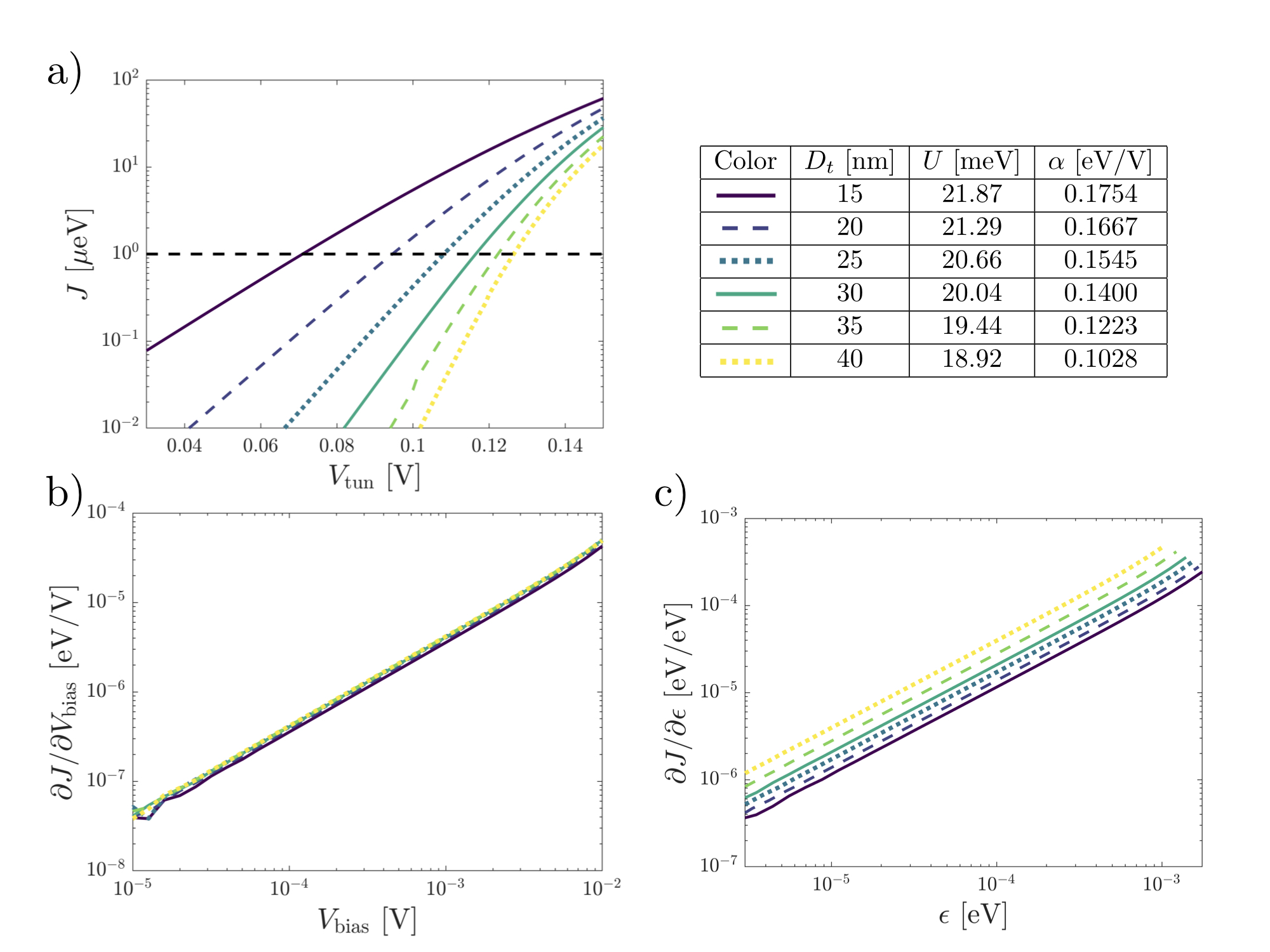}
    \caption{Dependence of $J$ on a varying tunnel gate width $D_t$. a) $J$ versus $V_{\rm tun}$, as the latter is varied from 0.03-0.150 V. The fixed device parameters are $V_{p}=0.150$ V, $V_{\rm bias}=0$ V, $D_x=$ $D_y=$ 40 nm, and $T=$ 15 nm. The dashed black line indicates $J=$ 1 $\mu$eV. b) Derivative of $J$ with respect to $V_{\rm bias}$. For each device geometry, $V_{\rm tun}$ is tuned so that $J=$ 1 $\mu$eV at $V_{\rm bias}=0$. c) Derivative of $J$ with respect to $\epsilon$, where $\epsilon = \alpha V_{\rm bias}$ and $\alpha$ is the gate lever arm. The upper-right table provides a legend and includes the charging energy $U$ and lever arm $\alpha$ versus $D_t$.}
    \label{fig:tun_gate}
\end{figure}

\subsubsection{Tunnel gate width}

Next, we study how varying the tunnel gate width $D_t$ affects the sensitivity to charge noise. The results are summarized in Figure~\ref{fig:tun_gate}, where $D_t$ is varied from 15 nm (purple) to 40 nm (yellow) in increments of 5 nm. The fixed device parameters are $D_x=$ $D_y=$ 40 nm, $T=$ 15 nm, and $V_p=$ 0.150 V.
Figure~\ref{fig:tun_gate}a shows the variation of $J$ versus $V_{\rm tun}$. For all curves, $J$ increases with $V_{\rm tun}$, as expected due to the decreasing tunnel barrier height. At a fixed $V_{\rm tun}$, $J$ decreases as $D_t$ increases. This is because the tunnel barrier region gets wider, reducing the orbital overlap in the tunnel barrier region. The specific $V_{\rm tun}$ values at which $J=$ 1 $\mu$eV are given in Appendix~\ref{app:zero_bias_pots}.

Figures~\ref{fig:tun_gate}b-c show the derivative of $J$ with respect to $V_{\rm bias}$ and $\epsilon$ for different tunnel gate widths. For all curves, $J=$ 1 $\mu$eV at $V_{\rm bias}=0$.
Unlike the previous example, there is a clear qualitative difference between the $V_{\rm bias}$ and $\epsilon$ derivative curves. Although the plunger gate size remains fixed, $U$ decreases as the tunnel gate becomes wider. For larger $D_t$, $V_{\rm tun}$ must be tuned to higher voltages to reach $J=$ 1 $\mu$eV at zero bias voltage, as shown in Figure~\ref{fig:tun_gate}a.
Both the higher $V_{\rm tun}$ value and increased tunnel gate width cause the dot potentials to widen, effectively increasing the QD radius (see Figure~\ref{fig:zero_bias_pots}b in Appendix~\ref{app:zero_bias_pots}) and decreasing the charging energy. The table in Figure~\ref{fig:tun_gate} shows that the lever arm $\alpha$ also decreases with increasing tunnel gate width. We attribute this to a partial screening of the plunger gate by the tunnel gate.

The fact that both $U$ and $\alpha$ decrease with increasing $D_t$ has an interesting effect on $\partial J/\partial V_{\rm bias}$. Recall that $\partial J/\partial V_{\rm bias}$ increases for smaller $U$ and larger $\alpha$. Here, $\alpha$ is a decreasing function of $D_t$, reducing $\partial J/\partial V_{\rm bias}$. The counteracting effects of $U$ and $\alpha$ almost perfectly cancel in this device geometry so that $\partial J/\partial V_{\rm bias}$ is nearly independent of varying tunnel gate width. However, when the bias voltage is translated into effective dot detuning, $J$ becomes more sensitive to $\epsilon$ as the tunnel gate widens.
This is due to the dependence of $U$ on $D_t$ as discussed above. In summary, Figures~\ref{fig:tun_gate}b-c show that narrower tunnel gates give reduced susceptibility to fluctuations in $\epsilon$, and that tunnel gate size has minimal impact on noise in the applied bias. 

\begin{figure}[t]
    \centering
    \includegraphics[width = 0.9\textwidth]{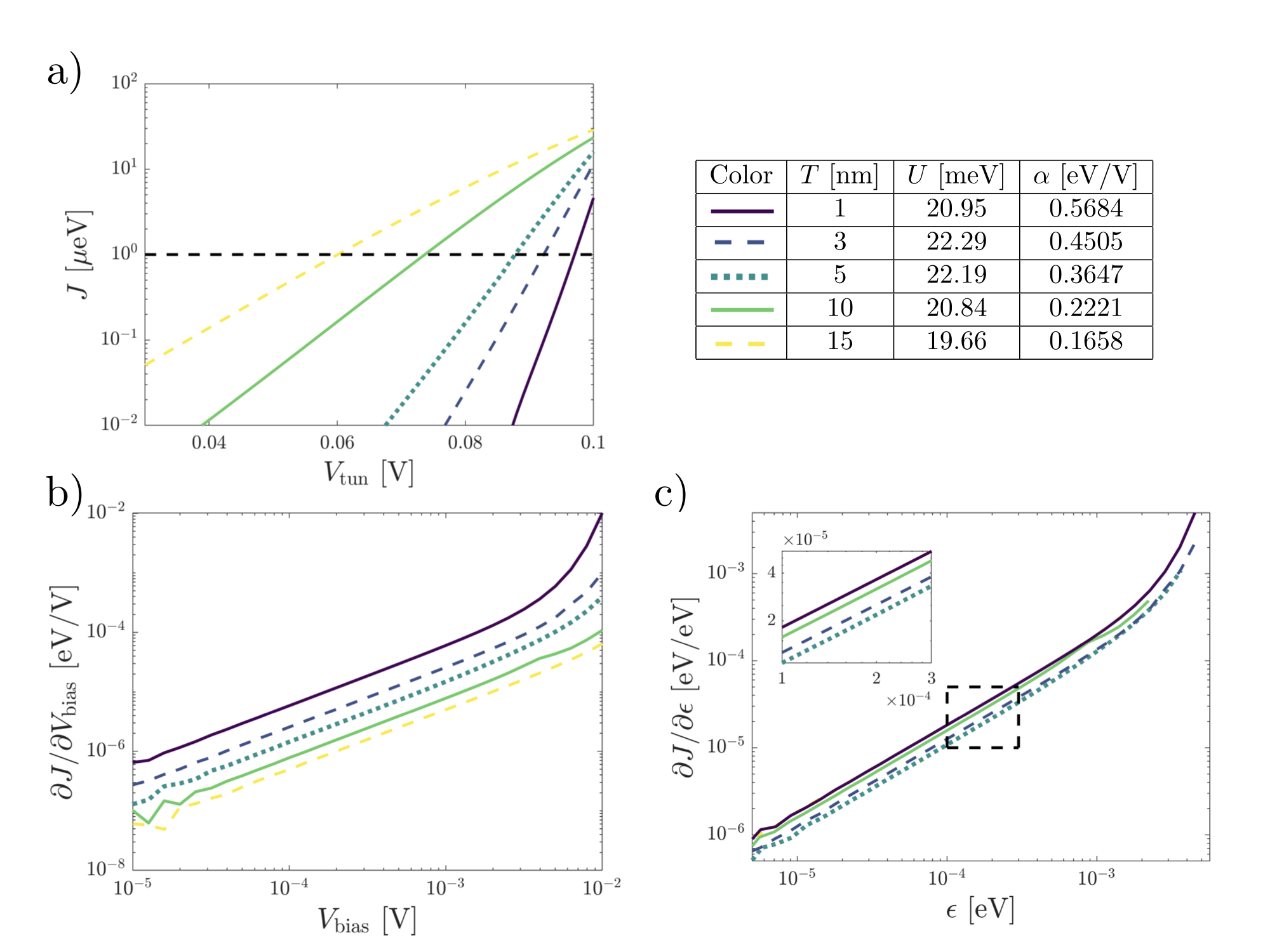}
    \caption{Dependence of $J$ on varying oxide thickness $T$. a) $J$ versus $V_{\rm tun}$, as the latter is varied from 0.0-0.100 V. The fixed device parameters are $V_{p}=0.100$ V, $V_{\rm bias}=0$ V, $D_x=$ $D_y=$ 40 nm, and $D_t=$ 20 nm. b) Derivative of $J$ with respect to $V_{\rm bias}$. Each device has $V_{\rm tun}$ tuned so that $J=$ 1 $\mu$eV at $V_{\rm bias}=0$, as indicated by the dashed black line in panel (a). c) Derivative of $J$ with respect to $\epsilon$, where $\epsilon = \alpha V_{\rm bias}$ and $\alpha$ is the lever arm between the applied bias voltage and the effective inter-dot detuning. The upper-right table provides a legend and includes the charging energies and lever arms versus $T$.}
    \label{fig:oxide}
\end{figure}

\subsubsection{Oxide thickness}

Next, we study the effects of variation in the oxide thickness $T$. The results are shown in Figure~\ref{fig:oxide}, where $T$ ranges from 1 nm (purple) to 15 nm (yellow). The fixed device parameters are $D_x=$ $D_y=$ 40 nm, $D_t=$ 20 nm, and $V_p=$ 0.100 V.
Figure~\ref{fig:oxide}a shows the dependence of $J$ on $V_{\rm tun}$. For all oxide thicknesses, $J$ increases with $V_{\rm tun}$ as expected. At a fixed $V_{\rm tun}$, $J$ decreases as the oxide becomes thinner. This occurs because the plunger and tunnel gates acquire a stronger capacitive coupling to the QDs as they move closer to the Si/SiO$_2$ interface. Compared to the tunnel gates, the plunger gates have a stronger effect due to their larger size, and because $V_{\rm tun} < V_p$. The net effect is a larger tunnel barrier height, and thus weaker exchange interaction, for a thinner oxide. The $V_{\rm tun}$ values that yield $J=$ 1~$\mu$eV are given in Appendix~\ref{app:zero_bias_pots}.

\begin{figure}
    \centering
    \includegraphics[width = 0.9\textwidth]{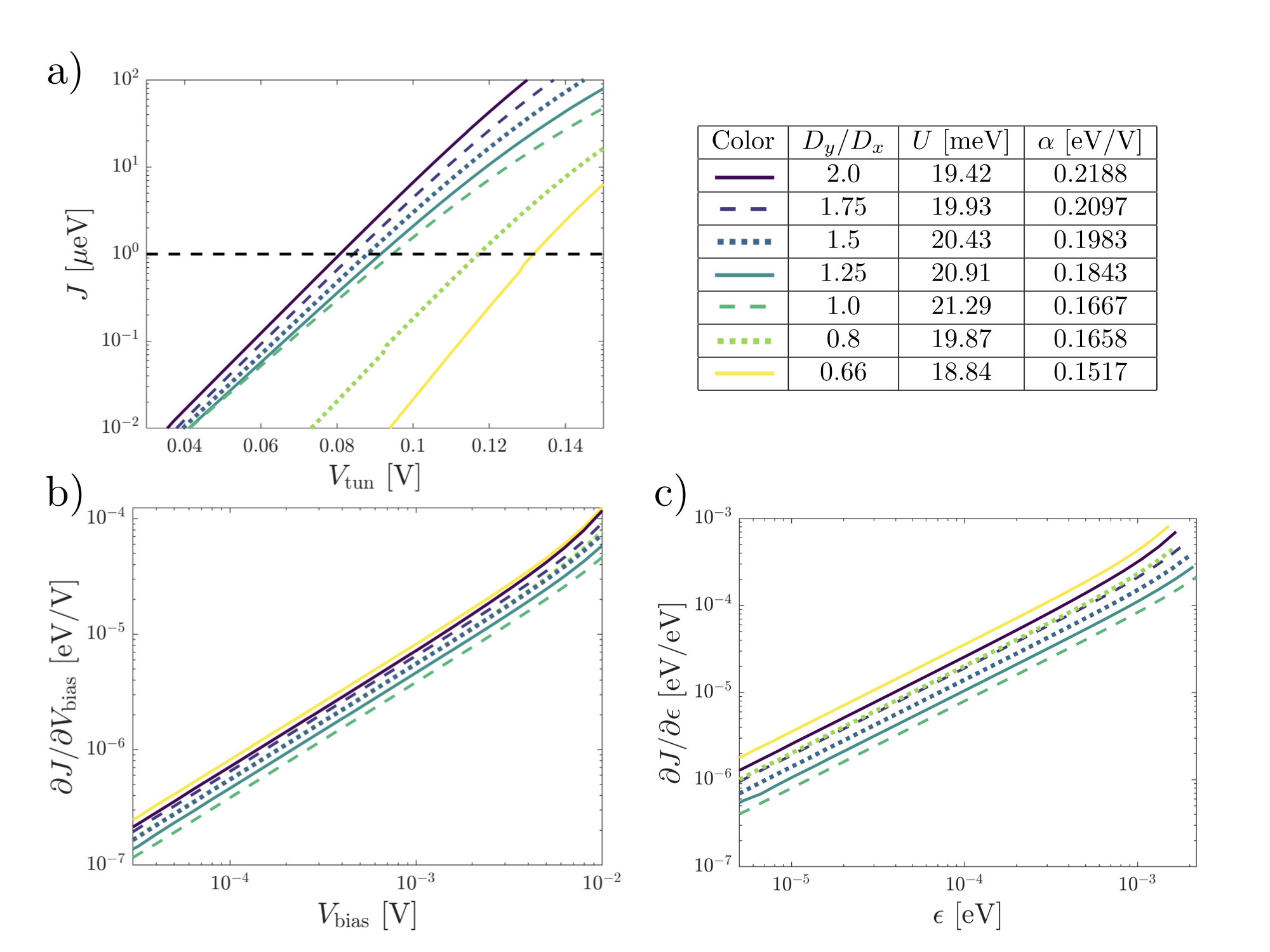}
    \caption{Dependence of $J$ on the plunger gate eccentricity, $D_y/D_x$. For $D_y/D_x>$ 1.0, $D_x=$ 40 nm, while for $D_y/D_x<$ 1.0, $D_y=$ 40 nm. a) $J$ versus $V_{\rm tun}$, as the latter is varied from 0.03-0.150 V. The fixed device parameters are $V_{p}=0.150$ V, $V_{\rm bias}=0$ V, $D_t=$ 20 nm, and $T=$ 15 nm. b) Derivative of $J$ with respect to $V_{\rm bias}$. Each device has $V_{\rm tun}$ tuned so that $J=$ 1 $\mu$eV at $V_{\rm bias}=0$, as indicated by the dashed black line in panel (a). c) Derivative of $J$ with respect to $\epsilon$, where $\epsilon = \alpha V_{\rm bias}$ and $\alpha$ is the lever arm between the applied bias voltage and the effective inter-dot detuning. The upper-right table provides a legend and includes the charging energy and lever arm versus dot eccentricity.}
    \label{fig:eccentricity}
\end{figure}

Figures~\ref{fig:oxide}b-c show the derivatives of $J$ with respect to $V_{\rm bias}$ and $\epsilon$ as $T$ is varied. For all curves, $J=$ 1 $\mu$eV at zero bias. The derivative $\partial J/\partial V_{\rm bias}$ depends more strongly on the oxide thickness than $\partial J/\partial \epsilon$. This is because the plunger gate lever arm $\alpha$ increases significantly as $T$ decreases, and this overshadows the impact of the charging energy $U$, which varies non-monotonically with $T$ (see upper-right table). The individual $\partial J/\partial \epsilon$ curves displayed in Figure~\ref{fig:oxide}c are ordered according to $U$, as can be seen in the inset panel. The non-monotonic behavior of $U$ with respect to $T$ is surprising. As $T$ decreases, the images of the gates are mapped more clearly onto the underlying potential landscape. In the limit $T\rightarrow 0$, the 2D potential would be two square wells connected by a square tunnel barrier. One might expect that as $T$ decreases, the well should become broader, reducing the confinement strength (and $U$) as the potential wells transition from a $\cup$-like to a $\sqcup$-like shape. However, we see that $U$ increases as the oxide is thinned from 15 nm to 3 nm. Below 3 nm, $U$ decreases. The potentials at zero-bias are shown in Appendix~\ref{app:zero_bias_pots}. As $T$ is reduced, the tunnel barrier transitions from $\cap$-like to $\sqcap$-like as expected. However, the slopes of the outer edges of the QD wells remain mostly unaffected until $T\approx 3$ nm. From $T=$ 15 nm to 5 nm, the outer edges of the potential move toward the center of the QD, creating tighter confinement and raising $U$. For $T=$ 3 nm to 1 nm, the outer edges of the potential shift outward and begin opening into the expected $\sqcup$-like shape, reversing the trend and lowering $U$.

The non-monotonic behavior of $U$ could be due to choices made in the device model. For example, the detailed shape of the potential landscape is affected by our choices to ground the outer tunnel barrier gates and to perform the self-consistent 3D Poisson calculation in the zero-electron regime. However, we expect that the qualitative behavior of $\partial J/\partial V_{\rm bias}$ should be robust to these details, since the dominant mechanism determining sensitivity with respect to $V_{\rm bias}$ is the lever arm, not $U$. On the other hand, the behavior of $\partial J/\partial \epsilon$ may differ under another set of model choices, as it is expected to depend mainly on the charging energy. We have not explicitly modeled charge traps, but note in passing that the number of charge fluctuators in the bulk of the oxide is reduced for thinner oxides. It is not clear, however, whether the dominant charge noise in experiments originates from bulk or interface defects. In summary, using our device model, it is found that reducing the oxide thickness increases the sensitivity to noise in $V_{\rm bias}$, but has a relatively minor effect on the sensitivity to charge noise caused by fluctuations in $\epsilon$.

\subsubsection{Plunger gate eccentricity}

Lastly, we study the effects of the eccentricity of the plunger gate dimensions, $D_y/D_x$. The results are summarized in Figure~\ref{fig:eccentricity}. Eccentricities $D_y/D_x>1.0$ indicate the plunger gate is elongated along the $y$-axis, while values $<1.0$ mean the plunger gate is elongated along the $x$-axis. For $D_y/D_x>1.0$, $D_x=$ 40 nm while $D_y$ is varied.
Conversely, for $D_y/D_x<1.0$, $D_y=$ 40 nm while $D_x$ is varied.
At $D_y/D_x=1.0$, $D_x=D_y=$ 40 nm. For all cases, $D_t=$ 20 nm, $T=$ 15 nm, and $V_p=$ 0.150 V. Figure~\ref{fig:eccentricity}a shows $J$ versus $V_{\rm tun}$, where $J$ increases with $V_{\rm tun}$ as expected. Interestingly, at a fixed value of $V_{\rm tun}$, $J$ decreases monotonically with $D_y/D_x$. When $D_y/D_x>1.0$, the dots are elongated along the $y$ direction and the net effect of the electric field distribution is to reduce the tunnel barrier height relative to the well minima. When $D_y/D_x<1.0$, the dots are elongated along the $x$-axis and the QD well minima are lowered relative to the tunnel barrier. The dashed black line shows where $J=$ 1 $\mu$eV, and the corresponding $V_{\rm tun}$ values are given in Appendix~\ref{app:zero_bias_pots}.

Figures~\ref{fig:eccentricity}b-c show the derivatives of $J$ with respect to $V_{\rm bias}$ and $\epsilon$ as $D_y/D_x$ is varied. For all curves, $J=$ 1 $\mu$eV at zero bias. Both derivatives $\partial J/\partial V_{\rm bias}$ and $\partial J/\partial \epsilon$ show a similar dependence on the dot eccentricity.
The lowest sensitivity is found for $D_y/D_x=1.0$, where the QDs have the smallest area, and therefore the largest charging energy $U$. When $D_y/D_x\neq1.0$, the larger surface area reduces $U$ and increases the susceptibility to fluctuations in $V_{\rm bias}$ and $\epsilon$. This trend for $\partial J/\partial V_{\rm bias}$ occurs even though the lever arm $\alpha$ decreases monotonically with $D_y/D_x$, showing that $U$ is the dominant factor. 
In summary, we find that symmetric QDs with no eccentricity are the least susceptible to charge noise from both $V_{\rm bias}$ and $\epsilon$, however, this appears to be mainly an area/size effect. It would be interesting to study the case when the plunger gate area is held constant as eccentricity is varied.

\section{Conclusion}
\label{sec:conclusion}

In summary, a modified LCHO-CI method was presented for calculating the many-electron states and energy spectra of a quantum dot network. Using an orthogonal basis of harmonic orbitals to approximate single-electron orbitals, the evaluation of the Coulomb matrix elements requires significantly less computational resources. Additionally, we demonstrate an efficient method for optimizing the choice of harmonic orbital basis to better approximate the single-electron orbitals and improve the accuracy of the CI calculation. Our modified LCHO-CI approach provides a significant reduction in computation time that can be exploited to obtain a large number of accurate energy spectra as a function of varying model parameters. The energy spectra can be mapped to an effective Heisenberg Hamiltonian to obtain the pairwise exchange interaction energies $J_{ij}$ in arbitrary quantum dot networks.
While the modified LCHO-CI method works in principle for general dot networks, we note that the computational efficiency of this approach for larger dot networks is a subject for future research. More dots will require larger orbital basis sets in order to achieve convergence of the many-electron energy spectra.
We expect it will be important to choose the eccentricity of the harmonic orbital basis properly. 
For square or triangular dot networks, a radially symmetric harmonic orbital basis, such as the one used throughout this work, would be appropriate. For linear dot chains, on the other hand, an eccentric harmonic orbital basis will most likely provide the fastest convergence of the LCHO-CI calculation. Investigating the convergence of the modified LCHO-CI method for different network geometries and topologies is left for future work.

The modified LCHO-CI approach was then used to investigate how the physical gate geometry of a quantum dot device impacts the sensitivity of exchange to charge noise. A 3D model of a MOSFET (Si/SiO$_2$) double quantum dot device was simulated using a self-consistent Poisson calculation. From these 3D simulations, planar 2D potentials were used in the LCHO-CI calculations to determine how $J$ varies with respect to an applied bias voltage between the plunger gates, and also with respect to the effective inter-dot detuning. The charge noise sensitivity was calculated as a function of plunger gate size, tunnel gate width, SiO$_2$ (gate oxide) thickness and dot eccentricity. Generally, device geometries that maximize the dot charging energy and reduce the lever arm of the plunger gates are found to be less sensitive to charge noise. For the device layout chosen in this work, this means that small and symmetric plunger gates, narrow tunnel gates, and suitably thick SiO$_2$ will improve the robustness to charge noise. Future work includes obtaining better approximations the electronic potential landscape by using self-consistent Schr\"{o}dinger-Poisson calculations; however, we expect that this will not substantially change the qualitative trends observed in this work. Many additional device geometries are ripe for exploration using our methods, such as asymmetric double quantum dots \cite{hiltunen2015charge}. We envision these methods enabling the realistic modeling of two-qubit quantum logic gates in quantum dot networks, such as multi-qubit processor nodes. 

\section*{Acknowledgements}
This research was undertaken thanks in part to funding from the Canada First Research Excellence Fund (CFREF) and the National Sciences and Engineering Research Council of Canada (NSERC). We thank S. Birner and Z. Wasilewski for assistance with software setup.

\appendix
\section{Derivation of analytical formula for Coulomb matrix elements}
\label{app:CMEs_symHOs}

In this section we give the derivation for the Coulomb matrix elements in the harmonic oscillator basis.  
The derivation follows similarly to Chapter 3 of Korkusinski's thesis \cite{korkusinski2004correlations}.
To begin, we rewrite the Coulomb potential into plane waves by using the inverse Fourier transform
\begin{align}
\label{eq:appEquation1}
    \bra{\alpha\beta}\frac{1}{|\vec{r}_2 - \vec{r}_1|}\ket{\gamma\delta} &= \bra{\alpha\beta} \frac{1}{8\pi^2} \int d\vec{q} \, \frac{4\pi}{q} e^{i\vec{q}(\vec{r}_1 - \vec{r}_2)}\ket{\gamma\delta} \nonumber \\
    &= \frac{1}{2\pi} \int d\vec{q} \, \frac{1}{q} \bra{\alpha}e^{i\vec{q}\vec{r}_1}\ket{\delta} \bra{\beta}e^{-i\vec{q}\vec{r}_2}\ket{\gamma}
\end{align}
where $\ket{\alpha}$ corresponds to a 2D harmonic orbital state.
The 2D harmonic orbital states are products of the 1D harmonic orbital states $\ket{\alpha} = \ket{n_\alpha m_\alpha}$ where $n_\alpha$ and $m_\alpha$ are the harmonic oscillator modes along the $x$ and $y$ axes respectively.
Next, we rewrite the position coordinates in terms of the canonical harmonic oscillator ladder operators
\begin{equation}
    x_i = \frac{1}{\sqrt{2\omega}}(a_i + a_i^\dagger)   \,\,\,\,\,\,\,\,\,\,   y_i = \frac{1}{\sqrt{2\omega}}(b_i + b_i^\dagger)
\end{equation}
where $\omega$ is the harmonic oscillator frequency.

We now focus on the first matrix element in Eq.~\ref{eq:appEquation1}, $\bra{\alpha}e^{i\vec{q}\vec{r}_1}\ket{\delta}$. 
After rewriting this matrix element in terms of ladder operators, it becomes
\begin{equation}
    \bra{\alpha}e^{i\vec{q}\vec{r}_1}\ket{\delta} = \bra{\alpha} e^{\frac{iq_x}{\sqrt{2\omega}}(a_1^\dagger + a_1)} e^{\frac{iq_y}{\sqrt{2\omega}}(b_1^\dagger + b_1)}\ket{\delta}.
\end{equation}
Using the fact that $a$ and $b$ commute as well as the Baker-Campbell-Hausdorff formula $e^{X+Y}$ $= e^X e^Y e^{-\frac{1}{2}[X,Y]}$ (valid when $[X,[X,Y]]$ $=$ $[Y,[X,Y]] = 0$), the matrix element can be written as
\begin{equation*}
    \bra{\alpha}e^{i\vec{q}\vec{r}_1}\ket{\delta} = e^{-\frac{1}{4\omega}(q_x^2 + q_y^2)} \bra{\alpha} e^{\frac{iq_x}{\sqrt{2\omega}} a_1^\dagger} e^{\frac{iq_y}{\sqrt{2\omega}} b_1^\dagger} e^{\frac{iq_x}{\sqrt{2\omega}}a_1} e^{\frac{iq_y}{\sqrt{2\omega}} b_1}\ket{\delta}
\end{equation*}
with a similar expression for the second matrix element $\bra{\beta}e^{-i\vec{q}\vec{r}_2}\ket{\gamma}$ where $i \rightarrow -i$.

Next, each 2D harmonic orbital state $\ket{\alpha}$ can be rewritten as the repeated application of the creation operator on the vacuum state $\ket{00}$
\begin{equation}
    \ket{\alpha} = \ket{n_\alpha m_\alpha} = \frac{1}{\sqrt{n_\alpha ! m_\alpha !}}\left(a^\dagger\right)^{n_\alpha}\left(b^\dagger\right)^{m_\alpha}\ket{00}.
\end{equation}
We also insert the following identity operator into the middle of the matrix element expression
\begin{equation}
    \hat{\mathbf{I}} = \sum_{p_1 = 0}^\infty \sum_{p_2 = 0}^\infty \ket{p_1p_2}\bra{p_2p_1} = \frac{1}{p_1!p_2!} \sum_{p_1 = 0}^\infty \sum_{p_2 = 0}^\infty \left(a_1^\dagger\right)^{p_1} \left(b_1^\dagger\right)^{p_2} \ket{00}\bra{00} b_1^{p_2} a_1^{p_1}
\end{equation}
where $p_1$ and $p_2$ are two dummy indices swept over.
Using these above expressions, as well as Taylor expanding the exponential operator terms, the matrix element becomes
\begin{align}
\label{eq:appEquation6}
    \bra{\alpha}e^{i\vec{q}\vec{r}_1}\ket{\delta} &= \frac{e^{-\frac{1}{4\omega}(q_x^2 + q_y^2)}}{\sqrt{n_\alpha! m_\alpha! n_\delta! m_\delta!}} \sum_{p_1 = 0}^\infty \sum_{p_2 = 0}^\infty \frac{1}{p_1!p_2!} \sum_{s_1, s_2, s_3, s_4 = 0}^\infty \frac{\left(\frac{iq_x}{\sqrt{2\omega}}\right)^{s_1 + s_3}}{s_1!s_3!} \frac{\left(\frac{iq_y}{\sqrt{2\omega}}\right)^{s_2 + s_4}}{s_2!s_4!} \nonumber \\
    &\quad\times \bra{00} a_1^{n_\alpha} \left(a_1^\dagger\right)^{p_1 + s_1} b_1^{m_\alpha} \left(b_1^\dagger\right)^{p_2 + s_2} \ket{00} \bra{00} a_1^{p_1 + s_3} \left(a_1^\dagger\right)^{n_\delta} b_1^{p_2 + s_4} \left(b_1^\dagger\right)^{m_\delta} \ket{00}
\end{align}
%
% Expression with taylor expansion yet (for sanity checking)
%\begin{align*}
%\bra{\alpha}e^{i\vec{q}\vec{r}_1}\ket{\delta} &= \frac{e^{-\frac{1}{4\omega}(q_x^2 + q_y^2)}}{\sqrt{n_\alpha! m_\alpha! n_\delta! m_\delta!}} \sum_{p_1 = 0}^\infty \sum_{p_2 = 0}^\infty \frac{1}{p_1!p_2!}\\
%&\times \bra{00} a_1^{n_\alpha} e^{\frac{iq_x}{\sqrt{2\omega}} a_1^\dagger}  \left(a_1^\dagger\right)^{p_1} b_1^{m_\alpha} e^{\frac{iq_y}{\sqrt{2\omega}} b_1^\dagger} \left(b_1^\dagger\right)^{p_2} \ket{00} \\
%&\times \bra{00} a_1^{p_1} e^{\frac{iq_x}{\sqrt{2\omega}}a_1} \left(a_1^\dagger\right)^{n_\delta} b_1^{p_2} e^{\frac{iq_y}{\sqrt{2\omega}} b_1} \left(b_1^\dagger\right)^{m_\delta} \ket{00}\\
%\end{align*}
where $s_i$ are the indices in the Taylor expansions. 

There are two things of note which simplify Eq.~\ref{eq:appEquation6}.
The first is that $p_1$ and $p_2$ cannot sweep all the way to $\infty$ but to only $\min(n_\alpha,n_\delta)$ and $\min(m_\alpha,m_\delta)$ respectively.
Above these limits, there will be indices of $s_i$ which result in the annihilation operator being applied onto the vacuum state.
The second thing is that the matrix elements are nonzero only when $p_1  + s_1 = n_\alpha$ with analogous relationships for the other three $s_i$ and $p_i$ indexing pairs.
These observations reduce the matrix element to
\begin{align}
    \bra{\alpha}e^{i\vec{q}\vec{r}_1}\ket{\delta} &= \frac{e^{-\frac{1}{4\omega}(q_x^2 + q_y^2)}}{\sqrt{n_\alpha! m_\alpha! n_\delta! m_\delta!}} \nonumber \\
    &\quad\times \sum_{p_1 = 0}^{\min(n_\alpha,n_\delta)} \sum_{p_2 = 0}^{\min(m_\alpha,m_\delta)} \frac{1}{p_1!p_2!} \frac{\left(\frac{iq_x}{\sqrt{2\omega}}\right)^{n_\alpha + n_\delta - 2p_1}}{(n_\alpha - p_1)!(n_\delta - p_1)!} \frac{\left(\frac{iq_y}{\sqrt{2\omega}}\right)^{m_\alpha + m_\delta - 2p_2}}{(m_\alpha - p_2)!(m_\delta - p_2)!} \nonumber\\
    &\quad\times \bra{00} a_1^{n_\alpha} \left(a_1^\dagger\right)^{n_\alpha} b_1^{m_\alpha} \left(b_1^\dagger\right)^{m_\alpha} \ket{00} \bra{00} a_1^{n_\delta} \left(a_1^\dagger\right)^{n_\delta} b_1^{m_\delta} \left(b_1^\dagger\right)^{m_\delta} \ket{00}.
\end{align}
Applying the ladder operators introduces a scalar term $n_\alpha!m_\alpha!n_\delta!m_\delta!$, and the expression simplifies to
\begin{align}
    \bra{\alpha}e^{i\vec{q}\vec{r}_1}\ket{\delta} &= \frac{e^{-\frac{1}{4\omega}(q_x^2 + q_y^2)}}{\sqrt{n_\alpha! m_\alpha! n_\delta! m_\delta!}} \sum_{p_1 = 0}^{\min(n_\alpha,n_\delta)} \sum_{p_2 = 0}^{\min(m_\alpha,m_\delta)} \left(\frac{iq_x}{\sqrt{2\omega}}\right)^{n_\alpha + n_\delta - 2p_1} \left(\frac{iq_y}{\sqrt{2\omega}}\right)^{m_\alpha + m_\delta - 2p_2}\nonumber\\
    &\quad\times p_1!p_2! \binom{n_\alpha}{p_1} \binom{n_\delta}{p_1} \binom{m_\alpha}{p_2} \binom{m_\delta}{p_2}
\end{align}
with an analogous expression for the second matrix element $\bra{\beta}e^{-i\vec{q}\vec{r}_2}\ket{\gamma}$ where $i \rightarrow -i$ and $p_1,p_2 \rightarrow p_3,p_4$.

Assembling everything together gives 
\begin{equation}
\label{eq:appAEquation9}
   \begin{split}
        \bra{\alpha\beta}v\ket{\gamma\delta} &= \frac{1}{2\pi} \int d\vec{q} \, \frac{1}{q} \bra{\alpha}e^{i\vec{q}\vec{r}_1}\ket{\delta} \bra{\beta}e^{-i\vec{q}\vec{r}_2}\ket{\gamma} \\
        &= \frac{1}{2\pi\sqrt{n_\alpha! m_\alpha! n_\delta! m_\delta! n_\beta! m_\beta! n_\gamma! m_\gamma!}} \sum_{p_1 = 0}^{\min(n_\alpha,n_\delta)} p_1! \binom{n_\alpha}{p_1} \binom{mn_\delta}{p_1} \\
        &\quad\times \sum_{p_2 = 0}^{\min(m_\alpha,m_\delta)} p_2! \binom{m_\alpha}{p_2} \binom{m_\delta}{p_2} \sum_{p_3 = 0}^{\min(n_\beta,n_\gamma)} p_3! \binom{n_\beta}{p_3} \binom{m_\gamma}{p_3} \\
        &\quad\times \sum_{p_4 = 0}^{\min(m_\beta,m_\gamma)} p_4! \binom{m_\beta}{p_4} \binom{m_\gamma}{p_4} I_{p_1p_2p_3p_4}
    \end{split} 
\end{equation}
where the $I_{p_1p_2p_3p_4}$ is
\begin{equation}
    \begin{split}
        I_{p_1p_2p_3p_4} &= \int \frac{d\vec{q}}{q} e^{-\frac{1}{2\omega}(q_x^2 + q_y^2)} \left(\frac{iq_x}{\sqrt{2\omega}}\right)^{n_\alpha + n_\delta - 2p_1} \left(\frac{iq_y}{\sqrt{2\omega}}\right)^{m_\alpha + m_\delta - 2p_2} \\ 
&\quad\times \left(\frac{-iq_x}{\sqrt{2\omega}}\right)^{n_\beta + n_\gamma - 2p_3} \left(\frac{-iq_y}{\sqrt{2\omega}}\right)^{m_\beta + m_\gamma - 2p_4}
    \end{split}.
\end{equation}

We will now focus on evaluating $I_{p_1p_2p_3p_4}$.
After converting to polar coordinates and using the change of variables $x = q/\sqrt{2\omega}$, the integral can be reduced to 
\begin{equation}
    I_{p_1p_2p_3p_4} = (-1)^{n_\beta + m_\beta + n_\gamma + m_\gamma + p} \frac{\sqrt{\omega}}{\sqrt{2}} \Gamma\left(p + \frac{1}{2}\right) \int_0^{2\pi} d\theta \left(\cos\theta \right)^{a} \left(\sin\theta \right)^{2p - a}
\end{equation}
where $\Gamma$ is the gamma function and the notation is condensed using parameters
\begin{align*}
    2p &= n_\alpha + m_\alpha + n_\delta + m_\delta +n_\beta + m_\beta + n_\gamma + m_\gamma - 2p_1 - 2p_2 - 2p_3 - 2p_4 \\
    a &= n_\alpha + n_\delta+ n_\beta + n_\gamma -2p_1-2p_3.
\end{align*}
The integral over $\theta$ can be evaluated analytically using the beta function
\begin{equation*}
    {\rm B}(x,y) = 2\int_0^{\frac{\pi}{2}} d\theta \, (\sin\theta)^{2x - 1} (\cos\theta)^{2y - 1} = \frac{\Gamma(x)\Gamma(y)}{\Gamma(x+y)}
\end{equation*}
giving
\begin{equation}
   \int_0^{2\pi} d\theta \left(\cos\theta \right)^{a} \left(\sin\theta \right)^{2p - a} =  \begin{cases} 
      2{\rm B}\left(p - \frac{a-1}{2}, \frac{a+1}{2}\right) & {\rm if\,\,} a {\rm\,\,and\,\,} 2p {\rm\,\,are\,\,even}  \\
      0 & {\rm otherwise} \\
   \end{cases}.
\end{equation}
Lastly, using the fact that $\Gamma(x) = x\,\Gamma(x-1)$ and $\Gamma(1/2) = \sqrt{\pi}$, the final form of the full CME is 
\begin{align}
\label{eq:CMEEquationAppendix}
    \bra{\alpha\beta}v\ket{\gamma\delta} &= \bra{n_\alpha m_\alpha n_\beta m_\beta}v\ket{n_\gamma m_\gamma n_\delta m_\delta} \nonumber \\
    &= \sqrt{\omega}\frac{\sqrt{\pi}(-1)^{n_\beta + m_\beta + n_\gamma + m_\gamma}}{\sqrt{n_\alpha! m_\alpha! n_\delta! m_\delta! n_\beta! m_\beta! n_\gamma! m_\gamma!}} \sum_{p_1 = 0}^{\min(n_\alpha,n_\delta)} p_1! \binom{n_\alpha}{p_1} \binom{n_\delta}{p_1} \nonumber \\
    &\quad\times \sum_{p_2 = 0}^{\min(m_\alpha,m_\delta)} p_2! \binom{m_\alpha}{p_2} \binom{m_\delta}{p_2} \sum_{p_3 = 0}^{\min(n_\beta,n_\gamma)} p_3! \binom{n_\beta}{p_3} \binom{n_\gamma}{p_3} \\
    &\quad\times \sum_{p_4 = 0}^{\min(m_\beta,m_\gamma)} p_4! \binom{m_\beta}{p_4} \binom{m_\gamma}{p_4} (-1)^p \, \frac{(2p-1)!! (2p-a-1)!! (a-1)!!}{2^{2p} \, p!} \nonumber
\end{align}
where $k!! = k(k-2)\cdots 3\cdot 1$ is the double factorial for odd $k$.

%**************************************%
%**************************************%
%**************************************%
\section{Extension of of analytical formula for Coulomb matrix elements to elliptical harmonic orbitals}
\label{app:CMEs_elipHOs}

In this section we generalize the formula for the Coulomb matrix elements in the case of elliptical harmonic orbitals.
To start, the coordinate operators are defined as follows:
\begin{equation}
	x_i = \frac{A}{\sqrt{2}}(a_i + a_i^\dagger), \quad  A=\sqrt{\frac{1}{\omega_x}};  \qquad
	y_i = \frac{B}{\sqrt{2}}(b_i + b_i^\dagger),  \quad  B=\sqrt{\frac{1}{\omega_y}}.
	\label{eq:definitions}
\end{equation}
%
%This is a natural extension of the previously obtained results for the case of elongated quantum dot arrays.
The first part of the derivation is analogous to Appendix \ref{app:CMEs_symHOs} and leads to almost the same expression for the Coulomb matrix elements as given in Eq.~\ref{eq:appAEquation9}. 
For elliptical harmonic orbitals, the integral $I_{p_1p_2p_3p_4}$ is now given by the formula:
% with the difference  following expression for the Coulomb matrix element:
% %
% \begin{align}
%     \braket{\alpha\beta|v|\gamma\delta} &= \frac{1}{4\pi^2} \int d\vec{q}\, \frac{2\pi}{q} \braket{\alpha|e^{i\vec{q}\vec{r}_1}|\delta} \braket{\beta|e^{-i\vec{q}\vec{r}_2}|\gamma} \nonumber\\
%     &= \frac{1}{2\pi\sqrt{n_\alpha! m_\alpha! n_\delta! m_\delta! n_\beta! m_\beta! n_\gamma! m_\gamma!}} \sum_{p_1 = 0}^{\min(n_\alpha,n_\delta)} p_1! {n_\alpha \choose p_1} {n_\delta \choose p_1} \sum_{p_2 = 0}^{\min(m_\alpha,m_\delta)} p_2! {m_\alpha \choose p_2} {m_\delta \choose p_2} \nonumber \\
%     &\times \sum_{p_3 = 0}^{\min(n_\beta,n_\gamma)} p_3! {n_\beta \choose p_3} {n_\gamma \choose p_3} \sum_{p_4 = 0}^{\min(m_\beta,m_\gamma)} p_4! {m_\beta \choose p_4} {m_\gamma \choose p_4} I_{p_1p_2p_3p_4}.  
% \label{eq:matrix_element}
% \end{align}
%
% The integral $I_{p_1p_2p_3p_4}$ is given by the formula:
%
\begin{align}
    I_{p_1p_2p_3p_4} &= \int \frac{d\vec{q}}{q} e^{-(\frac{A^2}{2}q_x^2+\frac{B^2}{2}q_y^2)} \left(\frac{iA}{\sqrt{2}}q_x\right)^{n_\alpha + n_\delta - 2p_1} \left(\frac{iB}{\sqrt{2}}q_y\right)^{m_\alpha + m_\delta - 2p_2} \nonumber \\ 
    &\quad\times \left(\frac{-iA}{\sqrt{2}}q_x\right)^{n_\beta + n_\gamma - 2p_3} \left(\frac{-iB}{\sqrt{2}}q_y\right)^{m_\beta + m_\gamma - 2p_4} \\
	&=C A^a B^{2p-a}\int_0^{2\pi} d\theta \int_{0}^{\infty} dq \, q^{2p} \left(\cos\theta\right)^a \left(\sin\theta\right)^{2p-a} e^{-\frac{q^2}{2} (A^2\cos^2\theta+B^2\sin^2\theta)},
\label{eq:I_unsimplified}
\end{align}
where the following parameters are introduced:
\begin{align*}
    2p &= n_\alpha + m_\alpha + n_\delta + m_\delta +n_\beta + m_\beta + n_\gamma + m_\gamma - 2p_1 - 2p_2 - 2p_3 - 2p_4 \\
    a&=n_\alpha + n_\delta+ n_\beta + n_\gamma -2p_1-2p_3 \\
    C &= (-1)^{p+n_\beta+n_\gamma+m_\beta+m_\gamma}.
\end{align*}
After converting to polar coordinates and substituting $x = q\sqrt{A^2\cos^2\theta + B^2\sin^2\theta}/\sqrt{2}$, the integration over $q$ immediately yields the Gamma function giving:
\begin{equation}
I_{p_1p_2p_3p_4}(A,B) = 2\sqrt{2} \; C\;\Gamma\left(p+\frac{1}{2}\right)A^a B^{2p-a} \underbrace{\int_{0}^{\frac{\pi}{2}} \frac{\left(\cos\theta\right)^a \left(\sin\theta\right)^{2p-a}}{\left(A^2\cos^2\theta+B^2\sin^2\theta \right)^{p+\frac{1}{2}}} d\theta }_{\tilde{I}_{[0,\frac{\pi}{2}]}(A,B)}
\label{eq:q_integrated}
\end{equation}
if $a$ and $2p$ are even, and zero otherwise. 
To calculate the integral $\tilde{I}_{[0,\frac{\pi}{2}]}(A,B),$ we consider a generating function: 
\begin{equation}
	G(A,B)=\int_{0}^{\frac{\pi}{2}} \frac{\mathrm{d}\theta}{\sqrt{\left(A^2\cos^2\theta+B^2\sin^2\theta \right)}}=\frac{\pi}{2M(A,B)},
	\label{eq:generator}
\end{equation}
where $M(A,B)$ is the arithmetic-geometric mean of the numbers $A,B$. 
The derivative of Eq.~\ref{eq:generator} with respect to $A^2$ and $B^2$:
\begin{equation}
        \frac{\partial^{k+l}G(A,B)}{\partial^k (A^2)\partial^l (B^2)} =
		\int_{0}^{\frac{\pi}{2}} d\theta \frac{\left(\cos^2\theta\right)^k \left(\sin^2\theta\right)^l }{\left(A^2\cos^2\theta+B^2\sin^2\theta \right)^{k+l+\frac{1}{2}}} {\frac{(-1)^{k+l}(2k+2l-1)!!}{2^{k+l}}}
		\label{eq:diff_general}
\end{equation}
is clearly within a constant factor from $\tilde{I}_{[0,\frac{\pi}{2}]}(A,B)$ when $k=\frac{a}{2}$ and $l=p-\frac{a}{2}$. 
This leads us to the following formula:
\begin{equation}
    I_{p_1 p_2 p_3 p_4}(A,B)
    = \pi\sqrt{2\pi}(-1)^{n_\beta+n_\gamma+m_\beta+m_\gamma} \left(\frac{\partial }{\partial (A^2)}\right)^\frac{a}{2} \left(\frac{\partial }{\partial (B^2)}\right)^{p-\frac{a}{2}}\frac{1}{M(A,B)}.
    \label{eq:total_integral_calculated}
\end{equation}
Here we also utilized the identity $\Gamma\left(p+\frac{1}{2}\right)=
	\frac{\sqrt{\pi}(2p-1)!!}{2^p}$ that holds for integer $p$.
	
Recollecting the definition of $A,B$ from Eq.~\ref{eq:definitions}, we are going to obtain the final expression in terms of $\omega_x, \omega_y$.
To achieve this, we note that the following equality follows from Eq.~\ref{eq:q_integrated} by factoring out constant terms from the denominator:
\begin{equation*}
	I_{p_1p_2p_3p_4}(A,B) = \frac{1}{AB} I_{p_1p_2p_3p_4}\left(\frac{1}{B},\frac{1}{A}\right)
\end{equation*}
Then, the Eq.~\eqref{eq:total_integral_calculated} takes the form:
\begin{align}
    I_{p_1 p_2 p_3 p_4}(A,B) &= \frac{\pi\sqrt{2\pi}(-1)^{n_\beta+n_\gamma+m_\beta+m_\gamma}}{A^{2p-a+1} B^{a+1}}  \left(\frac{\partial }{\partial (B^{-2})}\right)^\frac{a}{2} \left(\frac{\partial }{\partial (A^{-2})}\right)^{p-\frac{a}{2}}\frac{1}{M\left(\frac{1}{B},\frac{1}{A}\right)} \nonumber \\
    &=\pi\sqrt{2\pi}(-1)^{n_\beta+n_\gamma+m_\beta+m_\gamma} \omega_x^{p-\frac{a-1}{2}}\omega_y^{\frac{a+1}{2}} \nonumber \\
    &\quad\times \left(\frac{\partial }{\partial \omega_y}\right)^\frac{a}{2} \left(\frac{\partial }{\partial \omega_x}\right)^{p-\frac{a}{2}}
    \frac{1}{M\left( \sqrt{\omega_x\mathstrut}, \sqrt{\omega_y\mathstrut}\right) }
\label{eq:total_integral_transformed}
\end{align}
In the end, after substituting Eq.~\ref{eq:total_integral_transformed} into Eq.~\ref{eq:appAEquation9}, we obtain the final expression for Coulomb matrix elements in case of elliptical harmonic orbitals: 
\begin{multline}
    \braket{\alpha\beta|v|\gamma\delta} 
    =  \sqrt{\frac{\pi \omega_x\omega_y}{2}}\frac{(-1)^{n_\beta+n_\gamma+m_\beta+m_\gamma}}{\sqrt{n_\alpha! m_\alpha! n_\delta! m_\delta! n_\beta! m_\beta! n_\gamma! m_\gamma!}} \sum_{p_1 = 0}^{\min(n_\alpha,n_\delta)} p_1! \binom{n_\alpha}{p_1} \binom{n_\delta}{p_1} \\
    \times \sum_{p_2 = 0}^{\min(m_\alpha,m_\delta)} p_2! \binom{m_\alpha}{p_2} \binom{m_\delta}{p_2}
    \sum_{p_3 = 0}^{\min(n_\beta,n_\gamma)} p_3! \binom{n_\beta}{p_3} \binom{n_\gamma}{p_3} \sum_{p_4 = 0}^{\min(m_\beta,m_\gamma)} p_4! \binom{m_\beta}{p_4} \binom{m_\gamma}{p_4} \\
    \times \omega_x^{p-\frac{a}{2}}
    \left(\frac{\partial }{\partial \omega_x}\right)^{p-\frac{a}{2}}
    \omega_y^{\frac{a}{2}} 
    \left(\frac{\partial }{\partial \omega_y}\right)^\frac{a}{2} 
    \frac{1}{M\left( \sqrt{\omega_x\mathstrut}, \sqrt{\omega_y\mathstrut}\right) },
\label{eq:matrix_element_transformed}
\end{multline}
where $a = n_\alpha + n_\delta + n_\beta + n_\gamma - 2p_1 - 2p_3$, $2p = a + m_\alpha  + m_\delta + m_\beta + m_\gamma - 2p_2 - 2p_4$, and $2p$ and $a$ are even. Otherwise, the matrix elements are equal to zero.

In summary, we find the closed analytic formulas for Coulomb matrix elements in the cases of circular and elliptical orbitals.
However, unlike in the expression from Appx.~\ref{app:CMEs_symHOs}, the terms dependent on $\omega_x$ and $\omega_y$ do not factor out from Eq.~\ref{eq:matrix_element_transformed}. 
This does not allow us to simply scale the preliminary calculated library of CMEs for unit frequencies and achieve the desired computational efficiency. 
For this reason, only symmetric harmonic orbitals are used for all simulations described in the main text of the paper.

However, improvements can be made to Eq.~\ref{eq:matrix_element_transformed} to make it more useful for full LCHO-CI calculations, even if both $\omega_x$ and $\omega_y$ do not factor out of the expression.
We do this by rewriting Eq.~\ref{eq:q_integrated} as
\begin{equation}
    I_{p_1p_2p_3p_4}(A,\kappa) = \frac{1}{A}\,2\sqrt{2} \; C\;\Gamma\left(p+\frac{1}{2}\right) \kappa^{2p-a} \int_{0}^{\frac{\pi}{2}} \frac{\left(\cos\theta\right)^a \left(\sin\theta\right)^{2p-a}}{\left(\cos^2\theta+\kappa^2\sin^2\theta \right)^{p+\frac{1}{2}}} d\theta
\end{equation}
where $\kappa = B/A = \omega_x/\omega_y$ is the eccentricity of the harmonic orbitals.
$A$ easily factors out of the entire expression, and we can write the Coulomb matrix elements as
\begin{multline}
    \bra{\alpha\beta}v\ket{\gamma\delta} = \sqrt{\omega_x}\frac{\sqrt{2}(-1)^{n_\beta + n_\gamma + m_\beta + m_\gamma}}{\pi\sqrt{n_\alpha! m_\alpha! n_\delta! m_\delta! n_\beta! m_\beta! n_\gamma! m_\gamma!}} \sum_{p_1 = 0}^{\min(n_\alpha,n_\delta)} p_1! \binom{n_\alpha}{p_1} \binom{mn_\delta}{p_1} \\
    \times \sum_{p_2 = 0}^{\min(m_\alpha,m_\delta)} p_2! \binom{m_\alpha}{p_2} \binom{m_\delta}{p_2} \sum_{p_3 = 0}^{\min(n_\beta,n_\gamma)} p_3! \binom{n_\beta}{p_3} \binom{m_\gamma}{p_3} \sum_{p_4 = 0}^{\min(m_\beta,m_\gamma)} p_4! \binom{m_\beta}{p_4} \binom{m_\gamma}{p_4} \\
    \times (-1)^p \, \Gamma\left(p+\frac{1}{2}\right) \, \kappa^{2p-a} \int_{0}^{\frac{\pi}{2}} \frac{\left(\cos\theta\right)^a \left(\sin\theta\right)^{2p-a}}{\left(\cos^2\theta+\kappa^2\sin^2\theta \right)^{p+\frac{1}{2}}} d\theta
\end{multline} 
The term within the summation depends only on $\kappa$ and can be easily evaluated numerically. 
A discrete collection of full harmonic orbital Coulomb matrix elements $C_{\rm HO,1,\kappa}$ can be calculated for $\omega_x=1$ and a select set of $\kappa$ values (e.g. $\kappa=$ 0.1, 0.5, 2.0, 10.0).
The harmonic orbital basis can be optimized over a continuous choice of $\omega_x$ and a discrete set of $\kappa$, and the desired harmonic orbital Coulomb matrix elements are calculated simply as $C_{{\rm HO},\omega_x,\kappa}=$ $\sqrt{\omega_x}C_{\rm HO,1,\kappa}$.
Utilizing this approach requires the storage of several $C_{\rm, 1, \kappa}$, and since these matrices can be quite large (i.e. $M=16^2$ gives $C_{\rm HO,1,\kappa}$ size of 65536$\times$65536), it may not be feasible to store several sets of $C_{\rm HO,1,\kappa}$.
However, this approach will be critical to apply this method on 3D potential as the confinement along the $z$-axis is generally much smaller than the confinements along the $x$- and $y$- axes.
It will be difficult to get converged approximations of the 3D single electron orbitals if $\omega_x=$ $\omega_y=$ $\omega_z$ for the harmonic orbital basis.

%**************************************%
%**************************************%
%**************************************%
\section{Charging energies and lever arms}
\label{app:parameter_extraction}

Here we show how the charging energies $U$ and lever arms $\alpha$ are extracted for the device geometries discussed in the main text.
When evaluating $U$, we only use double QD potentials when $V_{\rm bias}=0$ V.
For a symmetric double QD system, the charging energy $U$ of each QD can be evaluated as \cite{sarma2011hubbard}
\begin{equation}
    U = \bra{\psi_{L/R}(\vec{r}_1) \psi_{L/R}(\vec{r}_2) }v\ket{\psi_{L/R}(\vec{r}_2) \psi_{L/R}(\vec{r}_1)}
\end{equation}
where $v$ is the standard Coulomb potential as in Eq.~\ref{eq:2ndQHam} of the main text, and $\ket{\psi_{L/R}}$ are the localized electron orbitals in the left/right QDs.
The localized orbitals are found by taking the symmetric and anti-symmetric combinations of the ground and first excited orbital eigenstates of the double QD potential: $\ket{\psi_{L/R}} = \frac{1}{\sqrt{2}}(\ket{\psi_0} \pm \ket{\psi_1})$.
The charging energy $U$ can be found by following the procedure outlined in Section~\ref{sec:mathMethods} to find the Coulomb matrix elements of these new single electron orbitals $\ket{\psi_{L/R}}$.
The specific $U$ values calculated for each device geometry are given in the tables of Figures~\ref{fig:dot_size}, \ref{fig:tun_gate}, \ref{fig:oxide}, and \ref{fig:eccentricity} in the main text.

\begin{figure}
    \centering
    \includegraphics[width = 0.6\textwidth]{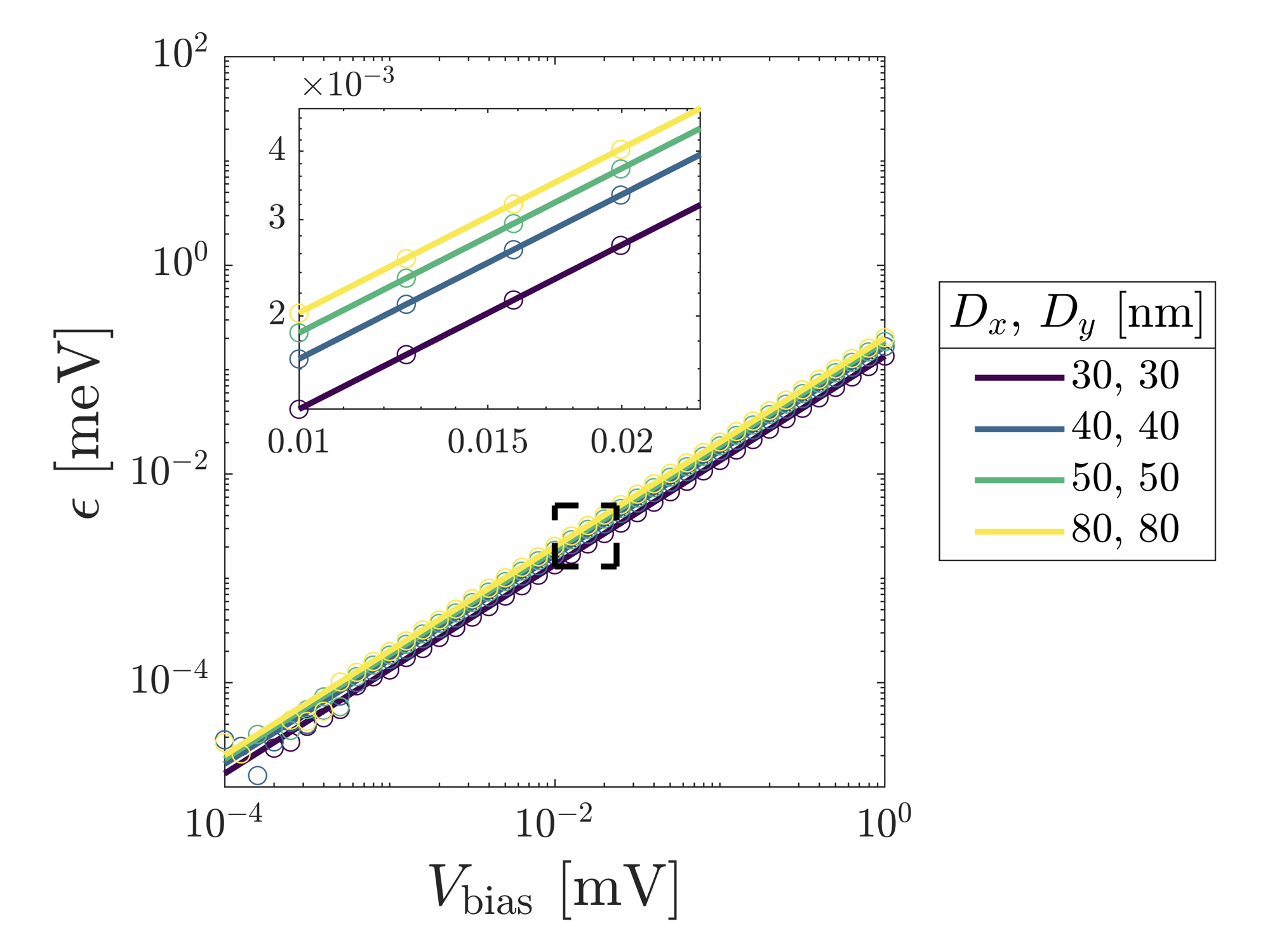}
    \caption{Relationship between bias voltage $V_{\rm bias}$ and inter-dot detuning $\epsilon$ for different dot sizes. The calculated data points are indicated by circles and the solid line is a fit to the equation $\epsilon = \alpha V_{\rm bias}$ where $\alpha$ is the lever arm. The region in the dashed black box is enlarged and shown in the inset to demonstrate the linearity of the fitted data.}
    \label{fig:detuning_fit}
\end{figure}

Next, we show how to evaluate the lever arm $\alpha$ for each individual device geometry.
The lever arm connects the plunger gate bias voltage to the effective inter-dot detuning $\epsilon = \epsilon_R - \epsilon_L$ as $\epsilon =\alpha V_{\rm bias}$ where $\epsilon_{L/R}$ is the localized ground state energy in the left/right QD.
To find $\alpha$, $V_{\rm bias}$ is varied and the tunnel gate voltage $V_{\rm tun}$ remains fixed.
We model the double QD system with the simple two-level Hamiltonian
\begin{equation}
    H = \begin{bmatrix}
    \epsilon_L & t_c \\
    t_c & \epsilon_R
    \end{bmatrix}.
\end{equation}
where the basis states are $\{\ket{\psi_L}, \ket{\psi_R}\}$.
The energy difference between the ground and first excited energy levels of $H$ is given as
\begin{equation}
\label{eq:detuningeq}
    \epsilon_1 - \epsilon_0 = \sqrt{\epsilon^2 + 4\,t_c^2}.
\end{equation}
The eigenenergies $\epsilon_0$ and $\epsilon_1$ are found by solving the Schr\"{o}dinger equation from Eq.~\ref{eq:basisHam} of the main text.
The inter-dot tunnel coupling $t_c$ is found when $V_{\rm bias}=0$ which corresponds to an effective inter-dot detuning of $\epsilon=0$.
$t_c$ is assumed to remain fixed as $V_{\rm bias}$ is varied.
Eq.~\ref{eq:detuningeq} is used to find the corresponding $\epsilon$ value for each $V_{\rm bias}$ data point.

The resulting data are fit to the linear relationship $\epsilon=\alpha V_{\rm bias}$ to find $\alpha$ for the corresponding device geometry.
Figure~\ref{fig:detuning_fit} shows data of detuning versus bias voltage for a few different device geometries where the dot size $D_x=D_y$ was varied.
The other device parameters are $D_t=$ 20 nm, $T=$ 15 nm, and $V_p=$ 0.150 V.
Circles are actual data points, and solid lines correspond to $\epsilon=\alpha V_{\rm bias}$ fits.
For all data sets, the data is only fit in the interval $V_{\rm bias}=$ $[10^{-3}, 10^{-1}]$ mV to avoid noise at low bias voltage values and prevent fitting in any non-linear regimes which can occur at high bias.
The region shown in the figure inset corresponds to the dashed black box in the main figure and demonstrates the accuracy of the linear fit to the data.
Only a handful of device geometries are presented here for visual clarity, but all fitted data sets used throughout this chapter show similar behavior and fit quality.
The specific $\alpha$ values extracted for each device geometry are given in the tables of Figures~\ref{fig:dot_size}, \ref{fig:tun_gate}, \ref{fig:oxide}, and \ref{fig:eccentricity} in the main text.

%**************************************%
%**************************************%
%**************************************%
\section{Visualizing the 1D zero-bias potentials}
\label{app:zero_bias_pots}

\begin{figure}
    \centering
    \includegraphics[width = 0.9\textwidth]{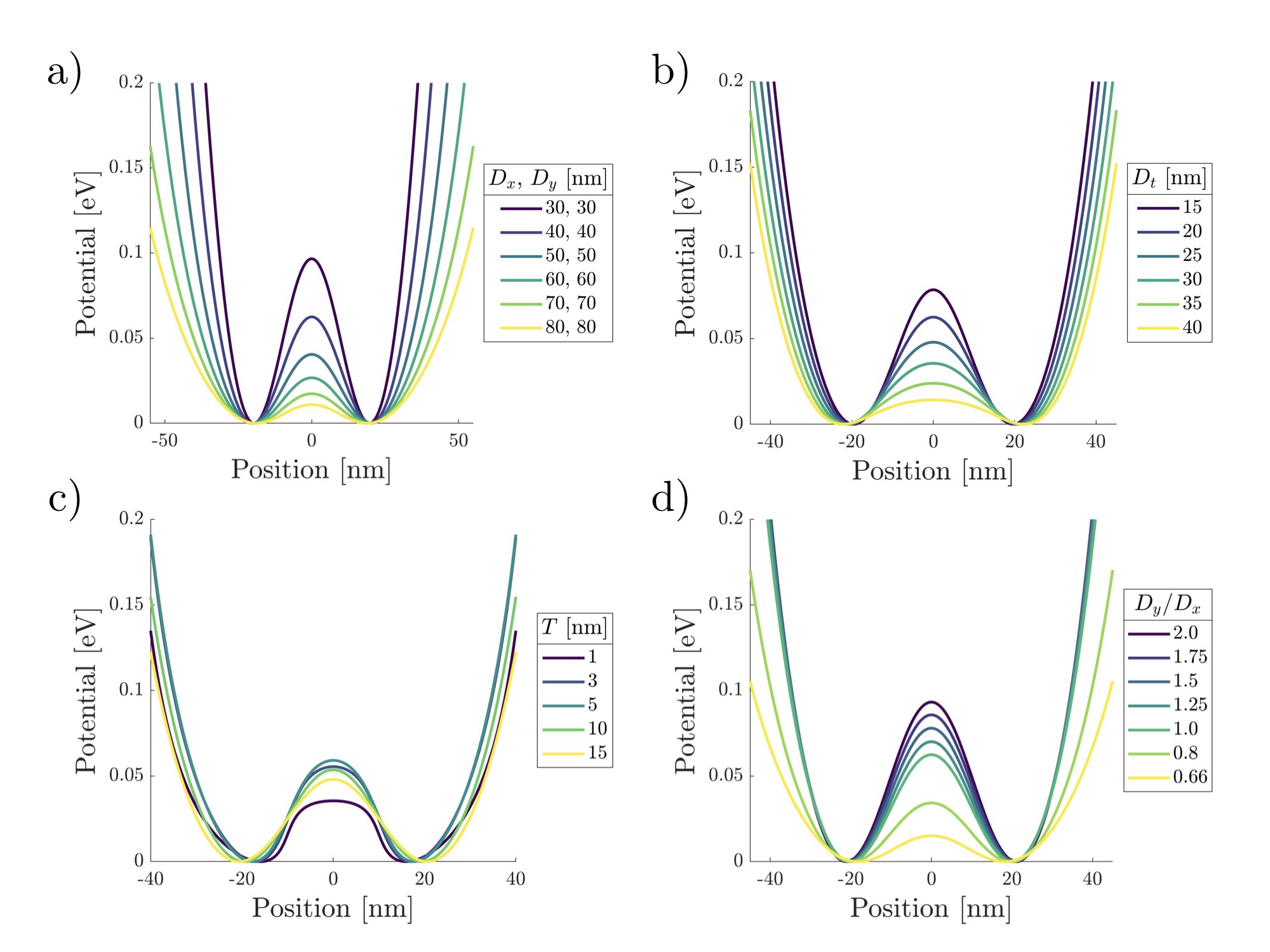}
    \caption{1D line cuts of the 2D potentials that give $J=$ 1~$\mu$eV at $V_{\rm bias}=$ 0 V. Unless varied, the default device parameters are $D_x=D_y=$ 40 nm, $D_t=$ 20 nm, and $T=$ 15 nm. For $D_y/D_x>1.0$, $D_x=$ 40 nm, and for $D_y/D_x<1.0$, $D_y=$ 40 nm. The corresponding gate voltages are given in Table~\ref{tab:zero_bias_volts}. The 1D slices are taken along the $x$-axis and pass through the lowest potential minima in the 2D potentials.}
    \label{fig:zero_bias_pots}
\end{figure}

Here we show the zero bias potentials for the varied device geometries discussed in the main text.
1D slices of the potentials along the $x$-axis are plotted in Figure~\ref{fig:zero_bias_pots}.
The $y$-coordinate at which the slice is taken is chosen such that the 1D slice goes through the minima of the double quantum dot potential.
Note from Figure~\ref{fig:device_geom}b that this point is not necessarily directly underneath the center of the plunger gate head and changes with the device geometry.
Table~\ref{tab:zero_bias_volts} lists the varied geometry parameters and the $V_{\rm tun}$ value where $J=$ 1 $\mu$eV at $V_{\rm bias}=$ 0 V.

\begin{table}
\centering
\begin{tabular}{|| c | c || c | c || c | c || c | c ||}
    \hline
     $D_x$, $D_y$ [nm] & $V_{\rm tun}$ [V] & $D_t$ [nm] & $V_{\rm tun}$ [V] & $T$ [nm] & $V_{\rm tun}$ [V] & $D_y/D_x$ & $V_{\rm tun}$ [V] \\
     \hline
     30, 30 & 0.057228 & 15 & 0.070977 & 3 & 0.060071& 2.0 & 0.080796 \\
     \hline
     40, 40 & 0.094548 & 20 & 0.094548 & 5 & 0.073663 & 1.75 & 0.084305 \\
     \hline
     50, 50 & 0.114617 & 25 & 0.108097 & 10 & 0.087739 & 1.5 & 0.087954 \\
     \hline
     60, 60 & 0.126211 & 30 & 0.116451 & 15 & 0.092118 & 1.25 & 0.091506 \\
     \hline
     70, 70 & 0.133647 & 35 & 0.122298 & & & 1.0 & 0.094548 \\
     \hline
     80, 80 & 0.138619 & 40 & 0.126458 & & & 0.8 & 0.117165 \\
     \hline
      &  & & & & & 0.66 & 0.131638 \\
     \hline
\end{tabular}
\caption{Corresponding tunnel gate voltages $V_{\rm tun}$ which yield $J=$ 1 $\mu$eV at $V_{\rm bias}=0$ for different device geometries. The default device geometry parameters unless varied are $D_x=$ $D_y=$ 40 nm, $D_t=$ 20 nm, and $T=$ 15 nm. For $D_y/D_x > 1.0$, $D_x=$ 40 nm, and for $D_y/D_x < 1.0$, $D_y=$ 40 nm. $V_{p}=$ 0.150 V for all geometries except when $T$ is varied where $V_p=$ 0.100 V. 1D slices of the corresponding potentials are shown in Figure~\ref{fig:zero_bias_pots}.}
\label{tab:zero_bias_volts}
\end{table}

Figure~\ref{fig:zero_bias_pots}a shows 1D potential slices as the dot size $D_x$, $D_y$ is varied.
The other geometry parameters are $D_t=$ 20 nm, $T=$ 15 nm, and $V_p=$ 0.150 V.
As the dot size increases, the tunnel barrier flattens and the potential confinement decreases causing a larger charging energy $U$ as discussed in the main text.
Interestingly, the minima of the potential wells seem to remain in the same $x$-coordinate location even as the plunger gate get larger.

Figure~\ref{fig:zero_bias_pots}b shows 1D potential slices as the tunnel gate width $D_t$ is varied.
The other geometry parameters are $D_x=$ $D_y=$ 40 nm, $T=$ 15 nm, and $V_p=$ 0.150 V.
As the tunnel gate widens the tunnel barrier flattens.
This reduces the potential confinement of each QD thereby increasing the charging energy $U$ as described in the main text.

Figure~\ref{fig:zero_bias_pots}c shows 1D potential slices as the oxide thickness $T$ is varied.
The other geometry parameters are $D_x=$ $D_y=$ 40 nm, $D_t=$ 20 nm, and $V_p=$ 0.100 V.
As the oxide thickness decreases, the image of the square plunger gate head is mapped more strongly onto the potential landscape.
In the limit where $T=0$, the double QD potential would be two square wells with a square tunnel barrier between them.
As this transition towards a more `square' potential occurs, the slope of the tunnel barrier increases and slightly widens.
This effect suppresses the wavefunction overlap in the tunnel barrier region as the electrons are more localized to each QD.
Additionally, the effective confinement of each QD increases which slightly increases the charging energy $U$ as shown in the main text.

Figure~\ref{fig:zero_bias_pots}d shows 1D potential slices as the oxide thickness $T$ is varied.
The other geometry parameters are $D_t=$ 20 nm, $T=$ 15 nm, and $V_p=$ 0.150 V.
For $D_y/D_x>1.0$, $D_x=$ 40 nm, while for $D_y/D_x<1.0$, $D_y=$ 40 nm.
It is clear from these potentials that $U$ decreases when $D_y/D_x<1.0$ as the QD potentials open.
However due to the fact that we are taking 1D slices along the $x$-axis, it is difficult to see the same effect when $D_y/D_x>1.0$ even though $U$ is decreasing as well.
The potentials do show that the tunnel barrier height increases as the plunger gate eccentricity increases due to the additional plunger gate material along the $y$-axis which pushes the QD minima lower.
For sensitivity to charge noise, the relevant parameter is the charging energy $U$ which increases whether or not the plunger gates are elongated along $x$ or $y$.

%\input{EllipticOrbitals}
%\nocite{*} 
%\printbibliography


\begin{thebibliography}{54}%
\makeatletter
\providecommand \@ifxundefined [1]{%
 \@ifx{#1\undefined}
}%
\providecommand \@ifnum [1]{%
 \ifnum #1\expandafter \@firstoftwo
 \else \expandafter \@secondoftwo
 \fi
}%
\providecommand \@ifx [1]{%
 \ifx #1\expandafter \@firstoftwo
 \else \expandafter \@secondoftwo
 \fi
}%
\providecommand \natexlab [1]{#1}%
\providecommand \enquote  [1]{``#1''}%
\providecommand \bibnamefont  [1]{#1}%
\providecommand \bibfnamefont [1]{#1}%
\providecommand \citenamefont [1]{#1}%
\providecommand \href@noop [0]{\@secondoftwo}%
\providecommand \href [0]{\begingroup \@sanitize@url \@href}%
\providecommand \@href[1]{\@@startlink{#1}\@@href}%
\providecommand \@@href[1]{\endgroup#1\@@endlink}%
\providecommand \@sanitize@url [0]{\catcode `\\12\catcode `\$12\catcode
  `\&12\catcode `\#12\catcode `\^12\catcode `\_12\catcode `\%12\relax}%
\providecommand \@@startlink[1]{}%
\providecommand \@@endlink[0]{}%
\providecommand \url  [0]{\begingroup\@sanitize@url \@url }%
\providecommand \@url [1]{\endgroup\@href {#1}{\urlprefix }}%
\providecommand \urlprefix  [0]{URL }%
\providecommand \Eprint [0]{\href }%
\providecommand \doibase [0]{https://doi.org/}%
\providecommand \selectlanguage [0]{\@gobble}%
\providecommand \bibinfo  [0]{\@secondoftwo}%
\providecommand \bibfield  [0]{\@secondoftwo}%
\providecommand \translation [1]{[#1]}%
\providecommand \BibitemOpen [0]{}%
\providecommand \bibitemStop [0]{}%
\providecommand \bibitemNoStop [0]{.\EOS\space}%
\providecommand \EOS [0]{\spacefactor3000\relax}%
\providecommand \BibitemShut  [1]{\csname bibitem#1\endcsname}%
\let\auto@bib@innerbib\@empty
%</preamble>
\bibitem [{\citenamefont {Veldhorst}\ \emph {et~al.}(2014)\citenamefont
  {Veldhorst}, \citenamefont {Hwang}, \citenamefont {Yang}, \citenamefont
  {Leenstra}, \citenamefont {de~Ronde}, \citenamefont {Dehollain},
  \citenamefont {Muhonen}, \citenamefont {Hudson}, \citenamefont {Itoh},
  \citenamefont {Morello} \emph {et~al.}}]{veldhorst2014addressable}%
  \BibitemOpen
  \bibfield  {author} {\bibinfo {author} {\bibfnamefont {M.}~\bibnamefont
  {Veldhorst}}, \bibinfo {author} {\bibfnamefont {J.}~\bibnamefont {Hwang}},
  \bibinfo {author} {\bibfnamefont {C.}~\bibnamefont {Yang}}, \bibinfo {author}
  {\bibfnamefont {A.}~\bibnamefont {Leenstra}}, \bibinfo {author}
  {\bibfnamefont {B.}~\bibnamefont {de~Ronde}}, \bibinfo {author}
  {\bibfnamefont {J.}~\bibnamefont {Dehollain}}, \bibinfo {author}
  {\bibfnamefont {J.}~\bibnamefont {Muhonen}}, \bibinfo {author} {\bibfnamefont
  {F.}~\bibnamefont {Hudson}}, \bibinfo {author} {\bibfnamefont {K.~M.}\
  \bibnamefont {Itoh}}, \bibinfo {author} {\bibfnamefont {A.}~\bibnamefont
  {Morello}}, \emph {et~al.},\ }\href@noop {} {\bibfield  {journal} {\bibinfo
  {journal} {Nature {N}anotechnology}\ }\textbf {\bibinfo {volume} {9}},\
  \bibinfo {pages} {981} (\bibinfo {year} {2014})}\BibitemShut {NoStop}%
\bibitem [{\citenamefont {Zajac}\ \emph {et~al.}(2018)\citenamefont {Zajac},
  \citenamefont {Sigillito}, \citenamefont {Russ}, \citenamefont {Borjans},
  \citenamefont {Taylor}, \citenamefont {Burkard},\ and\ \citenamefont
  {Petta}}]{zajac2018resonantly}%
  \BibitemOpen
  \bibfield  {author} {\bibinfo {author} {\bibfnamefont {D.~M.}\ \bibnamefont
  {Zajac}}, \bibinfo {author} {\bibfnamefont {A.~J.}\ \bibnamefont
  {Sigillito}}, \bibinfo {author} {\bibfnamefont {M.}~\bibnamefont {Russ}},
  \bibinfo {author} {\bibfnamefont {F.}~\bibnamefont {Borjans}}, \bibinfo
  {author} {\bibfnamefont {J.~M.}\ \bibnamefont {Taylor}}, \bibinfo {author}
  {\bibfnamefont {G.}~\bibnamefont {Burkard}},\ and\ \bibinfo {author}
  {\bibfnamefont {J.~R.}\ \bibnamefont {Petta}},\ }\href@noop {} {\bibfield
  {journal} {\bibinfo  {journal} {Science}\ }\textbf {\bibinfo {volume}
  {359}},\ \bibinfo {pages} {439} (\bibinfo {year} {2018})}\BibitemShut
  {NoStop}%
\bibitem [{\citenamefont {Watson}\ \emph {et~al.}(2018)\citenamefont {Watson},
  \citenamefont {Philips}, \citenamefont {Kawakami}, \citenamefont {Ward},
  \citenamefont {Scarlino}, \citenamefont {Veldhorst}, \citenamefont {Savage},
  \citenamefont {Lagally}, \citenamefont {Friesen}, \citenamefont {Coppersmith}
  \emph {et~al.}}]{watson2018programmable}%
  \BibitemOpen
  \bibfield  {author} {\bibinfo {author} {\bibfnamefont {T.}~\bibnamefont
  {Watson}}, \bibinfo {author} {\bibfnamefont {S.}~\bibnamefont {Philips}},
  \bibinfo {author} {\bibfnamefont {E.}~\bibnamefont {Kawakami}}, \bibinfo
  {author} {\bibfnamefont {D.}~\bibnamefont {Ward}}, \bibinfo {author}
  {\bibfnamefont {P.}~\bibnamefont {Scarlino}}, \bibinfo {author}
  {\bibfnamefont {M.}~\bibnamefont {Veldhorst}}, \bibinfo {author}
  {\bibfnamefont {D.}~\bibnamefont {Savage}}, \bibinfo {author} {\bibfnamefont
  {M.}~\bibnamefont {Lagally}}, \bibinfo {author} {\bibfnamefont
  {M.}~\bibnamefont {Friesen}}, \bibinfo {author} {\bibfnamefont
  {S.}~\bibnamefont {Coppersmith}}, \emph {et~al.},\ }\href@noop {} {\bibfield
  {journal} {\bibinfo  {journal} {Nature}\ }\textbf {\bibinfo {volume} {555}},\
  \bibinfo {pages} {633} (\bibinfo {year} {2018})}\BibitemShut {NoStop}%
\bibitem [{\citenamefont {Yoneda}\ \emph {et~al.}(2018)\citenamefont {Yoneda},
  \citenamefont {Takeda}, \citenamefont {Otsuka}, \citenamefont {Nakajima},
  \citenamefont {Delbecq}, \citenamefont {Allison}, \citenamefont {Honda},
  \citenamefont {Kodera}, \citenamefont {Oda}, \citenamefont {Hoshi} \emph
  {et~al.}}]{yoneda2018quantum}%
  \BibitemOpen
  \bibfield  {author} {\bibinfo {author} {\bibfnamefont {J.}~\bibnamefont
  {Yoneda}}, \bibinfo {author} {\bibfnamefont {K.}~\bibnamefont {Takeda}},
  \bibinfo {author} {\bibfnamefont {T.}~\bibnamefont {Otsuka}}, \bibinfo
  {author} {\bibfnamefont {T.}~\bibnamefont {Nakajima}}, \bibinfo {author}
  {\bibfnamefont {M.~R.}\ \bibnamefont {Delbecq}}, \bibinfo {author}
  {\bibfnamefont {G.}~\bibnamefont {Allison}}, \bibinfo {author} {\bibfnamefont
  {T.}~\bibnamefont {Honda}}, \bibinfo {author} {\bibfnamefont
  {T.}~\bibnamefont {Kodera}}, \bibinfo {author} {\bibfnamefont
  {S.}~\bibnamefont {Oda}}, \bibinfo {author} {\bibfnamefont {Y.}~\bibnamefont
  {Hoshi}}, \emph {et~al.},\ }\href@noop {} {\bibfield  {journal} {\bibinfo
  {journal} {Nature nanotechnology}\ }\textbf {\bibinfo {volume} {13}},\
  \bibinfo {pages} {102} (\bibinfo {year} {2018})}\BibitemShut {NoStop}%
\bibitem [{\citenamefont {Xue}\ \emph {et~al.}(2019)\citenamefont {Xue},
  \citenamefont {Watson}, \citenamefont {Helsen}, \citenamefont {Ward},
  \citenamefont {Savage}, \citenamefont {Lagally}, \citenamefont {Coppersmith},
  \citenamefont {Eriksson}, \citenamefont {Wehner},\ and\ \citenamefont
  {Vandersypen}}]{xue2019benchmarking}%
  \BibitemOpen
  \bibfield  {author} {\bibinfo {author} {\bibfnamefont {X.}~\bibnamefont
  {Xue}}, \bibinfo {author} {\bibfnamefont {T.}~\bibnamefont {Watson}},
  \bibinfo {author} {\bibfnamefont {J.}~\bibnamefont {Helsen}}, \bibinfo
  {author} {\bibfnamefont {D.~R.}\ \bibnamefont {Ward}}, \bibinfo {author}
  {\bibfnamefont {D.~E.}\ \bibnamefont {Savage}}, \bibinfo {author}
  {\bibfnamefont {M.~G.}\ \bibnamefont {Lagally}}, \bibinfo {author}
  {\bibfnamefont {S.~N.}\ \bibnamefont {Coppersmith}}, \bibinfo {author}
  {\bibfnamefont {M.}~\bibnamefont {Eriksson}}, \bibinfo {author}
  {\bibfnamefont {S.}~\bibnamefont {Wehner}},\ and\ \bibinfo {author}
  {\bibfnamefont {L.}~\bibnamefont {Vandersypen}},\ }\href@noop {} {\bibfield
  {journal} {\bibinfo  {journal} {Physical Review X}\ }\textbf {\bibinfo
  {volume} {9}},\ \bibinfo {pages} {021011} (\bibinfo {year}
  {2019})}\BibitemShut {NoStop}%
\bibitem [{\citenamefont {Huang}\ \emph {et~al.}(2019)\citenamefont {Huang},
  \citenamefont {Yang}, \citenamefont {Chan}, \citenamefont {Tanttu},
  \citenamefont {Hensen}, \citenamefont {Leon}, \citenamefont {Fogarty},
  \citenamefont {Hwang}, \citenamefont {Hudson}, \citenamefont {Itoh} \emph
  {et~al.}}]{huang2019fidelity}%
  \BibitemOpen
  \bibfield  {author} {\bibinfo {author} {\bibfnamefont {W.}~\bibnamefont
  {Huang}}, \bibinfo {author} {\bibfnamefont {C.}~\bibnamefont {Yang}},
  \bibinfo {author} {\bibfnamefont {K.}~\bibnamefont {Chan}}, \bibinfo {author}
  {\bibfnamefont {T.}~\bibnamefont {Tanttu}}, \bibinfo {author} {\bibfnamefont
  {B.}~\bibnamefont {Hensen}}, \bibinfo {author} {\bibfnamefont
  {R.}~\bibnamefont {Leon}}, \bibinfo {author} {\bibfnamefont {M.}~\bibnamefont
  {Fogarty}}, \bibinfo {author} {\bibfnamefont {J.}~\bibnamefont {Hwang}},
  \bibinfo {author} {\bibfnamefont {F.}~\bibnamefont {Hudson}}, \bibinfo
  {author} {\bibfnamefont {K.~M.}\ \bibnamefont {Itoh}}, \emph {et~al.},\
  }\href@noop {} {\bibfield  {journal} {\bibinfo  {journal} {Nature}\ }\textbf
  {\bibinfo {volume} {569}},\ \bibinfo {pages} {532} (\bibinfo {year}
  {2019})}\BibitemShut {NoStop}%
\bibitem [{\citenamefont {Sigillito}\ \emph {et~al.}(2019)\citenamefont
  {Sigillito}, \citenamefont {Gullans}, \citenamefont {Edge}, \citenamefont
  {Borselli},\ and\ \citenamefont {Petta}}]{sigillito2019coherent}%
  \BibitemOpen
  \bibfield  {author} {\bibinfo {author} {\bibfnamefont {A.}~\bibnamefont
  {Sigillito}}, \bibinfo {author} {\bibfnamefont {M.}~\bibnamefont {Gullans}},
  \bibinfo {author} {\bibfnamefont {L.}~\bibnamefont {Edge}}, \bibinfo {author}
  {\bibfnamefont {M.}~\bibnamefont {Borselli}},\ and\ \bibinfo {author}
  {\bibfnamefont {J.}~\bibnamefont {Petta}},\ }\href@noop {} {\bibfield
  {journal} {\bibinfo  {journal} {npj Quantum Information}\ }\textbf {\bibinfo
  {volume} {5}},\ \bibinfo {pages} {1} (\bibinfo {year} {2019})}\BibitemShut
  {NoStop}%
\bibitem [{\citenamefont {Nichol}\ \emph {et~al.}(2017)\citenamefont {Nichol},
  \citenamefont {Orona}, \citenamefont {Harvey}, \citenamefont {Fallahi},
  \citenamefont {Gardner}, \citenamefont {Manfra},\ and\ \citenamefont
  {Yacoby}}]{nichol2017high}%
  \BibitemOpen
  \bibfield  {author} {\bibinfo {author} {\bibfnamefont {J.~M.}\ \bibnamefont
  {Nichol}}, \bibinfo {author} {\bibfnamefont {L.~A.}\ \bibnamefont {Orona}},
  \bibinfo {author} {\bibfnamefont {S.~P.}\ \bibnamefont {Harvey}}, \bibinfo
  {author} {\bibfnamefont {S.}~\bibnamefont {Fallahi}}, \bibinfo {author}
  {\bibfnamefont {G.~C.}\ \bibnamefont {Gardner}}, \bibinfo {author}
  {\bibfnamefont {M.~J.}\ \bibnamefont {Manfra}},\ and\ \bibinfo {author}
  {\bibfnamefont {A.}~\bibnamefont {Yacoby}},\ }\href@noop {} {\bibfield
  {journal} {\bibinfo  {journal} {npj Quantum Information}\ }\textbf {\bibinfo
  {volume} {3}},\ \bibinfo {pages} {1} (\bibinfo {year} {2017})}\BibitemShut
  {NoStop}%
\bibitem [{\citenamefont {Cerfontaine}\ \emph {et~al.}(2019)\citenamefont
  {Cerfontaine}, \citenamefont {Botzem}, \citenamefont {Ritzmann},
  \citenamefont {Humpohl}, \citenamefont {Ludwig}, \citenamefont {Schuh},
  \citenamefont {Bougeard}, \citenamefont {Wieck},\ and\ \citenamefont
  {Bluhm}}]{cerfontaine2019closed}%
  \BibitemOpen
  \bibfield  {author} {\bibinfo {author} {\bibfnamefont {P.}~\bibnamefont
  {Cerfontaine}}, \bibinfo {author} {\bibfnamefont {T.}~\bibnamefont {Botzem}},
  \bibinfo {author} {\bibfnamefont {J.}~\bibnamefont {Ritzmann}}, \bibinfo
  {author} {\bibfnamefont {S.~S.}\ \bibnamefont {Humpohl}}, \bibinfo {author}
  {\bibfnamefont {A.}~\bibnamefont {Ludwig}}, \bibinfo {author} {\bibfnamefont
  {D.}~\bibnamefont {Schuh}}, \bibinfo {author} {\bibfnamefont
  {D.}~\bibnamefont {Bougeard}}, \bibinfo {author} {\bibfnamefont {A.~D.}\
  \bibnamefont {Wieck}},\ and\ \bibinfo {author} {\bibfnamefont
  {H.}~\bibnamefont {Bluhm}},\ }\href@noop {} {\bibfield  {journal} {\bibinfo
  {journal} {arXiv preprint arXiv:1906.06169}\ } (\bibinfo {year}
  {2019})}\BibitemShut {NoStop}%
\bibitem [{\citenamefont {Loss}\ and\ \citenamefont
  {DiVincenzo}(1998)}]{loss1998quantum}%
  \BibitemOpen
  \bibfield  {author} {\bibinfo {author} {\bibfnamefont {D.}~\bibnamefont
  {Loss}}\ and\ \bibinfo {author} {\bibfnamefont {D.~P.}\ \bibnamefont
  {DiVincenzo}},\ }\href@noop {} {\bibfield  {journal} {\bibinfo  {journal}
  {Physical Review A}\ }\textbf {\bibinfo {volume} {57}},\ \bibinfo {pages}
  {120} (\bibinfo {year} {1998})}\BibitemShut {NoStop}%
\bibitem [{\citenamefont {DiVincenzo}\ \emph {et~al.}(2000)\citenamefont
  {DiVincenzo}, \citenamefont {Bacon}, \citenamefont {Kempe}, \citenamefont
  {Burkard},\ and\ \citenamefont {Whaley}}]{divincenzo2000universal}%
  \BibitemOpen
  \bibfield  {author} {\bibinfo {author} {\bibfnamefont {D.~P.}\ \bibnamefont
  {DiVincenzo}}, \bibinfo {author} {\bibfnamefont {D.}~\bibnamefont {Bacon}},
  \bibinfo {author} {\bibfnamefont {J.}~\bibnamefont {Kempe}}, \bibinfo
  {author} {\bibfnamefont {G.}~\bibnamefont {Burkard}},\ and\ \bibinfo {author}
  {\bibfnamefont {K.~B.}\ \bibnamefont {Whaley}},\ }\href@noop {} {\bibfield
  {journal} {\bibinfo  {journal} {nature}\ }\textbf {\bibinfo {volume} {408}},\
  \bibinfo {pages} {339} (\bibinfo {year} {2000})}\BibitemShut {NoStop}%
\bibitem [{\citenamefont {Petta}\ \emph {et~al.}(2005)\citenamefont {Petta},
  \citenamefont {Johnson}, \citenamefont {Taylor}, \citenamefont {Laird},
  \citenamefont {Yacoby}, \citenamefont {Lukin}, \citenamefont {Marcus},
  \citenamefont {Hanson},\ and\ \citenamefont {Gossard}}]{petta2005coherent}%
  \BibitemOpen
  \bibfield  {author} {\bibinfo {author} {\bibfnamefont {J.~R.}\ \bibnamefont
  {Petta}}, \bibinfo {author} {\bibfnamefont {A.~C.}\ \bibnamefont {Johnson}},
  \bibinfo {author} {\bibfnamefont {J.~M.}\ \bibnamefont {Taylor}}, \bibinfo
  {author} {\bibfnamefont {E.~A.}\ \bibnamefont {Laird}}, \bibinfo {author}
  {\bibfnamefont {A.}~\bibnamefont {Yacoby}}, \bibinfo {author} {\bibfnamefont
  {M.~D.}\ \bibnamefont {Lukin}}, \bibinfo {author} {\bibfnamefont {C.~M.}\
  \bibnamefont {Marcus}}, \bibinfo {author} {\bibfnamefont {M.~P.}\
  \bibnamefont {Hanson}},\ and\ \bibinfo {author} {\bibfnamefont {A.~C.}\
  \bibnamefont {Gossard}},\ }\href@noop {} {\bibfield  {journal} {\bibinfo
  {journal} {Science}\ }\textbf {\bibinfo {volume} {309}},\ \bibinfo {pages}
  {2180} (\bibinfo {year} {2005})}\BibitemShut {NoStop}%
\bibitem [{\citenamefont {Nowack}\ \emph {et~al.}(2011)\citenamefont {Nowack},
  \citenamefont {Shafiei}, \citenamefont {Laforest}, \citenamefont
  {Prawiroatmodjo}, \citenamefont {Schreiber}, \citenamefont {Reichl},
  \citenamefont {Wegscheider},\ and\ \citenamefont
  {Vandersypen}}]{nowack2011single}%
  \BibitemOpen
  \bibfield  {author} {\bibinfo {author} {\bibfnamefont {K.}~\bibnamefont
  {Nowack}}, \bibinfo {author} {\bibfnamefont {M.}~\bibnamefont {Shafiei}},
  \bibinfo {author} {\bibfnamefont {M.}~\bibnamefont {Laforest}}, \bibinfo
  {author} {\bibfnamefont {G.}~\bibnamefont {Prawiroatmodjo}}, \bibinfo
  {author} {\bibfnamefont {L.}~\bibnamefont {Schreiber}}, \bibinfo {author}
  {\bibfnamefont {C.}~\bibnamefont {Reichl}}, \bibinfo {author} {\bibfnamefont
  {W.}~\bibnamefont {Wegscheider}},\ and\ \bibinfo {author} {\bibfnamefont
  {L.}~\bibnamefont {Vandersypen}},\ }\href@noop {} {\bibfield  {journal}
  {\bibinfo  {journal} {Science}\ }\textbf {\bibinfo {volume} {333}},\ \bibinfo
  {pages} {1269} (\bibinfo {year} {2011})}\BibitemShut {NoStop}%
\bibitem [{\citenamefont {Kandel}\ \emph {et~al.}(2019)\citenamefont {Kandel},
  \citenamefont {Qiao}, \citenamefont {Fallahi}, \citenamefont {Gardner},
  \citenamefont {Manfra},\ and\ \citenamefont {Nichol}}]{kandel2019coherent}%
  \BibitemOpen
  \bibfield  {author} {\bibinfo {author} {\bibfnamefont {Y.~P.}\ \bibnamefont
  {Kandel}}, \bibinfo {author} {\bibfnamefont {H.}~\bibnamefont {Qiao}},
  \bibinfo {author} {\bibfnamefont {S.}~\bibnamefont {Fallahi}}, \bibinfo
  {author} {\bibfnamefont {G.~C.}\ \bibnamefont {Gardner}}, \bibinfo {author}
  {\bibfnamefont {M.~J.}\ \bibnamefont {Manfra}},\ and\ \bibinfo {author}
  {\bibfnamefont {J.~M.}\ \bibnamefont {Nichol}},\ }\href@noop {} {\bibfield
  {journal} {\bibinfo  {journal} {Nature}\ }\textbf {\bibinfo {volume} {573}},\
  \bibinfo {pages} {553} (\bibinfo {year} {2019})}\BibitemShut {NoStop}%
\bibitem [{\citenamefont {Veldhorst}\ \emph {et~al.}(2015)\citenamefont
  {Veldhorst}, \citenamefont {Yang}, \citenamefont {Hwang}, \citenamefont
  {Huang}, \citenamefont {Dehollain}, \citenamefont {Muhonen}, \citenamefont
  {Simmons}, \citenamefont {Laucht}, \citenamefont {Hudson}, \citenamefont
  {Itoh} \emph {et~al.}}]{veldhorst2015two}%
  \BibitemOpen
  \bibfield  {author} {\bibinfo {author} {\bibfnamefont {M.}~\bibnamefont
  {Veldhorst}}, \bibinfo {author} {\bibfnamefont {C.}~\bibnamefont {Yang}},
  \bibinfo {author} {\bibfnamefont {J.}~\bibnamefont {Hwang}}, \bibinfo
  {author} {\bibfnamefont {W.}~\bibnamefont {Huang}}, \bibinfo {author}
  {\bibfnamefont {J.}~\bibnamefont {Dehollain}}, \bibinfo {author}
  {\bibfnamefont {J.}~\bibnamefont {Muhonen}}, \bibinfo {author} {\bibfnamefont
  {S.}~\bibnamefont {Simmons}}, \bibinfo {author} {\bibfnamefont
  {A.}~\bibnamefont {Laucht}}, \bibinfo {author} {\bibfnamefont
  {F.}~\bibnamefont {Hudson}}, \bibinfo {author} {\bibfnamefont {K.~M.}\
  \bibnamefont {Itoh}}, \emph {et~al.},\ }\href@noop {} {\bibfield  {journal}
  {\bibinfo  {journal} {Nature}\ }\textbf {\bibinfo {volume} {526}},\ \bibinfo
  {pages} {410} (\bibinfo {year} {2015})}\BibitemShut {NoStop}%
\bibitem [{\citenamefont {Dial}\ \emph {et~al.}(2013)\citenamefont {Dial},
  \citenamefont {Shulman}, \citenamefont {Harvey}, \citenamefont {Bluhm},
  \citenamefont {Umansky},\ and\ \citenamefont {Yacoby}}]{dial2013charge}%
  \BibitemOpen
  \bibfield  {author} {\bibinfo {author} {\bibfnamefont {O.}~\bibnamefont
  {Dial}}, \bibinfo {author} {\bibfnamefont {M.~D.}\ \bibnamefont {Shulman}},
  \bibinfo {author} {\bibfnamefont {S.~P.}\ \bibnamefont {Harvey}}, \bibinfo
  {author} {\bibfnamefont {H.}~\bibnamefont {Bluhm}}, \bibinfo {author}
  {\bibfnamefont {V.}~\bibnamefont {Umansky}},\ and\ \bibinfo {author}
  {\bibfnamefont {A.}~\bibnamefont {Yacoby}},\ }\href@noop {} {\bibfield
  {journal} {\bibinfo  {journal} {Physical review letters}\ }\textbf {\bibinfo
  {volume} {110}},\ \bibinfo {pages} {146804} (\bibinfo {year}
  {2013})}\BibitemShut {NoStop}%
\bibitem [{\citenamefont {Paladino}\ \emph {et~al.}(2014)\citenamefont
  {Paladino}, \citenamefont {Galperin}, \citenamefont {Falci},\ and\
  \citenamefont {Altshuler}}]{paladino20141}%
  \BibitemOpen
  \bibfield  {author} {\bibinfo {author} {\bibfnamefont {E.}~\bibnamefont
  {Paladino}}, \bibinfo {author} {\bibfnamefont {Y.}~\bibnamefont {Galperin}},
  \bibinfo {author} {\bibfnamefont {G.}~\bibnamefont {Falci}},\ and\ \bibinfo
  {author} {\bibfnamefont {B.}~\bibnamefont {Altshuler}},\ }\href@noop {}
  {\bibfield  {journal} {\bibinfo  {journal} {Reviews of Modern Physics}\
  }\textbf {\bibinfo {volume} {86}},\ \bibinfo {pages} {361} (\bibinfo {year}
  {2014})}\BibitemShut {NoStop}%
\bibitem [{\citenamefont {Korkusinski}\ \emph {et~al.}(2007)\citenamefont
  {Korkusinski}, \citenamefont {Gimenez}, \citenamefont {Hawrylak},
  \citenamefont {Gaudreau}, \citenamefont {Studenikin},\ and\ \citenamefont
  {Sachrajda}}]{korkusinski2007topological}%
  \BibitemOpen
  \bibfield  {author} {\bibinfo {author} {\bibfnamefont {M.}~\bibnamefont
  {Korkusinski}}, \bibinfo {author} {\bibfnamefont {I.~P.}\ \bibnamefont
  {Gimenez}}, \bibinfo {author} {\bibfnamefont {P.}~\bibnamefont {Hawrylak}},
  \bibinfo {author} {\bibfnamefont {L.}~\bibnamefont {Gaudreau}}, \bibinfo
  {author} {\bibfnamefont {S.~A.}\ \bibnamefont {Studenikin}},\ and\ \bibinfo
  {author} {\bibfnamefont {A.~S.}\ \bibnamefont {Sachrajda}},\ }\href@noop {}
  {\bibfield  {journal} {\bibinfo  {journal} {Physical Review B}\ }\textbf
  {\bibinfo {volume} {75}},\ \bibinfo {pages} {115301} (\bibinfo {year}
  {2007})}\BibitemShut {NoStop}%
\bibitem [{\citenamefont {Deng}\ \emph {et~al.}(2018)\citenamefont {Deng},
  \citenamefont {Calderon-Vargas}, \citenamefont {Mayhall},\ and\ \citenamefont
  {Barnes}}]{deng2018negative}%
  \BibitemOpen
  \bibfield  {author} {\bibinfo {author} {\bibfnamefont {K.}~\bibnamefont
  {Deng}}, \bibinfo {author} {\bibfnamefont {F.}~\bibnamefont
  {Calderon-Vargas}}, \bibinfo {author} {\bibfnamefont {N.~J.}\ \bibnamefont
  {Mayhall}},\ and\ \bibinfo {author} {\bibfnamefont {E.}~\bibnamefont
  {Barnes}},\ }\href@noop {} {\bibfield  {journal} {\bibinfo  {journal}
  {Physical Review B}\ }\textbf {\bibinfo {volume} {97}},\ \bibinfo {pages}
  {245301} (\bibinfo {year} {2018})}\BibitemShut {NoStop}%
\bibitem [{\citenamefont {Burkard}\ \emph {et~al.}(1999)\citenamefont
  {Burkard}, \citenamefont {Loss},\ and\ \citenamefont
  {DiVincenzo}}]{burkard1999coupled}%
  \BibitemOpen
  \bibfield  {author} {\bibinfo {author} {\bibfnamefont {G.}~\bibnamefont
  {Burkard}}, \bibinfo {author} {\bibfnamefont {D.}~\bibnamefont {Loss}},\ and\
  \bibinfo {author} {\bibfnamefont {D.~P.}\ \bibnamefont {DiVincenzo}},\
  }\href@noop {} {\bibfield  {journal} {\bibinfo  {journal} {Physical Review
  B}\ }\textbf {\bibinfo {volume} {59}},\ \bibinfo {pages} {2070} (\bibinfo
  {year} {1999})}\BibitemShut {NoStop}%
\bibitem [{\citenamefont {Calder{\'o}n}\ \emph {et~al.}(2006)\citenamefont
  {Calder{\'o}n}, \citenamefont {Koiller},\ and\ \citenamefont
  {Sarma}}]{calderon2006exchange}%
  \BibitemOpen
  \bibfield  {author} {\bibinfo {author} {\bibfnamefont {M.}~\bibnamefont
  {Calder{\'o}n}}, \bibinfo {author} {\bibfnamefont {B.}~\bibnamefont
  {Koiller}},\ and\ \bibinfo {author} {\bibfnamefont {S.~D.}\ \bibnamefont
  {Sarma}},\ }\href@noop {} {\bibfield  {journal} {\bibinfo  {journal}
  {Physical Review B}\ }\textbf {\bibinfo {volume} {74}},\ \bibinfo {pages}
  {045310} (\bibinfo {year} {2006})}\BibitemShut {NoStop}%
\bibitem [{\citenamefont {Hu}\ and\ \citenamefont
  {Sarma}(2000)}]{hu2000hilbert}%
  \BibitemOpen
  \bibfield  {author} {\bibinfo {author} {\bibfnamefont {X.}~\bibnamefont
  {Hu}}\ and\ \bibinfo {author} {\bibfnamefont {S.~D.}\ \bibnamefont {Sarma}},\
  }\href@noop {} {\bibfield  {journal} {\bibinfo  {journal} {Physical Review
  A}\ }\textbf {\bibinfo {volume} {61}},\ \bibinfo {pages} {062301} (\bibinfo
  {year} {2000})}\BibitemShut {NoStop}%
\bibitem [{\citenamefont {van~der Wiel}\ \emph {et~al.}(2006)\citenamefont
  {van~der Wiel}, \citenamefont {Stopa}, \citenamefont {Kodera}, \citenamefont
  {Hatano},\ and\ \citenamefont {Tarucha}}]{van2006semiconductor}%
  \BibitemOpen
  \bibfield  {author} {\bibinfo {author} {\bibfnamefont {W.}~\bibnamefont
  {van~der Wiel}}, \bibinfo {author} {\bibfnamefont {M.}~\bibnamefont {Stopa}},
  \bibinfo {author} {\bibfnamefont {T.}~\bibnamefont {Kodera}}, \bibinfo
  {author} {\bibfnamefont {T.}~\bibnamefont {Hatano}},\ and\ \bibinfo {author}
  {\bibfnamefont {S.}~\bibnamefont {Tarucha}},\ }\href@noop {} {\bibfield
  {journal} {\bibinfo  {journal} {New journal of physics}\ }\textbf {\bibinfo
  {volume} {8}},\ \bibinfo {pages} {28} (\bibinfo {year} {2006})}\BibitemShut
  {NoStop}%
\bibitem [{\citenamefont {Hatano}\ \emph {et~al.}(2008)\citenamefont {Hatano},
  \citenamefont {Amaha}, \citenamefont {Kubo}, \citenamefont {Tokura},
  \citenamefont {Nishi}, \citenamefont {Hirayama},\ and\ \citenamefont
  {Tarucha}}]{hatano2008manipulation}%
  \BibitemOpen
  \bibfield  {author} {\bibinfo {author} {\bibfnamefont {T.}~\bibnamefont
  {Hatano}}, \bibinfo {author} {\bibfnamefont {S.}~\bibnamefont {Amaha}},
  \bibinfo {author} {\bibfnamefont {T.}~\bibnamefont {Kubo}}, \bibinfo {author}
  {\bibfnamefont {Y.}~\bibnamefont {Tokura}}, \bibinfo {author} {\bibfnamefont
  {Y.}~\bibnamefont {Nishi}}, \bibinfo {author} {\bibfnamefont
  {Y.}~\bibnamefont {Hirayama}},\ and\ \bibinfo {author} {\bibfnamefont
  {S.}~\bibnamefont {Tarucha}},\ }\href@noop {} {\bibfield  {journal} {\bibinfo
   {journal} {Physical Review B}\ }\textbf {\bibinfo {volume} {77}},\ \bibinfo
  {pages} {241301} (\bibinfo {year} {2008})}\BibitemShut {NoStop}%
\bibitem [{\citenamefont {Pedersen}\ \emph {et~al.}(2007)\citenamefont
  {Pedersen}, \citenamefont {Flindt}, \citenamefont {Mortensen},\ and\
  \citenamefont {Jauho}}]{pedersen2007failure}%
  \BibitemOpen
  \bibfield  {author} {\bibinfo {author} {\bibfnamefont {J.}~\bibnamefont
  {Pedersen}}, \bibinfo {author} {\bibfnamefont {C.}~\bibnamefont {Flindt}},
  \bibinfo {author} {\bibfnamefont {N.~A.}\ \bibnamefont {Mortensen}},\ and\
  \bibinfo {author} {\bibfnamefont {A.-P.}\ \bibnamefont {Jauho}},\ }\href@noop
  {} {\bibfield  {journal} {\bibinfo  {journal} {Physical Review B}\ }\textbf
  {\bibinfo {volume} {76}},\ \bibinfo {pages} {125323} (\bibinfo {year}
  {2007})}\BibitemShut {NoStop}%
\bibitem [{\citenamefont {Li}\ \emph {et~al.}(2010)\citenamefont {Li},
  \citenamefont {Cywi{\'n}ski}, \citenamefont {Culcer}, \citenamefont {Hu},\
  and\ \citenamefont {Sarma}}]{li2010exchange}%
  \BibitemOpen
  \bibfield  {author} {\bibinfo {author} {\bibfnamefont {Q.}~\bibnamefont
  {Li}}, \bibinfo {author} {\bibfnamefont {{\L}.}~\bibnamefont {Cywi{\'n}ski}},
  \bibinfo {author} {\bibfnamefont {D.}~\bibnamefont {Culcer}}, \bibinfo
  {author} {\bibfnamefont {X.}~\bibnamefont {Hu}},\ and\ \bibinfo {author}
  {\bibfnamefont {S.~D.}\ \bibnamefont {Sarma}},\ }\href@noop {} {\bibfield
  {journal} {\bibinfo  {journal} {Physical Review B}\ }\textbf {\bibinfo
  {volume} {81}},\ \bibinfo {pages} {085313} (\bibinfo {year}
  {2010})}\BibitemShut {NoStop}%
\bibitem [{\citenamefont {Delgado}\ \emph {et~al.}(2007)\citenamefont
  {Delgado}, \citenamefont {Shim}, \citenamefont {Korkusinski},\ and\
  \citenamefont {Hawrylak}}]{delgado2007theory}%
  \BibitemOpen
  \bibfield  {author} {\bibinfo {author} {\bibfnamefont {F.}~\bibnamefont
  {Delgado}}, \bibinfo {author} {\bibfnamefont {Y.-P.}\ \bibnamefont {Shim}},
  \bibinfo {author} {\bibfnamefont {M.}~\bibnamefont {Korkusinski}},\ and\
  \bibinfo {author} {\bibfnamefont {P.}~\bibnamefont {Hawrylak}},\ }\href@noop
  {} {\bibfield  {journal} {\bibinfo  {journal} {Physical Review B}\ }\textbf
  {\bibinfo {volume} {76}},\ \bibinfo {pages} {115332} (\bibinfo {year}
  {2007})}\BibitemShut {NoStop}%
\bibitem [{\citenamefont {Shim}\ and\ \citenamefont
  {Hawrylak}(2008)}]{shim2008gate}%
  \BibitemOpen
  \bibfield  {author} {\bibinfo {author} {\bibfnamefont {Y.-P.}\ \bibnamefont
  {Shim}}\ and\ \bibinfo {author} {\bibfnamefont {P.}~\bibnamefont
  {Hawrylak}},\ }\href@noop {} {\bibfield  {journal} {\bibinfo  {journal}
  {Physical Review B}\ }\textbf {\bibinfo {volume} {78}},\ \bibinfo {pages}
  {165317} (\bibinfo {year} {2008})}\BibitemShut {NoStop}%
\bibitem [{\citenamefont {Hsieh}\ and\ \citenamefont
  {Hawrylak}(2010)}]{hsieh2010quantum}%
  \BibitemOpen
  \bibfield  {author} {\bibinfo {author} {\bibfnamefont {C.-Y.}\ \bibnamefont
  {Hsieh}}\ and\ \bibinfo {author} {\bibfnamefont {P.}~\bibnamefont
  {Hawrylak}},\ }\href@noop {} {\bibfield  {journal} {\bibinfo  {journal}
  {Physical Review B}\ }\textbf {\bibinfo {volume} {82}},\ \bibinfo {pages}
  {205311} (\bibinfo {year} {2010})}\BibitemShut {NoStop}%
\bibitem [{\citenamefont {Nielsen}\ \emph {et~al.}(2010)\citenamefont
  {Nielsen}, \citenamefont {Young}, \citenamefont {Muller},\ and\ \citenamefont
  {Carroll}}]{nielsen2010implications}%
  \BibitemOpen
  \bibfield  {author} {\bibinfo {author} {\bibfnamefont {E.}~\bibnamefont
  {Nielsen}}, \bibinfo {author} {\bibfnamefont {R.~W.}\ \bibnamefont {Young}},
  \bibinfo {author} {\bibfnamefont {R.~P.}\ \bibnamefont {Muller}},\ and\
  \bibinfo {author} {\bibfnamefont {M.}~\bibnamefont {Carroll}},\ }\href@noop
  {} {\bibfield  {journal} {\bibinfo  {journal} {Physical Review B}\ }\textbf
  {\bibinfo {volume} {82}},\ \bibinfo {pages} {075319} (\bibinfo {year}
  {2010})}\BibitemShut {NoStop}%
\bibitem [{\citenamefont {Barnes}\ \emph {et~al.}(2011)\citenamefont {Barnes},
  \citenamefont {Kestner}, \citenamefont {Nguyen},\ and\ \citenamefont
  {Sarma}}]{barnes2011screening}%
  \BibitemOpen
  \bibfield  {author} {\bibinfo {author} {\bibfnamefont {E.}~\bibnamefont
  {Barnes}}, \bibinfo {author} {\bibfnamefont {J.}~\bibnamefont {Kestner}},
  \bibinfo {author} {\bibfnamefont {N.}~\bibnamefont {Nguyen}},\ and\ \bibinfo
  {author} {\bibfnamefont {S.~D.}\ \bibnamefont {Sarma}},\ }\href@noop {}
  {\bibfield  {journal} {\bibinfo  {journal} {Physical Review B}\ }\textbf
  {\bibinfo {volume} {84}},\ \bibinfo {pages} {235309} (\bibinfo {year}
  {2011})}\BibitemShut {NoStop}%
\bibitem [{\citenamefont {Deng}\ and\ \citenamefont
  {Barnes}(2020)}]{deng2020interplay}%
  \BibitemOpen
  \bibfield  {author} {\bibinfo {author} {\bibfnamefont {K.}~\bibnamefont
  {Deng}}\ and\ \bibinfo {author} {\bibfnamefont {E.}~\bibnamefont {Barnes}},\
  }\href@noop {} {\bibfield  {journal} {\bibinfo  {journal} {arXiv preprint
  arXiv:2003.03416}\ } (\bibinfo {year} {2020})}\BibitemShut {NoStop}%
\bibitem [{\citenamefont {Buterakos}\ \emph {et~al.}(2018)\citenamefont
  {Buterakos}, \citenamefont {Throckmorton},\ and\ \citenamefont
  {Sarma}}]{buterakos2018crosstalk}%
  \BibitemOpen
  \bibfield  {author} {\bibinfo {author} {\bibfnamefont {D.}~\bibnamefont
  {Buterakos}}, \bibinfo {author} {\bibfnamefont {R.~E.}\ \bibnamefont
  {Throckmorton}},\ and\ \bibinfo {author} {\bibfnamefont {S.~D.}\ \bibnamefont
  {Sarma}},\ }\href@noop {} {\bibfield  {journal} {\bibinfo  {journal}
  {Physical Review B}\ }\textbf {\bibinfo {volume} {97}},\ \bibinfo {pages}
  {045431} (\bibinfo {year} {2018})}\BibitemShut {NoStop}%
\bibitem [{\citenamefont {Setiawan}\ \emph {et~al.}(2014)\citenamefont
  {Setiawan}, \citenamefont {Hui}, \citenamefont {Kestner}, \citenamefont
  {Wang},\ and\ \citenamefont {Sarma}}]{setiawan2014robust}%
  \BibitemOpen
  \bibfield  {author} {\bibinfo {author} {\bibfnamefont {F.}~\bibnamefont
  {Setiawan}}, \bibinfo {author} {\bibfnamefont {H.-Y.}\ \bibnamefont {Hui}},
  \bibinfo {author} {\bibfnamefont {J.}~\bibnamefont {Kestner}}, \bibinfo
  {author} {\bibfnamefont {X.}~\bibnamefont {Wang}},\ and\ \bibinfo {author}
  {\bibfnamefont {S.~D.}\ \bibnamefont {Sarma}},\ }\href@noop {} {\bibfield
  {journal} {\bibinfo  {journal} {Physical Review B}\ }\textbf {\bibinfo
  {volume} {89}},\ \bibinfo {pages} {085314} (\bibinfo {year}
  {2014})}\BibitemShut {NoStop}%
\bibitem [{\citenamefont {Wang}\ \emph {et~al.}(2014)\citenamefont {Wang},
  \citenamefont {Bishop}, \citenamefont {Barnes}, \citenamefont {Kestner},\
  and\ \citenamefont {Sarma}}]{wang2014robust}%
  \BibitemOpen
  \bibfield  {author} {\bibinfo {author} {\bibfnamefont {X.}~\bibnamefont
  {Wang}}, \bibinfo {author} {\bibfnamefont {L.~S.}\ \bibnamefont {Bishop}},
  \bibinfo {author} {\bibfnamefont {E.}~\bibnamefont {Barnes}}, \bibinfo
  {author} {\bibfnamefont {J.}~\bibnamefont {Kestner}},\ and\ \bibinfo {author}
  {\bibfnamefont {S.~D.}\ \bibnamefont {Sarma}},\ }\href@noop {} {\bibfield
  {journal} {\bibinfo  {journal} {Physical Review A}\ }\textbf {\bibinfo
  {volume} {89}},\ \bibinfo {pages} {022310} (\bibinfo {year}
  {2014})}\BibitemShut {NoStop}%
\bibitem [{\citenamefont {Zhang}\ \emph {et~al.}(2017)\citenamefont {Zhang},
  \citenamefont {Throckmorton}, \citenamefont {Yang}, \citenamefont {Wang},
  \citenamefont {Barnes},\ and\ \citenamefont {Sarma}}]{zhang2017randomized}%
  \BibitemOpen
  \bibfield  {author} {\bibinfo {author} {\bibfnamefont {C.}~\bibnamefont
  {Zhang}}, \bibinfo {author} {\bibfnamefont {R.~E.}\ \bibnamefont
  {Throckmorton}}, \bibinfo {author} {\bibfnamefont {X.-C.}\ \bibnamefont
  {Yang}}, \bibinfo {author} {\bibfnamefont {X.}~\bibnamefont {Wang}}, \bibinfo
  {author} {\bibfnamefont {E.}~\bibnamefont {Barnes}},\ and\ \bibinfo {author}
  {\bibfnamefont {S.~D.}\ \bibnamefont {Sarma}},\ }\href@noop {} {\bibfield
  {journal} {\bibinfo  {journal} {Physical Review Letters}\ }\textbf {\bibinfo
  {volume} {118}},\ \bibinfo {pages} {216802} (\bibinfo {year}
  {2017})}\BibitemShut {NoStop}%
\bibitem [{\citenamefont {Reed}\ \emph {et~al.}(2016)\citenamefont {Reed},
  \citenamefont {Maune}, \citenamefont {Andrews}, \citenamefont {Borselli},
  \citenamefont {Eng}, \citenamefont {Jura}, \citenamefont {Kiselev},
  \citenamefont {Ladd}, \citenamefont {Merkel}, \citenamefont {Milosavljevic}
  \emph {et~al.}}]{reed2016reduced}%
  \BibitemOpen
  \bibfield  {author} {\bibinfo {author} {\bibfnamefont {M.}~\bibnamefont
  {Reed}}, \bibinfo {author} {\bibfnamefont {B.}~\bibnamefont {Maune}},
  \bibinfo {author} {\bibfnamefont {R.}~\bibnamefont {Andrews}}, \bibinfo
  {author} {\bibfnamefont {M.}~\bibnamefont {Borselli}}, \bibinfo {author}
  {\bibfnamefont {K.}~\bibnamefont {Eng}}, \bibinfo {author} {\bibfnamefont
  {M.}~\bibnamefont {Jura}}, \bibinfo {author} {\bibfnamefont {A.}~\bibnamefont
  {Kiselev}}, \bibinfo {author} {\bibfnamefont {T.}~\bibnamefont {Ladd}},
  \bibinfo {author} {\bibfnamefont {S.}~\bibnamefont {Merkel}}, \bibinfo
  {author} {\bibfnamefont {I.}~\bibnamefont {Milosavljevic}}, \emph {et~al.},\
  }\href@noop {} {\bibfield  {journal} {\bibinfo  {journal} {Physical review
  letters}\ }\textbf {\bibinfo {volume} {116}},\ \bibinfo {pages} {110402}
  (\bibinfo {year} {2016})}\BibitemShut {NoStop}%
\bibitem [{\citenamefont {Martins}\ \emph {et~al.}(2016)\citenamefont
  {Martins}, \citenamefont {Malinowski}, \citenamefont {Nissen}, \citenamefont
  {Barnes}, \citenamefont {Fallahi}, \citenamefont {Gardner}, \citenamefont
  {Manfra}, \citenamefont {Marcus},\ and\ \citenamefont
  {Kuemmeth}}]{martins2016noise}%
  \BibitemOpen
  \bibfield  {author} {\bibinfo {author} {\bibfnamefont {F.}~\bibnamefont
  {Martins}}, \bibinfo {author} {\bibfnamefont {F.~K.}\ \bibnamefont
  {Malinowski}}, \bibinfo {author} {\bibfnamefont {P.~D.}\ \bibnamefont
  {Nissen}}, \bibinfo {author} {\bibfnamefont {E.}~\bibnamefont {Barnes}},
  \bibinfo {author} {\bibfnamefont {S.}~\bibnamefont {Fallahi}}, \bibinfo
  {author} {\bibfnamefont {G.~C.}\ \bibnamefont {Gardner}}, \bibinfo {author}
  {\bibfnamefont {M.~J.}\ \bibnamefont {Manfra}}, \bibinfo {author}
  {\bibfnamefont {C.~M.}\ \bibnamefont {Marcus}},\ and\ \bibinfo {author}
  {\bibfnamefont {F.}~\bibnamefont {Kuemmeth}},\ }\href@noop {} {\bibfield
  {journal} {\bibinfo  {journal} {Physical review letters}\ }\textbf {\bibinfo
  {volume} {116}},\ \bibinfo {pages} {116801} (\bibinfo {year}
  {2016})}\BibitemShut {NoStop}%
\bibitem [{\citenamefont {Yang}\ and\ \citenamefont
  {Wang}(2017)}]{yang2017suppression}%
  \BibitemOpen
  \bibfield  {author} {\bibinfo {author} {\bibfnamefont {X.-C.}\ \bibnamefont
  {Yang}}\ and\ \bibinfo {author} {\bibfnamefont {X.}~\bibnamefont {Wang}},\
  }\href@noop {} {\bibfield  {journal} {\bibinfo  {journal} {Physical Review
  A}\ }\textbf {\bibinfo {volume} {96}},\ \bibinfo {pages} {012318} (\bibinfo
  {year} {2017})}\BibitemShut {NoStop}%
\bibitem [{\citenamefont {Khaneja}\ \emph {et~al.}(2005)\citenamefont
  {Khaneja}, \citenamefont {Reiss}, \citenamefont {Kehlet}, \citenamefont
  {Schulte-Herbr{\"u}ggen},\ and\ \citenamefont {Glaser}}]{khaneja2005optimal}%
  \BibitemOpen
  \bibfield  {author} {\bibinfo {author} {\bibfnamefont {N.}~\bibnamefont
  {Khaneja}}, \bibinfo {author} {\bibfnamefont {T.}~\bibnamefont {Reiss}},
  \bibinfo {author} {\bibfnamefont {C.}~\bibnamefont {Kehlet}}, \bibinfo
  {author} {\bibfnamefont {T.}~\bibnamefont {Schulte-Herbr{\"u}ggen}},\ and\
  \bibinfo {author} {\bibfnamefont {S.~J.}\ \bibnamefont {Glaser}},\
  }\href@noop {} {\bibfield  {journal} {\bibinfo  {journal} {Journal of
  magnetic resonance}\ }\textbf {\bibinfo {volume} {172}},\ \bibinfo {pages}
  {296} (\bibinfo {year} {2005})}\BibitemShut {NoStop}%
\bibitem [{\citenamefont {Yang}\ \emph {et~al.}(2019)\citenamefont {Yang},
  \citenamefont {Chan}, \citenamefont {Harper}, \citenamefont {Huang},
  \citenamefont {Evans}, \citenamefont {Hwang}, \citenamefont {Hensen},
  \citenamefont {Laucht}, \citenamefont {Tanttu}, \citenamefont {Hudson} \emph
  {et~al.}}]{yang2019silicon}%
  \BibitemOpen
  \bibfield  {author} {\bibinfo {author} {\bibfnamefont {C.}~\bibnamefont
  {Yang}}, \bibinfo {author} {\bibfnamefont {K.}~\bibnamefont {Chan}}, \bibinfo
  {author} {\bibfnamefont {R.}~\bibnamefont {Harper}}, \bibinfo {author}
  {\bibfnamefont {W.}~\bibnamefont {Huang}}, \bibinfo {author} {\bibfnamefont
  {T.}~\bibnamefont {Evans}}, \bibinfo {author} {\bibfnamefont
  {J.}~\bibnamefont {Hwang}}, \bibinfo {author} {\bibfnamefont
  {B.}~\bibnamefont {Hensen}}, \bibinfo {author} {\bibfnamefont
  {A.}~\bibnamefont {Laucht}}, \bibinfo {author} {\bibfnamefont
  {T.}~\bibnamefont {Tanttu}}, \bibinfo {author} {\bibfnamefont
  {F.}~\bibnamefont {Hudson}}, \emph {et~al.},\ }\href@noop {} {\bibfield
  {journal} {\bibinfo  {journal} {Nature Electronics}\ }\textbf {\bibinfo
  {volume} {2}},\ \bibinfo {pages} {151} (\bibinfo {year} {2019})}\BibitemShut
  {NoStop}%
\bibitem [{\citenamefont {Haas}\ \emph {et~al.}(2019)\citenamefont {Haas},
  \citenamefont {Puzzuoli}, \citenamefont {Zhang},\ and\ \citenamefont
  {Cory}}]{haas2019engineering}%
  \BibitemOpen
  \bibfield  {author} {\bibinfo {author} {\bibfnamefont {H.}~\bibnamefont
  {Haas}}, \bibinfo {author} {\bibfnamefont {D.}~\bibnamefont {Puzzuoli}},
  \bibinfo {author} {\bibfnamefont {F.}~\bibnamefont {Zhang}},\ and\ \bibinfo
  {author} {\bibfnamefont {D.~G.}\ \bibnamefont {Cory}},\ }\href@noop {}
  {\bibfield  {journal} {\bibinfo  {journal} {New Journal of Physics}\ }\textbf
  {\bibinfo {volume} {21}},\ \bibinfo {pages} {103011} (\bibinfo {year}
  {2019})}\BibitemShut {NoStop}%
\bibitem [{\citenamefont {Gimenez}\ \emph {et~al.}(2007)\citenamefont
  {Gimenez}, \citenamefont {Korkusinski},\ and\ \citenamefont
  {Hawrylak}}]{gimenez2007linear}%
  \BibitemOpen
  \bibfield  {author} {\bibinfo {author} {\bibfnamefont {I.~P.}\ \bibnamefont
  {Gimenez}}, \bibinfo {author} {\bibfnamefont {M.}~\bibnamefont
  {Korkusinski}},\ and\ \bibinfo {author} {\bibfnamefont {P.}~\bibnamefont
  {Hawrylak}},\ }\href@noop {} {\bibfield  {journal} {\bibinfo  {journal}
  {Physical Review B}\ }\textbf {\bibinfo {volume} {76}},\ \bibinfo {pages}
  {075336} (\bibinfo {year} {2007})}\BibitemShut {NoStop}%
\bibitem [{\citenamefont {Hawrylak}(1993)}]{hawrylak1993far}%
  \BibitemOpen
  \bibfield  {author} {\bibinfo {author} {\bibfnamefont {P.}~\bibnamefont
  {Hawrylak}},\ }\href@noop {} {\bibfield  {journal} {\bibinfo  {journal}
  {Solid state communications}\ }\textbf {\bibinfo {volume} {88}},\ \bibinfo
  {pages} {475} (\bibinfo {year} {1993})}\BibitemShut {NoStop}%
\bibitem [{\citenamefont {Kyriakidis}\ \emph {et~al.}(2002)\citenamefont
  {Kyriakidis}, \citenamefont {Pioro-Ladriere}, \citenamefont {Ciorga},
  \citenamefont {Sachrajda},\ and\ \citenamefont
  {Hawrylak}}]{kyriakidis2002voltage}%
  \BibitemOpen
  \bibfield  {author} {\bibinfo {author} {\bibfnamefont {J.}~\bibnamefont
  {Kyriakidis}}, \bibinfo {author} {\bibfnamefont {M.}~\bibnamefont
  {Pioro-Ladriere}}, \bibinfo {author} {\bibfnamefont {M.}~\bibnamefont
  {Ciorga}}, \bibinfo {author} {\bibfnamefont {A.}~\bibnamefont {Sachrajda}},\
  and\ \bibinfo {author} {\bibfnamefont {P.}~\bibnamefont {Hawrylak}},\
  }\href@noop {} {\bibfield  {journal} {\bibinfo  {journal} {Physical Review
  B}\ }\textbf {\bibinfo {volume} {66}},\ \bibinfo {pages} {035320} (\bibinfo
  {year} {2002})}\BibitemShut {NoStop}%
\bibitem [{\citenamefont {Lieb}\ and\ \citenamefont
  {Mattis}(1962)}]{lieb1962ordering}%
  \BibitemOpen
  \bibfield  {author} {\bibinfo {author} {\bibfnamefont {E.}~\bibnamefont
  {Lieb}}\ and\ \bibinfo {author} {\bibfnamefont {D.}~\bibnamefont {Mattis}},\
  }\href@noop {} {\bibfield  {journal} {\bibinfo  {journal} {Journal of
  Mathematical Physics}\ }\textbf {\bibinfo {volume} {3}},\ \bibinfo {pages}
  {749} (\bibinfo {year} {1962})}\BibitemShut {NoStop}%
\bibitem [{\citenamefont {Fletcher}(2013)}]{fletcher2013practical}%
  \BibitemOpen
  \bibfield  {author} {\bibinfo {author} {\bibfnamefont {R.}~\bibnamefont
  {Fletcher}},\ }\href@noop {} {\emph {\bibinfo {title} {Practical methods of
  optimization}}}\ (\bibinfo  {publisher} {John Wiley \& Sons},\ \bibinfo
  {year} {2013})\BibitemShut {NoStop}%
\bibitem [{\citenamefont {Culcer}\ \emph {et~al.}(2010)\citenamefont {Culcer},
  \citenamefont {Cywi{\'n}ski}, \citenamefont {Li}, \citenamefont {Hu},\ and\
  \citenamefont {Sarma}}]{culcer2010quantum}%
  \BibitemOpen
  \bibfield  {author} {\bibinfo {author} {\bibfnamefont {D.}~\bibnamefont
  {Culcer}}, \bibinfo {author} {\bibfnamefont {{\L}.}~\bibnamefont
  {Cywi{\'n}ski}}, \bibinfo {author} {\bibfnamefont {Q.}~\bibnamefont {Li}},
  \bibinfo {author} {\bibfnamefont {X.}~\bibnamefont {Hu}},\ and\ \bibinfo
  {author} {\bibfnamefont {S.~D.}\ \bibnamefont {Sarma}},\ }\href@noop {}
  {\bibfield  {journal} {\bibinfo  {journal} {Physical Review B}\ }\textbf
  {\bibinfo {volume} {82}},\ \bibinfo {pages} {155312} (\bibinfo {year}
  {2010})}\BibitemShut {NoStop}%
\bibitem [{\citenamefont {Tanttu}\ \emph {et~al.}(2019)\citenamefont {Tanttu},
  \citenamefont {Hensen}, \citenamefont {Chan}, \citenamefont {Yang},
  \citenamefont {Huang}, \citenamefont {Fogarty}, \citenamefont {Hudson},
  \citenamefont {Itoh}, \citenamefont {Culcer}, \citenamefont {Laucht} \emph
  {et~al.}}]{tanttu2019controlling}%
  \BibitemOpen
  \bibfield  {author} {\bibinfo {author} {\bibfnamefont {T.}~\bibnamefont
  {Tanttu}}, \bibinfo {author} {\bibfnamefont {B.}~\bibnamefont {Hensen}},
  \bibinfo {author} {\bibfnamefont {K.~W.}\ \bibnamefont {Chan}}, \bibinfo
  {author} {\bibfnamefont {C.~H.}\ \bibnamefont {Yang}}, \bibinfo {author}
  {\bibfnamefont {W.~W.}\ \bibnamefont {Huang}}, \bibinfo {author}
  {\bibfnamefont {M.}~\bibnamefont {Fogarty}}, \bibinfo {author} {\bibfnamefont
  {F.}~\bibnamefont {Hudson}}, \bibinfo {author} {\bibfnamefont
  {K.}~\bibnamefont {Itoh}}, \bibinfo {author} {\bibfnamefont {D.}~\bibnamefont
  {Culcer}}, \bibinfo {author} {\bibfnamefont {A.}~\bibnamefont {Laucht}},
  \emph {et~al.},\ }\href@noop {} {\bibfield  {journal} {\bibinfo  {journal}
  {Physical Review X}\ }\textbf {\bibinfo {volume} {9}},\ \bibinfo {pages}
  {021028} (\bibinfo {year} {2019})}\BibitemShut {NoStop}%
\bibitem [{\citenamefont {Birner}\ \emph {et~al.}(2007)\citenamefont {Birner},
  \citenamefont {Zibold}, \citenamefont {Andlauer}, \citenamefont {Kubis},
  \citenamefont {Sabathil}, \citenamefont {Trellakis},\ and\ \citenamefont
  {Vogl}}]{birner2007nextnano}%
  \BibitemOpen
  \bibfield  {author} {\bibinfo {author} {\bibfnamefont {S.}~\bibnamefont
  {Birner}}, \bibinfo {author} {\bibfnamefont {T.}~\bibnamefont {Zibold}},
  \bibinfo {author} {\bibfnamefont {T.}~\bibnamefont {Andlauer}}, \bibinfo
  {author} {\bibfnamefont {T.}~\bibnamefont {Kubis}}, \bibinfo {author}
  {\bibfnamefont {M.}~\bibnamefont {Sabathil}}, \bibinfo {author}
  {\bibfnamefont {A.}~\bibnamefont {Trellakis}},\ and\ \bibinfo {author}
  {\bibfnamefont {P.}~\bibnamefont {Vogl}},\ }\href@noop {} {\bibfield
  {journal} {\bibinfo  {journal} {IEEE Transactions on Electron Devices}\
  }\textbf {\bibinfo {volume} {54}},\ \bibinfo {pages} {2137} (\bibinfo {year}
  {2007})}\BibitemShut {NoStop}%
\bibitem [{\citenamefont {Buonacorsi}\ \emph {et~al.}(2020)\citenamefont
  {Buonacorsi}, \citenamefont {Shaw},\ and\ \citenamefont
  {Baugh}}]{buonacorsi2020simulated}%
  \BibitemOpen
  \bibfield  {author} {\bibinfo {author} {\bibfnamefont {B.}~\bibnamefont
  {Buonacorsi}}, \bibinfo {author} {\bibfnamefont {B.}~\bibnamefont {Shaw}},\
  and\ \bibinfo {author} {\bibfnamefont {J.}~\bibnamefont {Baugh}},\
  }\href@noop {} {\bibfield  {journal} {\bibinfo  {journal} {Physical Review
  B}\ }\textbf {\bibinfo {volume} {102}},\ \bibinfo {pages} {125406} (\bibinfo
  {year} {2020})}\BibitemShut {NoStop}%
\bibitem [{\citenamefont {Ramon}\ and\ \citenamefont
  {Hu}(2010)}]{ramon2010decoherence}%
  \BibitemOpen
  \bibfield  {author} {\bibinfo {author} {\bibfnamefont {G.}~\bibnamefont
  {Ramon}}\ and\ \bibinfo {author} {\bibfnamefont {X.}~\bibnamefont {Hu}},\
  }\href@noop {} {\bibfield  {journal} {\bibinfo  {journal} {Physical Review
  B}\ }\textbf {\bibinfo {volume} {81}},\ \bibinfo {pages} {045304} (\bibinfo
  {year} {2010})}\BibitemShut {NoStop}%
\bibitem [{\citenamefont {Kouwenhoven}\ \emph {et~al.}(1997)\citenamefont
  {Kouwenhoven}, \citenamefont {Marcus}, \citenamefont {McEuen}, \citenamefont
  {Tarucha}, \citenamefont {Westervelt},\ and\ \citenamefont
  {Wingreen}}]{kouwenhoven1997electron}%
  \BibitemOpen
  \bibfield  {author} {\bibinfo {author} {\bibfnamefont {L.~P.}\ \bibnamefont
  {Kouwenhoven}}, \bibinfo {author} {\bibfnamefont {C.~M.}\ \bibnamefont
  {Marcus}}, \bibinfo {author} {\bibfnamefont {P.~L.}\ \bibnamefont {McEuen}},
  \bibinfo {author} {\bibfnamefont {S.}~\bibnamefont {Tarucha}}, \bibinfo
  {author} {\bibfnamefont {R.~M.}\ \bibnamefont {Westervelt}},\ and\ \bibinfo
  {author} {\bibfnamefont {N.~S.}\ \bibnamefont {Wingreen}},\ }in\ \href@noop
  {} {\emph {\bibinfo {booktitle} {Mesoscopic electron transport}}}\ (\bibinfo
  {publisher} {Springer},\ \bibinfo {year} {1997})\ pp.\ \bibinfo {pages}
  {105--214}\BibitemShut {NoStop}%
\bibitem [{\citenamefont {Hiltunen}\ \emph {et~al.}(2015)\citenamefont
  {Hiltunen}, \citenamefont {Bluhm}, \citenamefont {Mehl},\ and\ \citenamefont
  {Harju}}]{hiltunen2015charge}%
  \BibitemOpen
  \bibfield  {author} {\bibinfo {author} {\bibfnamefont {T.}~\bibnamefont
  {Hiltunen}}, \bibinfo {author} {\bibfnamefont {H.}~\bibnamefont {Bluhm}},
  \bibinfo {author} {\bibfnamefont {S.}~\bibnamefont {Mehl}},\ and\ \bibinfo
  {author} {\bibfnamefont {A.}~\bibnamefont {Harju}},\ }\href@noop {}
  {\bibfield  {journal} {\bibinfo  {journal} {Physical Review B}\ }\textbf
  {\bibinfo {volume} {91}},\ \bibinfo {pages} {075301} (\bibinfo {year}
  {2015})}\BibitemShut {NoStop}%
\bibitem [{\citenamefont {Korkusinski}(2004)}]{korkusinski2004correlations}%
  \BibitemOpen
  \bibfield  {author} {\bibinfo {author} {\bibfnamefont {M.}~\bibnamefont
  {Korkusinski}},\ }\href@noop {} {\emph {\bibinfo {title} {Correlations in
  semiconductor quantum dots}}}\ (\bibinfo  {publisher} {University of
  Ottawa},\ \bibinfo {year} {2004})\BibitemShut {NoStop}%
\bibitem [{\citenamefont {Sarma}\ \emph {et~al.}(2011)\citenamefont {Sarma},
  \citenamefont {Wang},\ and\ \citenamefont {Yang}}]{sarma2011hubbard}%
  \BibitemOpen
  \bibfield  {author} {\bibinfo {author} {\bibfnamefont {S.~D.}\ \bibnamefont
  {Sarma}}, \bibinfo {author} {\bibfnamefont {X.}~\bibnamefont {Wang}},\ and\
  \bibinfo {author} {\bibfnamefont {S.}~\bibnamefont {Yang}},\ }\href@noop {}
  {\bibfield  {journal} {\bibinfo  {journal} {Physical Review B}\ }\textbf
  {\bibinfo {volume} {83}},\ \bibinfo {pages} {235314} (\bibinfo {year}
  {2011})}\BibitemShut {NoStop}%
\end{thebibliography}
\end{document}